\documentclass[11pt,a4paper]{article}
\usepackage{jheppub}
\usepackage{microtype}

\usepackage{booktabs}
\usepackage{amsmath}
\usepackage{amssymb}
\usepackage{bbold}
\usepackage[flushleft]{threeparttable}
\usepackage{lscape}
\usepackage{rotating}
\usepackage{upgreek}
\usepackage[english]{babel}
\usepackage{multirow}
\usepackage[section]{placeins}
\usepackage[dvipsnames]{xcolor}
\definecolor{linkcolor}{rgb}{.17578125,.1875,.5703125}
\usepackage{hyperref}
\hypersetup{
    colorlinks=true,
    linkcolor=linkcolor,
    filecolor=magenta,      
    urlcolor=linkcolor,
    citecolor=Maroon
}
\usepackage{braket}
\usepackage{slashed}
\usepackage[toc]{appendix}
\usepackage{float}
\usepackage{caption}
\usepackage{subcaption}
\usepackage[titles]{tocloft}
\usepackage{graphicx}
\graphicspath{{./figures/}}

\newcommand{\vus}{V_\mathrm{us}}
\newcommand{\vud}{V_\mathrm{ud}}
\newcommand{\nn}{\nonumber}
\newcommand{\order}{\mathrm{O}}
\newcommand{\aem}{\alpha_\mathrm{em}}
\newcommand{\Nf}{N_\mathrm{f}}

\newcommand{\inclrate}[2]{\Gamma( #1^+ \rightarrow #2^+\nu_{#2}[\gamma])}
\newcommand{\dRKPi}{\delta R_{K\pi}}
\newcommand{\mell}{m_{\ell}}
\newcommand{\vell}{\mathbf{v}_{\ell}}
\newcommand{\pslash}{\slashed{p}}
\newcommand{\rl}[1][2]{r^{#1}_\ell}
\newcommand{\QCD}{\mathrm{QCD}}
\newcommand{\QED}{\mathrm{QED}}
\newcommand{\QEDL}{\QED_\mathrm{L}}
\newcommand{\chipt}{\chi\text{PT}}
\newcommand{\e}{\mathrm{e}}
\newcommand{\ii}{\mathrm{i}}
\newcommand{\rmu}{\mathrm{u}}
\newcommand{\rmd}{\mathrm{d}}
\newcommand{\rmud}{\mathrm{ud}}
\newcommand{\rms}{\mathrm{s}}
\newcommand{\Mcal}{\mathcal{M}}
\newcommand{\Acal}{\mathcal{A}}
\newcommand{\Lcal}{\mathcal{L}}
\newcommand{\Scal}{\mathcal{S}}
\newcommand{\mvec}{\mathbf{m}}
\newcommand{\pvec}{\mathbf{p}}
\newcommand{\kvec}{\mathbf{k}}
\newcommand{\Mvec}{\mathbf{M}}
\newcommand{\sigmavec}{\boldsymbol{\sigma}}
\newcommand{\xvec}{\mathbf{x}}
\newcommand{\yvec}{\mathbf{y}}
\newcommand{\iso}{{\textrm{\tiny (0)}}}
\newcommand{\qcd}{{\textrm{\tiny QCD}}}
\newcommand{\bmw}{{\textrm{\tiny BMW}}}
\newcommand{\pdg}{{\textrm{\tiny PDG}}}
\newcommand{\diff}{\mathrm{d}}
\newcommand{\dd}[2]{\diff^{#1}{#2}\,}

\usepackage{cleveref}
\crefformat{section}{section~#2#1#3}
\crefmultiformat{section}{sections~#2#1#3}
    { and~#2#1#3}{, #2#1#3}{ and~#2#1#3}
\crefrangeformat{section}{sections~#3#1#4\,--\,#5#2#6}
\crefformat{subsection}{section~#2#1#3}
\crefmultiformat{subsection}{sections~#2#1#3}
    { and~#2#1#3}{, #2#1#3}{ and~#2#1#3}
\crefrangeformat{subsection}{sections~#3#1#4\,--\,#5#2#6}
\crefformat{subsubsection}{section~#2#1#3}
\crefmultiformat{subsubsection}{sections~#2#1#3}
    { and~#2#1#3}{, #2#1#3}{ and~#2#1#3}
\crefrangeformat{subsubsection}{sections~#3#1#4\,--\,#5#2#6}
\crefformat{figure}{figure~#2#1#3}
\crefmultiformat{figure}{figures~#2#1#3}
    { and~#2#1#3}{, #2#1#3}{ and~#2#1#3}
\crefrangeformat{figure}{figures~#3#1#4\,--\,#5#2#6}
\crefformat{table}{table~#2#1#3}
\crefmultiformat{table}{tables~#2#1#3}
    { and~#2#1#3}{, #2#1#3}{ and~#2#1#3}
\crefrangeformat{table}{tables~#3#1#4\,--\,#5#2#6}
\crefformat{equation}{eq.~(#2#1#3)}
\crefmultiformat{equation}{eqs.~(#2#1#3)}{ and~(#2#1#3)}{, (#2#1#3)}{ and~(#2#1#3)}
\crefrangeformat{equation}{eqs.~(#3#1#4)--(#5#2#6)}

\DeclareFontFamily{U}{mathx}{\hyphenchar\font45}
\DeclareFontShape{U}{mathx}{m}{n}{<-> mathx10}{}
\DeclareSymbolFont{mathx}{U}{mathx}{m}{n}
\DeclareMathAccent{\widebar}{0}{mathx}{"73}

\newcommand{\uoe}{School of Physics and Astronomy, University of Edinburgh, Edinburgh EH9 3FD, United Kingdom}
\newcommand{\bern}{Albert Einstein Center for Fundamental Physics, Institute for Theoretical Physics, Universit\"at Bern,  Sidlerstrasse 5, CH-3012 Bern, Switzerland}
\newcommand{\cern}{CERN, Theoretical Physics Department, CH-1211 Geneva, Switzerland}
\newcommand{\bnl}{Physics Department, Brookhaven National Laboratory, Upton NY 11973, USA}
\newcommand{\lund}{Department of Astronomy and Theoretical Physics, Lund University, S\"olvegatan 14A, 223 62 Lund, Sweden}
\newcommand{\uos}{Physics and Astronomy, University of Southampton, Southampton SO17 1BJ, United Kingdom}
\newcommand{\epcc}{EPCC, University of Edinburgh, EH8 9BT, Edinburgh, United Kingdom}

\title{Isospin-breaking corrections to light-meson leptonic decays from lattice simulations at physical quark masses}

\author[a,b]{Peter Boyle,}
\author[b]{Matteo Di Carlo,}
\author[b]{Felix Erben,}
\author[b]{Vera G\"ulpers,}
\author[b]{Maxwell T. Hansen,}
\author[b]{Tim Harris,}
\author[c,d]{Nils Hermansson-Truedsson,}
\author[b]{Raoul Hodgson,}
\author[e,f]{Andreas J\"uttner,}
\author[b]{Fionn \'O h\'Og\'ain,}
\author[b]{Antonin Portelli,}
\author[b,e,g]{James Richings,}
\author[b]{and Andrew Zhen Ning Yong}

\affiliation[a]{\bnl}
\affiliation[b]{\uoe} 
\affiliation[c]{\bern}
\affiliation[d]{\lund}
\affiliation[e]{\uos} 
\affiliation[f]{\cern}
\affiliation[g]{\epcc}

\emailAdd{pboyle@bnl.gov}
\emailAdd{matteo.dicarlo@ed.ac.uk}
\emailAdd{felix.erben@ed.ac.uk}
\emailAdd{vera.guelpers@ed.ac.uk}
\emailAdd{maxwell.hansen@ed.ac.uk}
\emailAdd{tharris@ed.ac.uk}
\emailAdd{nils.hermansson-truedsson@thep.lu.se}
\emailAdd{raoul.hodgson@ed.ac.uk}
\emailAdd{andreas.juttner@cern.ch}
\emailAdd{fionn.o.hogain@ed.ac.uk}
\emailAdd{antonin.portelli@ed.ac.uk}
\emailAdd{j.richings@epcc.ed.ac.uk}
\emailAdd{andrew.yong@ed.ac.uk}

\preprint{CERN-TH-2022-193,\ LU-TP 22-59}

\abstract{
    The decreasing uncertainties in theoretical predictions and experimental measurements of several hadronic observables related to weak processes, which in many cases are now smaller than $\order(1\%)$, require theoretical calculations to include subleading corrections that were neglected so far. 
    Precise determinations of leptonic and semi-leptonic decay rates, including QED and strong isospin-breaking effects, can play a central role in solving the current tensions in the first-row unitarity of the CKM matrix.
    In this work we present the first RBC/UKQCD lattice calculation of the isospin-breaking corrections to the ratio of leptonic decay rates of kaons and pions into muons and neutrinos. 
    The calculation is performed  with $\Nf=2+1$ dynamical quarks close to the physical point and domain wall fermions in the M\"obius formulation are employed. Long-distance QED interactions are included according to the $\QEDL$ prescription and the crucial role of finite-volume electromagnetic corrections in the determination of leptonic decay rates, which produce a large systematic uncertainty, is extensively discussed. 
    Finally, we study the different sources of uncertainty on $|\vus|/|\vud|$ and observe that, if finite-volume systematics can be reduced, the error from isospin-breaking corrections is potentially sub-dominant in the final precision of the ratio of the CKM matrix elements. 
    }

\begin{document}

\maketitle

\section{Introduction}
\label{sec:introduction}
Flavour physics offers a unique opportunity in the search for new physics at the precision frontier of the Standard Model (SM). Discrepancies between SM predictions and experimental observations of processes where yet undiscovered particles or fields may play a tiny but measurable role can in fact be signals of new physics beyond the SM.
In the hadronic sector, the study of processes mediated by the weak force
gives access to the elements of the Cabibbo-Kobayashi-Maskawa (CKM) matrix
describing quark-flavour mixing.
The accurate determination of the CKM matrix elements $\vud$ and $\vus$ is of
crucial importance to test the first-row unitarity
$|\vud|^2+|\vus|^2+|V_\mathrm{ub}|^2=1$ imposed by the SM and to probe
emerging tensions that are approaching the $3\sigma$
confidence level~\cite{Workman:2022ynf,Aoki:2021kgd,Cirigliano:2022yyo}.
A complete understanding of SM processes like the leptonic and semi-leptonic
decay modes of pseudoscalar mesons or nuclear beta decays which underpin these
constraints is therefore necessary to test CKM unitarity and eventually put
bounds on the new physics energy scale and couplings.

In particular, in this work we are concerned with the precision determination of the ratio $|\vus|/|\vud|$ obtained by combining the experimental leptonic decay rates of the pion~($\pi_{\mu 2}$) and kaon~($K_{\mu 2}$) into a muon and a neutrino with hadronic matrix elements which parameterize the SM prediction. 
Given the non-perturbative dynamics of strong interactions at low energies, these theoretical determinations can be obtained in a reliable and systematically improvable way from first principles lattice field theory computations.
Lattice QCD has now entered the precision era and is able to provide many hadronic quantities with percent precision, e.g.~the ratio of kaon and pion leptonic decay constants $f_K/f_\pi$ and the kaon semi-leptonic decay ($K_{\ell 3}$) vector form factor $f^{+}(0)$, which play a central role in the determination of the CKM quantities $|\vus|/|\vud|$ and $|\vus|$, respectively~\cite{Aoki:2021kgd}. 
To date, most lattice QCD computations in flavour physics neglect isospin-breaking (IB) effects,
namely the inclusion of electromagnetism and the difference of the up and down
quark masses, which are required to go beyond percent level precision.
These contributions have been historically included using effective field theories such as chiral perturbation theory~($\chi\mathrm{PT}$)~\cite{Ananthanarayan:2004qk,Descotes-Genon:2005wrq,Cirigliano:2007ga}, where, however, it can be difficult to systematically assess uncertainties emerging from effective expansions.
The RM123+Southampton~(RM123S) collaboration pioneered the first lattice calculations beyond the QCD isospin limit~\cite{Giusti:2017dwk,DiCarlo:2019thl}, although with an extrapolation of the result from unphysical quark masses.
In this work we provide a first determination of the IB effects in $f_K/f_\pi$ using ab initio computations of lattice QCD and QED using a regularization with good chiral properties directly at physical quark masses, and examine its impact on the determination of $|\vus|/|\vud|$.

When electromagnetism is included, leptonic decay amplitudes can no longer be factorised into QCD and non-QCD contributions, as the lepton can interact with the pseudoscalar meson. Additionally, this new interaction generates infrared (IR) divergences which only cancel when summing diagrams containing virtual and real photon corrections~\cite{Bloch:1937pw}.
Thus, the decay rate can be properly written including these effects as
\begin{align}
    \label{eq:bn}
    \Gamma(P^\pm\to\ell^\pm\nu[\gamma]) 
    &= \lim_{\Lambda_\mathrm{IR}\to 0} \left[ 
        \Gamma_0(\Lambda_\mathrm{IR}) + \Gamma_1(\Lambda_\mathrm{IR}) \right],
\end{align}
where we indicate the contribution to the decay rate with virtual photon
corrections as $\Gamma_0$ (virtual decay rate), and that with one real photon in the final state as
$\Gamma_1$ (real decay rate), and where $\Lambda_\mathrm{IR}$ is an arbitrary IR energy cutoff (e.g. a photon mass).
A practical strategy for non-perturbative computations was put forward by the
RM123S group in ref.~\cite{Carrasco:2015xwa} (and applied in
successive calculations~\cite{Giusti:2017dwk,DiCarlo:2019thl}), which consists
in defining the inclusive rate as the sum of two contributions which are
separately IR safe, namely
\begin{align}
    \label{eq:RM123S_0}
    \Gamma(P^\pm\to\ell^\pm\nu[\gamma]) 
    &= \lim_{\Lambda_\mathrm{IR}\to 0} \left[
        \Gamma_0(\Lambda_\mathrm{IR}) - \Gamma_0^\mathrm{uni}(\Lambda_\mathrm{IR}) \right]
        + \lim_{\Lambda'_\mathrm{IR}\to 0} \left[ 
            \Gamma_0^\mathrm{uni}(\Lambda'_\mathrm{IR}) +
            \Gamma_1(\Lambda'_\mathrm{IR}) \right],
\end{align}
where we note that different IR regulators can be used in both terms, as will
be the case in practice.
The quantity $\Gamma_0^\mathrm{uni}(\Lambda_\mathrm{IR})$ corresponds to the
\emph{universal} (structure-independent) IR-divergent part of the virtual
decay rate, which can be computed perturbatively assuming the decaying meson
to be a point-like particle. This term also exactly cancels the divergence in $\Gamma_1(\Lambda'_\mathrm{IR})$, when evaluated with $\Lambda'_\mathrm{IR}$ as an IR regulator.

In principle, both virtual and real decay rates should be computed
non-perturbatively since photons with sufficiently high energy can resolve the
internal structure of the decaying meson.
While there is no choice for virtual corrections, as all photon modes
contribute to the rate, one can impose a cut on the real photon energy, 
$\omega_\gamma^\mathrm{cut}$, such that its sensitivity to the structure of the meson is suppressed.
In this case $\Gamma_1(\Lambda_\mathrm{IR})$ can be
computed analytically in perturbation theory in the point-like approximation,
namely $\Gamma_1^\mathrm{pt}(\omega_\gamma^\mathrm{cut},\Lambda_\mathrm{IR})$.
As predicted by $\chi\mathrm{PT}$~\cite{Cirigliano:2007ga} and confirmed by lattice
calculations~\cite{Carrasco:2015xwa,Desiderio:2020oej}, structure-dependent
contributions are negligible for the decay channels studied in this work, namely
the decay of pions or kaons into muons and neutrinos, and therefore at our level of precision we can reliably consider
$\Gamma_1(\Lambda_\mathrm{IR})\simeq \Gamma_1^\mathrm{pt}(\omega_\gamma^\mathrm{max},\Lambda_\mathrm{IR})$, where $\omega_\gamma^\mathrm{max}$ is the maximum photon energy kinematically allowed.
We will focus on the non-perturbative calculation of the virtual decay rate
$\Gamma_0$ and will use the finite spatial extent of the lattice, $L$, along with a
suitable prescription for QED in a finite volume called QED$_\mathrm{L}$, as an
IR regulator~\cite{Hayakawa:2008an}.
The real decay rate will be evaluated in the well-motivated point-like approximation
and directly in infinite volume, regularizing the IR divergence with a photon mass $\Lambda_\mathrm{IR}'= m_\gamma$ in~\cref{eq:RM123S_0}.
Thus, we follow the approach outlined in
refs.~\cite{Carrasco:2015xwa,DiCarlo:2019thl}.

Going beyond the isospin-symmetric limit, the leading corrections to the
pseudoscalar decay rates from the electromagnetic fine structure constant
$\aem\approx1/137$ and from the renormalized (e.g.~in $\overline{\mathrm{MS}}$ at 2~GeV) up-down quark mass difference
$(m^\mathrm{R}_\mathrm{d}-m^\mathrm{R}_\mathrm{u})/\Lambda_\QCD\sim1\%$ are
both of the order of $1\%$, which we denote universally by
$\order(\varepsilon)=\order[\aem,(m^\mathrm{R}_\mathrm{d}-m^\mathrm{R}_\mathrm{u})/\Lambda_\QCD]$.
The correction to the ratio of kaon and pion decay rates is then parameterized by $\delta
R_{K\pi}$, which can be expressed through the relation~\cite{Workman:2022ynf}
\begin{equation}
    \label{eq:deltaRP}
    \frac{\Gamma(K^+\to \mu^+\nu_\mu[\gamma])}{\Gamma(\pi^+\to \mu^+\nu_\mu[\gamma])}
    = 
    \frac{|\vus|^2}{|\vud|^2}
    \frac{m_\pi}{m_K} \frac{(m_K^2-m_\mu^2)}{(m_\pi^2-m_\mu^2)}
    \frac{f_K^2}{f_\pi^2}(1+\delta R_{K\pi})
    + \mathrm O(\varepsilon^2)\,,
\end{equation}
where $\order(\varepsilon^2)$ is understood as a second-order correction in $(\aem,(m^\mathrm{R}_\mathrm{d}-m^\mathrm{R}_\mathrm{u})/\Lambda_\QCD)$.
The specific choice defining the isospin-symmetric theory, implicit in the
definition of $f_K/f_\pi$, will be the subject of~\cref{sec:IB_effects}, where we discuss
the consistency with other choices in the literature and the advantage of
simulating with close-to-physical quark masses.
With that in mind, we anticipate our final result for the leading IB corrections to
$f_K^2/f_\pi^2$ as 
\begin{align}
    \delta R_{K\pi} = -
    0.0086\,(3)_{\mathrm{stat.}}({}^{+11}_{-4})_\mathrm{fit}(5)_{\mathrm{disc.}}(5)_{\mathrm{quench.}}(39)_\mathrm{vol.}
    \,,
    \label{eq:finaldRkpi}
\end{align}
where the first error is statistical, while the others are systematic uncertainties and will be discussed in~\cref{sec:results}.
We note that our result is compatible with the only other lattice determination by the
RM123S collaboration~\cite{DiCarlo:2019thl} and with $\chi\mathrm{PT}$~\cite{Cirigliano:2011tm}
\begin{align}
    \delta R^\mathrm{RM123S}_{K\pi} &= -0.0126\,(14),\\
    \delta R^\mathrm{\chi PT}_{K\pi} &= -0.0112\,(21).
    \label{eq:finalrm123chipt}
\end{align}
All these numbers are dependent on a choice of scheme to define the separation of isospin-breaking effects. However, as we will discuss later in this paper, one can provide quantitative evidence that the prescriptions used in the results above are close enough so that the prescription-dependence lies well below the quoted uncertainties.
In addition to our numerical result, one of the main findings of this work is a refined investigation of the large power-law finite-size corrections which are induced by the QED$_\mathrm{L}$ treatment of electromagnetism in a finite volume, and are reflected in a large systematic error in our result.
As the leading structure-dependent finite-size effects are now known to be negligible~\cite{DiCarlo:2021apt}, the dominant point-like corrections which are investigated in~\cref{sec:matrixelements} exhibit unexpectedly large higher-order corrections.
In that section, the connection between the Euclidean correlation functions and the hadronic matrix elements of interest is outlined.
The details of the lattice implementation using domain wall fermions and the gauge ensemble  generated by the RBC/UKQCD collaboration may be found in~\cref{sec:methodology}, while the analysis of the numerical data including the estimation of the systematic
effects is detailed in~\cref{sec:numeval}.
The discussion of the result and the implications for the extraction of
$|\vus|/|\vud|$ are found in~\cref{sec:results}, before the conclusion.

\section{Isospin-breaking corrections}
\label{sec:IB_effects}

Leptonic decays of pions and kaons are low-energy processes that can be studied in an effective Fermi theory where the $W$-boson is integrated out, and the process is mediated by a local four-fermion interaction. At first order in the Fermi constant $G_F$, we can then assume that low-energy observables can be predicted to a high degree of precision within a theory of QCD+QED. We will refer to this as the full or physical theory in the rest of the paper. In the full theory, quantities like the meson decay constant $f_{P}$ entering~\cref{eq:deltaRP} are ambiguous as they are defined in the unphysical iso-symmetric limit of QCD (iso-QCD) where $m_\rmu=m_\rmd$ and $\aem=0$. Such unphysical definitions are related to the fact that QCD and QCD+QED interactions generate different ultraviolet (UV) divergences and hence require different renormalization procedures to fix the bare parameters of the respective actions. In order to give a meaning to (iso-)QCD observables within the full QCD+QED theory, additional renormalization conditions are then required. 

The discussion in this section is divided in two: first, we will discuss how to non-perturbatively renormalize the full theory on the lattice and make well-defined physical predictions in the continuum limit. Then, we will define QCD and its iso-symmetric limit, which is employed in our numerical lattice calculation and fixes the definition of IB corrections. However, the discussion here is general and not restricted to lattice QCD calculations. In fact, the definitions of the QCD+QED, QCD and iso-QCD theory discussed in the rest of the section hold for any other non-perturbative approach (like e.g.~effective field theories).

\subsection{Renormalizing the full theory}
\label{sec:ren_full_theory}
In the full QCD+QED theory with $\Nf$ flavours of quarks, once a UV regulator is introduced (in our case the lattice spacing $a$), the action depends only on the bare quark masses in lattice units ${\hat{\mvec}=(\hat{m}_1,\dots,\hat{m}_{\Nf})}$ and the bare strong and electromagnetic couplings, $g$ and $e$, respectively. In this theory every physical observable can be predicted once the bare parameters of the action are defined and the regulator is removed. 
Note that, since we are only working at first order $\mathrm{O}(\varepsilon)$ in the IB effects, we can neglect the running of the electromagnetic coupling and safely fix it to its Thomson limit, $e^\phi=(4\pi\aem^\phi)^{1/2}$  with $\aem^\phi=1/137.035999084\,(21)$~\cite{RevModPhys.93.025010}, without the need to impose a specific renormalization condition. Moreover, when working at first order in $\aem$ and when a lepton is also included in the theory, its mass $\mell$ can be renormalized perturbatively in the usual way by imposing that its on-shell value coincides with the experimental one, i.e. $\mell^\phi=\mell^{\textrm{\tiny PDG}}=105.6583755\,(23)$~MeV~\cite{Workman:2022ynf}. The superscript $\phi$ is used to denote quantities evaluated in the physical (QCD+QED) theory.

At a fixed value of the bare strong coupling~$g$, we define the bare lattice quark masses in the full theory $\hat{\mvec}^\phi$ by identifying $\Nf+1$ dimensionful quantities, which we assume without loss of generality\footnote{In practice one only requires $\Nf+1$ quantities that can be used to form $\Nf$ independent dimensionless ratios.} to have mass dimension 1, namely $\mathrm{M}_1^\phi,\dots, \mathrm{M}_{\Nf}^\phi, \Lambda^\phi$, and requiring the following ratios to take on the correct values when $e$ is also at its physical value $e^\phi$,
\begin{equation}
\label{eq:ren_cond_full}
\bigg[\frac{\hat{\mathrm{M}}_j }{\hat{\Lambda} }\bigg]^2\!\!(g, e^\phi, \hat{\mvec}^\phi) = \bigg(\frac{\mathrm{M}_j^\phi}{\Lambda^\phi}\bigg)^2 \,, \qquad \text{for} \  j = 1, \cdots, \Nf \,. 
\end{equation}
Here, the $\hat{\mathrm{M}}_j$ and $\hat{\Lambda}$ denote the same quantities evaluated in lattice units at the physical point $(g, e^\phi, \hat{\mvec}^\phi)$. For later use, we define this point as $\sigmavec^\phi=(g,e^\phi,\hat\mvec^\phi)$.
Note that the procedure for fixing $\hat\mvec^\phi$ must be performed at every value of the coupling $g$, so in this sense we can think of the bare quark masses as a function of this coupling, $ \hat \mvec^\phi(g)$. Moreover, the lattice quantities appearing on the left-hand side of~\cref{eq:ren_cond_full} are considered to be evaluated in the infinite volume limit. In practice, in the full QCD+QED theory, electromagnetic interactions can generate sizeable power-like finite-volume effects and should be removed, as discussed in~\cref{sec:FVE}. Once the bare quark masses $\hat \mvec^\phi$ are determined, we can predict any other quantity $\hat{X}^\phi$ in lattice units and in the full theory as a function of $g$, namely
\begin{equation}
    \hat X^\phi(g) = \hat X(g, e^\phi, \hat \mvec^\phi(g))\,.
\end{equation}
From this, we can give a physical value to $a$ as a function of $g$ using a suitable dimensionful external input. For concreteness, we envision using $\Lambda^\phi$
\begin{equation}
\label{eq:latt_spacing_full}
    a (g) = \frac{\hat \Lambda^\phi(g) }{\Lambda^\phi} \,.
\end{equation}
With this definition, one can predict dimensionful quantities as
\begin{equation}
    X^{\phi}(g)= a(g)^{-[X]}\hat{X}^{\phi}(g)\,,
\end{equation}
where $[X]$ is the mass dimension of $X$. As a consequence of asymptotic freedom, the limit $g \to 0$ implies $a(g) \to 0$, and for renormalized quantities the equation above has a $g\to 0 $ limit which is cutoff independent. At non-zero $g$ and $a$ there is a family of choices that have the same continuum limit. For a given discretization of the QCD+QED action, this family is defined by (i) the set of $\Nf$ ratios that we use to define $\hat \mvec(g)$, and (ii) the physical quantity $\Lambda^\phi$ that we use to set the scale.

In this calculation we employ three flavours of quarks, so we require four hadronic observables to fix the bare quark masses $\mvec=(m_\rmu,m_\rmd,m_\rms)$ and the scale, which we choose to be $\Mvec=(\mathrm{M}_1,\mathrm{M}_2,\mathrm{M}_3)=(m_{\pi^+},m_{K^+},m_{K^0})$ and $\Lambda=m_{\Omega^-}$. The physical values of such hadronic masses are taken as their experimental measurements, reported in the PDG~\cite{Workman:2022ynf}, i.e. $\Mvec^\phi=(m_{\pi^+}^\pdg,m_{K^+}^\pdg,m_{K^0}^\pdg) = (139.57039\,(18),\, 493.677\,(16),\, 497.611\,(13))~\mathrm{MeV}$ and $\Lambda^\phi=m_{\Omega^-}^\pdg=1672.45\,(29)~\mathrm{MeV}$.

\subsection{Defining QCD and its isospin-symmetric limit}
\label{sec:isoqcd}
The calculation of IB corrections requires the definition of an isospin-symmetric limit of QCD.
If we write the parameters of the bare QCD+QED Lagrangian as $\sigmavec=(g,e,\hat\mvec)$, then the point $\sigmavec^\phi=(g,e^\phi,\hat\mvec^\phi)$ identifies the full theory defined above in~\cref{sec:ren_full_theory}. Points where $e=0$ correspond instead to the bare parameters for a QCD Lagrangian, that we denote as ${\sigmavec^\qcd=(g^\qcd,0,\hat\mvec^\qcd)}$, where the vector $\hat\mvec^\qcd=(\hat{m}^\qcd_\rmu,\hat{m}^\qcd_\rmd,\hat{m}^\qcd_{\rms})$ contains the QCD bare quark masses, with $\hat{m}^\qcd_\rmu\neq \hat{m}^\qcd_\rmd$. Therefore,  we see that defining QCD within the full QCD+QED theory consists of choosing \emph{one} point by imposing an additional renormalization condition. The same holds for iso-QCD theories, all belonging to the set identified by  ${\sigmavec^\iso=(g^\iso,0,\hat\mvec^\iso)}$, with $\hat\mvec^\iso=(\hat m^\iso_\rmud,\hat m^\iso_\rmud,\hat{m}^\iso_{\rms})$, i.e. with $\delta m=0$. Here we denoted the average light quark mass as ${m}_\rmud=({m}_\rmu+{m}_\rmd)/2$ and the up-down quark mass difference as $\delta m=m_\rmu-m_\rmd$. Again, the definition of a given iso-QCD theory requires imposing specific renormalization conditions. The possibility of choosing different renormalization prescriptions for the definition of QCD and iso-QCD translates into a \emph{scheme dependence} in any observable computed in such theories.

Since there are many possible valid choices such that the IB corrections are small, there is no single generally accepted scheme and lattice collaborations have used in the past different prescriptions to define the iso-QCD theory~\cite{Aoki:2021kgd}. Although no significant differences have been observed so far in the use of different schemes~\cite{Aoki:2021kgd}, when aiming at sub-percent precision calculations the ambiguities related to the different renormalization prescriptions adopted might become relevant when comparing results for (scheme-dependent) iso-QCD observables or IB effects. We therefore advocate that lattice collaborations be as clear and transparent as possible in defining the scheme used to define iso-QCD in their calculations, and eventually agree on a common reference scheme. 

In this work we adopt a scheme very similar to that introduced by the BMW Collaboration in refs.~\cite{PhysRevLett.111.252001,Fodor:2016bgu,Borsanyi:2020mff}, which relies on the use of the neutral mesonic observables $\Mvec^\bmw=(M_\rmud,\,\Delta M,\,M_{K\chi})$, where
\begin{equation}
    M_\rmud^2 = \frac12 \left( M^2_{\rmu\rmu}+M^2_{\rmd\rmd}\right)\, , \quad
    \Delta M^2 =  M^2_{\rmu\rmu}-M^2_{\rmd\rmd}\, ,\quad
    M_{K\chi}^2 = \frac12 \left( m^2_{K^+}+m^2_{K^0}-m^2_{\pi^+}\right)\,,
\end{equation}
and $M_\mathrm{qq}^2$ denotes the squared mass of the connected $\bar{q}q$ neutral pseudoscalar mesons. One can show that the leading-order partially-quenched chiral corrections to such quantities are given by~\cite{PhysRevLett.111.252001, Bijnens:2006mk}
\begin{equation}
    M_\rmud^2 = 2B m_\rmud^{\qcd,\mathrm{R}} + \cdots\,, \quad
    \Delta M^2 = 2B \delta m^{\qcd,\mathrm{R}} + \cdots\,, \quad
    M^2_{K\chi} = 2B m_\rms^{\qcd,\mathrm{R}} +\cdots\,,
\end{equation}
where $B$ is the QCD chiral condensate, the superscript $\mathrm{R}$ denotes the QCD-renormalized quark masses in a given scheme (e.g.~$\overline{\mathrm{MS}}$ at 2~GeV) and the ellipses are next-to-leading order $\mathrm{SU}(3)$ chiral corrections~\cite{Bijnens:2006mk}. An important feature of this choice of variables is the systematic absence of $\order(\aem)$ corrections at leading order.
Therefore, these specific meson masses are expected to be dominated by the part proportional to the quark masses, allowing comparisons with quark-mass schemes like the Gasser-Rusetsky-Scimemi (GRS) one~\cite{Gasser:2003hk} without the need of using short-distance schemes (e.g.~the $\overline{\mathrm{MS}}$ scheme) to renormalize the quark masses.

In this scheme we assume that the bare strong coupling of any unphysical theory identified by the vector $\sigmavec^\star$ is kept equal to that of the full theory, i.e. $\sigmavec^\star=(g,e^\star,\mvec^\star)$. The bare quark masses of QCD and iso-QCD are then obtained by imposing the following renormalization conditions, respectively
\begin{equation}
  \label{eq:ren_cond_QCD}
  \resizebox{.9\hsize}{!}{$\displaystyle
  \bigg[\frac{ \hat{\mathbf{M}}^\bmw }{\hat{m}_\Omega }\bigg]^2\!\!(g, 0, \hat{\mvec}^\qcd)  =   \bigg( 
      \bigg[\frac{ \hat{M}_{\rmud}}{\hat{m}_\Omega}\bigg]^2 \!\!(g, e^\phi, \hat{\mvec}^\phi)\,, \, 
      \bigg[\frac{ \Delta\hat{M}}{\hat{m}_\Omega}\bigg]^2 \!\!(g, e^\phi, \hat{\mvec}^\phi)\,, \, 
      \bigg[\frac{ \hat{M}_{K\chi}}{\hat{m}_\Omega}\bigg]^2 \!\!(g, e^\phi, \hat{\mvec}^\phi) \bigg)\,,
      $}
\end{equation}
\begin{equation}
    \label{eq:ren_cond_iso}
    \resizebox{.9\hsize}{!}{ $\displaystyle
    \bigg[\frac{ \hat{\mathbf{M}}^\bmw }{\hat{m}_\Omega }\bigg]^2\!(g, 0, \hat{\mvec}^\iso)\ = \bigg( 
      \bigg[\frac{ \hat{M}_{\rmud}}{\hat{m}_\Omega}\bigg]^2 \!(g, e^\phi, \hat{\mvec}^\phi)\,, \, 
      0\,, \, 
      \bigg[\frac{ \hat{M}_{K\chi}}{\hat{m}_\Omega}\bigg]^2 \!(g, e^\phi, \hat{\mvec}^\phi) \bigg)\,.\hspace{2.8cm}
    $}
\end{equation}
The above conditions can be extended beyond $\Nf=3$ flavours by choosing ratios of hadron masses with a dependence on the heavier quarks.
The lattice spacings for the two theories can be evaluated by imposing the additional conditions
\begin{equation}
    a^\qcd(g) = \frac{\hat{m}_\Omega(g,0,\hat{\mvec}^\qcd)}{m_{\Omega^-}^\pdg}\quad \text{and} \quad a^\iso(g) = \frac{\hat{m}_\Omega(g,0,\hat{\mvec}^\iso)}{m_{\Omega^-}^\pdg}~.
\end{equation}
In principle, one could also impose the simpler condition $a^\qcd(g)=a^\iso(g)$. Since the UV divergences of the two theories do not depend on quark masses, the difference between the two approaches would result in cut-off effects that eventually vanish in the continuum limit.
However, in this work, the quantities we aim to compute are dimensionless, and therefore we will not need, in practice, to make dimensions explicit in the (iso-)QCD theory.

\subsection{Computing isospin-breaking effects on the lattice}
\label{sec:IBeffects_lattice}

Once the bare parameters of both the full theory and  iso-QCD have been determined through the renormalization procedure described above, one can define the expansion of a QCD+QED observable $\hat{X}(\sigmavec^\phi)$ around the iso-QCD point $\sigmavec^\iso$ as
\begin{equation}
    \hat{X}(\sigmavec^\phi) = \hat{X}(\sigmavec^\iso) + \delta \hat{X}(\sigmavec^\iso) + \order(\varepsilon^2)\,,
    \label{eq:isoexp}
\end{equation}
where the quantity $\delta \hat{X}(\sigmavec^\iso)$ encodes the leading order IB corrections of $\order(\varepsilon)$ relative to the specific iso-QCD point $\sigmavec^\iso$.
Suppose now that the lattice setup used for the calculation has been tuned to some iso-QCD point $\sigmavec=(g,0,\hat\mvec)$, close to the physical point $\sigmavec^{\phi}$. In practice this point is generally the result of a simulation-parameter tuning procedure and might differ from the iso-QCD point defined in the previous section, i.e.~$\sigmavec\neq\sigmavec^{(0)}$. However, we assume that this simulation point is sufficiently close to $\sigmavec^{\phi}$, $\sigmavec^{(0)}$, and $\sigmavec^\qcd$  that the value of $\hat{X}$ at any of those points can be described accurately enough by a linear correction to the simulation point. This is a fairly strong assumption that would not be valid in a number of lattice calculations, particularly when working at non-physical quark masses where non-linear corrections are expected to be sizeable. However, as we will demonstrate in~\cref{sec:tuning}, this applies to the close-to-physical point simulations used in this work. More explicitly, under this linearity assumption the physical value of $\hat{X}$ can be written as
\begin{equation}
    \hat{X}(\sigmavec^\phi) = \hat{X}(\sigmavec) + \aem^\phi \, \frac{\partial \hat{X}}{\partial \aem}(\sigmavec)+ \sum_{\mathrm{q}=1}^{\Nf} \, (\hat{m}^\phi_\mathrm{q}-\hat{m}_\mathrm{q})\,\frac{\partial \hat X}{\partial \hat{m}_\mathrm{q}}(\sigmavec)+\order(\bar{\varepsilon}^2)~,
    \label{eq:phys_to_sim}
\end{equation}
where the $\order(\bar{\varepsilon}^2)$ represent any second order corrections in $\hat{m}^\phi_\mathrm{q}-\hat{m}_\mathrm{q}$ and $\aem$, which we consider to be of similar size to higher-order isospin-breaking corrections $\order(\varepsilon^2)$ that are taken to be negligible and are discarded throughout this work. This expanded power-counting is necessary as some isospin-symmetric parameters like $m_{\rms}$ and $m_{\rmud}$ might be slightly mistuned at the simulation point. If we consider $\hat{X}$ to be the ratio of hadron masses in \cref{eq:ren_cond_full}, by using lattice data for $\hat{X}(\sigmavec)$ and its derivatives, and applying the linear equation above, we can solve the resulting system to obtain the lattice bare quark masses for the physical point $\hat{\mvec}^{\phi}$.

A similar equation is obtained by linearizing the first term on the right-hand side of~\cref{eq:isoexp},
\begin{equation}
    \hat{X}(\sigmavec^\iso) = \hat{X}(\sigmavec) + \sum_{\mathrm{q}=1}^{\Nf} \, (\hat{m}^\iso_\mathrm{q}-\hat{m}_\mathrm{q})\,\frac{\partial \hat X}{\partial \hat{m}_\mathrm{q}}(\sigmavec)+\order(\bar{\varepsilon}^2)~,
    \label{eq:iso_to_sim}
\end{equation}
and can be solved for $\hat{\mvec}^{\iso}$ applying the renormalization conditions in \cref{eq:ren_cond_iso}. Combining~\cref{eq:phys_to_sim,eq:iso_to_sim}, the isospin-breaking correction in \cref{eq:isoexp} can be identified as
\begin{equation}
    \delta \hat{X}(\sigmavec^\iso) =  \aem^\phi \, \frac{\partial \hat{X}}{\partial \aem}(\sigmavec)+ \sum_{\mathrm{q}=1}^{\Nf} \, (\hat{m}^\phi_\mathrm{q}-\hat{m}_\mathrm{q}^\iso)\,\frac{\partial \hat X}{\partial \hat{m}_\mathrm{q}}(\sigmavec)~.
\end{equation}
The QCD masses and $\hat{X}(\sigmavec^\qcd)$ can be determined using analogous linear expansions.

We can then define a separation of $\delta \hat{X}(\sigmavec^\iso)$ into a contribution due to the strong isospin breaking~(SIB) and another due to electromagnetic interactions, namely
\begin{align}
    \label{eq:X_SIB}
    \delta \hat{X}^{\textrm{\tiny SIB}}(\sigmavec^\iso)&=\hat{X}(\sigmavec^\qcd)-\hat{X}(\sigmavec^\iso)
    =\sum_{\mathrm{q}=1}^{\Nf} \, (\hat{m}^\qcd_\mathrm{q}-\hat{m}_\mathrm{q}^\iso)\,\frac{\partial \hat X}{\partial \hat{m}_\mathrm{q}}(\sigmavec)\,,\\
    \label{eq:X_gamma}
    \delta \hat{X}^\gamma(\sigmavec^\iso)&=\hat{X}(\sigmavec^\phi)-\hat{X}(\sigmavec^\qcd)=
    \aem^\phi \, \frac{\partial \hat{X}}{\partial \aem}(\sigmavec)+\sum_{\mathrm{q}=1}^{\Nf} \, (\hat{m}^\phi_\mathrm{q}-\hat{m}_\mathrm{q}^\qcd)\,\frac{\partial \hat X}{\partial \hat{m}_\mathrm{q}}(\sigmavec)\,,
\end{align}
such that
\begin{equation}
    \label{eq:obs_decomposition_iso_SIB_gamma}
    \hat{X}(\sigmavec^\phi) = \hat{X}(\sigmavec^\iso) + \delta \hat{X}^{\textrm{\tiny SIB}}(\sigmavec^\iso)+\delta \hat{X}^\gamma(\sigmavec^\iso) + \order(\varepsilon^2)\,.
\end{equation}

From the above discussion it is clear that in order to determine IB corrections one needs to evaluate numerically the linear coefficients of the expansion of a given observable in terms of $\aem$ and quark mass shifts. 
The QED effects have been computed in the past by different collaborations following two approaches. On the one hand, one can include QED gauge links in the fermion operator to be inverted and produce QCD+QED quark propagators to construct hadronic correlation functions, as introduced first in ref.~\cite{Duncan:1996xy}, and used in a wide range of lattice calculations~\citep{Aoki:2021kgd}. On the other hand, it is possible to obtain the same corrections by perturbatively expanding the path integral for $\hat{X}$ with respect to $\aem$, with the result of evaluating diagrams with electromagnetic current insertions, as  originally proposed by the RM123 collaboration in refs.~\cite{deDivitiis:2013xla,Giusti:2017dmp}. In the following, the latter method will be adopted to determine the IB corrections to hadron masses and to the ratio of the leptonic decay rates of kaons and pions into muons. The implementation of the method will be discussed in~\cref{sec:methodology}.

\subsection{Scheme ambiguities}
\label{sec:scheme_ambiguities}
The isospin decomposition in~\cref{eq:obs_decomposition_iso_SIB_gamma}, as discussed
in~\cref{sec:isoqcd}, depends on the prescriptions in~\cref{eq:ren_cond_iso,eq:ren_cond_QCD},
although the physical observable, $\hat{X}(\sigmavec^\phi)$, is not. Varying these prescriptions will
lead to different bare masses $\hat{\mvec}_\mathrm{q}^{\qcd}$ and $\hat{\mvec}_\mathrm{q}^{\iso}$,
which generate the scheme-dependence of $\delta \hat{X}^{\textrm{\tiny SIB}}$ and $\delta
\hat{X}^{\gamma}$ through~\cref{eq:X_SIB,eq:X_gamma}. In principle this ambiguity is an
$\order(\varepsilon)$ effect on $\hat{\mvec}_\mathrm{q}^{\qcd}$ and $\hat{\mvec}_\mathrm{q}^{\iso}$,
and therefore the ambiguity on the iso-symmetric component $\hat{X}(\sigmavec^\iso)$ can potentially
be as large as the isospin-breaking corrections $\delta \hat{X}(\sigmavec^\iso)$ themselves. It is
therefore important to identify classes of schemes which are phenomenologically relevant with a
minimal level of ambiguity.

For instance, the GRS scheme~\cite{Gasser:2003hk} assumes that the renormalized quark masses and the
renormalized strong coupling constant, in a given scheme and at a chosen renormalization scale, are
kept constant for any value of $\aem$. This scheme itself depends then on a choice of
renormalization procedure for these quantities. The GRS scheme is generally the prescription used in
phenomenological calculations, such as chiral perturbation theory predictions for weak decay rates
including the ones discussed in this paper~\cite{Ananthanarayan:2004qk,Descotes-Genon:2005wrq,Cirigliano:2007ga}. It is therefore attractive for lattice calculations to
use prescriptions which produce predictions close to the GRS scheme. Defining precisely such a class
of schemes is a rich technical topic which is relevant for precision physics, and will be the topic
of a future publication based on the lattice data presented in this paper. The BMW variables chosen
here are designed to be hadronic quantities providing an isospin decomposition close to the GRS
scheme, and in~\cref{sec:tuning} we will explicitly check that by comparing with existing results
from ref.~\cite{DiCarlo:2019thl}.

\section{Matrix elements from Euclidean correlation functions}
\label{sec:matrixelements}

As discussed in~\cref{sec:introduction}, the  inclusion of IB corrections in the calculation of decay rates is complicated by the appearance of IR divergences, generated by $\order(\aem)$ QED corrections to the decay amplitude. In this calculation we adopt the RM123S method of~\cref{eq:RM123S_0} to regularize IR divergences. We choose two separate regulators for the two terms. In particular, we compute the virtual decay rate on the lattice using the finite volume with the $\QEDL$ prescription as an IR regulator, while the real decay width is evaluated in perturbation theory using a photon mass, namely
\begin{equation}
\label{eq:RM123S_1}
\Gamma(P^\pm\to\ell^\pm\nu[\gamma])= \lim_{L\rightarrow \infty}\left[\Gamma_0(L) - \Gamma^\text{uni}_0(L)\right]
    + \lim_{m_\gamma \rightarrow 0} \left[\Gamma^\text{uni}_0(m_\gamma) + \Gamma_1(\omega_\gamma^\mathrm{max},m_\gamma)\right]\,.
\end{equation}
The first bracketed term removes the universal (structure-independent) logarithmic IR divergence and finite volume effects~(FVE) up to $\order(1/L)$~\cite{PhysRevD.95.034504,Tantalo:2016vxk}. Recently, the $\order({1}/{L^2})$  corrections to $\Gamma_0(L)$ have been calculated in $\QEDL$, including structure-dependent contributions, in ref.~\cite{DiCarlo:2021apt}. Thus, we can extend~\cref{eq:RM123S_1} to
\begin{align}
    \inclrate{P}{\mu}&= \lim_{L\rightarrow \infty}\big[\Gamma_0(L) - \Gamma^{(2)}_0(L)\big] 
    + \lim_{m_\gamma \rightarrow 0} \left[\Gamma^\text{uni}_0(m_\gamma) + \Gamma_1(\omega_\gamma^\mathrm{max},m_\gamma)\right]\,,
    \label{eq:strategy}
\end{align} 
where $\Gamma^{(2)}_0(L)$ contains the finite-volume effects up to $\order(1/L^2)$ and will be discussed in detail in~\cref{sec:FVE}. Here, we only note that the residual finite-volume effects in the quantity $\Gamma_0(L)-\Gamma^{(2)}_0(L)$ now begin at $\order ({1}/{L^3})$.
The second bracketed term in~\cref{eq:strategy} has been instead calculated analytically in ref.~\cite{Carrasco:2015xwa} and is reported below in~\cref{sec:real_photon}. 
For convenience, we distinguish the two contributions, the one computed on the lattice with a finite volume and the one evaluated in perturbation theory with a massive photon, respectively as
\begin{equation}
  \Gamma_P^\mathrm{latt}(L) \equiv \big[\Gamma_0(L) - \Gamma^{(2)}_0(L)\big] \quad \text{and} \quad
  \Gamma_P^\mathrm{pert}(m_\gamma) \equiv \big[\Gamma^\text{uni}_0(m_\gamma) + \Gamma_1(\omega_\gamma^\mathrm{max},m_\gamma)\big]\,.
\end{equation}
We can expand these expressions at leading order in the IB corrections 
\begin{align}
\label{eq:deltaRP_virt}
\Gamma_P^\mathrm{latt}(L) & =
 \Gamma_P^\mathrm{tree}\,\Big[1+\delta R_P^\mathrm{latt}(L)- \delta R_P^{(2)}(L)\Big]  + \order(\varepsilon^2)\,,&\\
\label{eq:deltaRP_real}
\Gamma_P^\mathrm{pert}(m_\gamma) & =  \Gamma_P^\mathrm{tree}\,\delta R_P^\mathrm{pert}(\omega_\gamma^\mathrm{max},m_\gamma)+\order(\varepsilon^2)\,,&
\end{align}
having defined
\begin{equation}
  \Gamma_0(L) = \Gamma_P^\mathrm{tree} \Big[1+\delta R_P^\mathrm{latt}(L)\Big]+\order(\varepsilon^2) \quad \text{and} \quad
  \Gamma^{(2)}_0(L) = \Gamma_P^\mathrm{tree}\Big[1 + \delta R_P^{(2)}(L)\Big]+\order(\varepsilon^2)\,.
  \label{eq:Gamma0_2_L}
\end{equation}
It follows that, writing the leptonic decay rate as 
\begin{equation}
  \label{eq:deltaRP_def}
      \inclrate{P}{\mu} = \Gamma_P^\mathrm{tree}\big(1+\delta R_P\big) + \order(\varepsilon^2)~
  \end{equation}
its leading IB correction is given by 
\begin{equation}
\label{eq:deltaRP_limits}
    \delta R_P = \lim_{L\rightarrow \infty} \Big[\delta R_P^\mathrm{latt}(L)- \delta R_P^{(2)}(L) \Big] + \lim_{m_\gamma \rightarrow 0} \ \delta R_P^\mathrm{pert}(\omega_\gamma^\mathrm{max},m_\gamma) \,.
\end{equation}
The outline of the rest of the section is as follows. In~\cref{sec:virtual} we give our definition of $\Gamma_P^\mathrm{tree}$ and derive the corresponding correction $\delta R_P^\mathrm{latt}(L)$ at finite volume in terms of IB corrections to matrix elements and meson masses. Then, in~\cref{sec:mel_from_corr} we describe how to obtain such corrections starting from Euclidean lattice correlation functions. The subtraction of $\QEDL$ finite-volume effects $\delta R_P^{(2)}(L)$ and the calculation of the real photon emission in perturbation theory are then discussed in~\cref{sec:FVE,sec:real_photon}, respectively.

\subsection{Virtual corrections to the leptonic decay rate}
\label{sec:virtual}

\begin{figure}[t]
  \centering
  \includegraphics[width=0.35\textwidth]{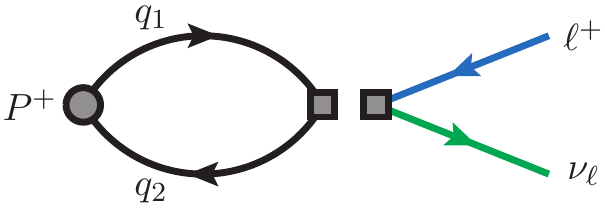}
  \caption{Feynman diagram of the tree-level contribution to the weak decay of a positive pseudoscalar meson $P^+ \in \{ \pi^+, K^+ \}$ into a lepton-neutrino pair. The double-square vertex represents the effective weak Hamiltonian \cref{eq:effweakH}.}
  \label{fig:tree diagram}
\end{figure}

We focus here on the determination of the matrix element associated to the virtual decay of a positive pseudoscalar meson, $P^+\to\ell^+\nu_\ell$, without including real photons in the final state. As this is an IR divergent quantity, we assume that an IR regulator is in place throughout the section. For concreteness, we regulate IR divergences on a finite volume of size $L$ adopting the $\QEDL$ prescription~\cite{Hayakawa:2008an} to remove the spatial zero modes of the lattice photon propagator (see~\cref{sec:FVE}).
At the lowest order in QED and QCD, pseudoscalar mesons decay into a lepton-neutrino pair via the exchange of a $W$-boson between the constituent quarks of the meson and the leptons. Since for both pions and kaons the process $P^+\to\ell^+\nu_\ell$ has a momentum transfer much smaller than the $W$-boson mass $m_W$, we can study it in an effective theory with a local four-fermion interaction described by the effective Hamiltonian
\begin{equation}
\label{eq:effweakH}
  \mathcal{H}_W 
  = \frac{G_F}{\sqrt{2}}  V_\mathrm{q_1q_2}^* \; {O}_W 
\end{equation}
where $G_F$ is the Fermi constant and $V_\mathrm{q_1q_2}^*$ the relevant CKM matrix element. The four-fermion operator mediating the process is
\begin{equation}
  \label{eq:OW}
{O}_W \equiv J_H^\rho \, J_L^\rho = 
\big(\,\bar{q}_2\,\gamma^\rho(1-\gamma_5)\,q_1\,\big)\, 
\big(\,\bar{\nu}_\ell\,\gamma^\rho(1-\gamma_5) \ell \,\big)~,
\end{equation}
with $q_1$ being a $u$-type quark and $q_2$ a $d$-type quark and $J_H^\rho$ and $J_L^\rho$ denoting the weak $(V-A)$ hadronic and leptonic currents, respectively. 
The Feynman diagram associated with this tree-level term is represented by~\cref{fig:tree diagram}.

When including QED at $\order(\aem)$, the UV corrections to matrix elements of the local operator ${O}_W$ differ from those of the Standard Model and a matching between the two theories is therefore needed. This is usually performed in the so-called $W$-regularization~\cite{Sirlin:1980nh,Sirlin:1981ie}, and we refer to refs.~\cite{Carrasco:2015xwa,DiCarlo:2019thl} for detailed discussions on the argument. After the inclusion of QCD and QED at  $\order(\aem)$ and assuming that chiral symmetry is preserved, the effective Hamiltonian reads\footnote{When including electromagnetic corrections at $\order(\aem)$, the Fermi constant $G_F$ has to be defined accordingly. This is conventionally obtained from the muon lifetime including one-loop electromagnetic corrections and reads $G_F=1.16634\times10^{-5}~\mathrm{GeV}^{-2}$~\cite{Berman:1958ti,Kinoshita:1958ru}. }
\begin{equation}
  \mathcal{H}_W 
  = \frac{G_F}{\sqrt{2}}  V_\mathrm{q_1q_2}^* \; \left(\mathcal{Z}_0+\frac{\aem}{4\pi}\, \delta \mathcal{Z} \right) {O}_W\,. 
  \label{eq:H_W_renormalized}
\end{equation}
Here $\mathcal{Z}_0$ is the non-perturbative QCD renormalization constant of the operator $O_W$. The quantity $\delta \mathcal{Z}$ encodes instead the short-distance matching between the effective theory in the $W$-regularization and the Standard Model, as well as the electromagnetic corrections to the matching of the four-fermion operator $O_W$ renormalized non-perturbatively in a given scheme to the $W$-regularization one. 
If $O_W$ is a lattice operator and the regularization used for the fermionic action introduces an explicit chiral symmetry breaking, then the operator $O_W$ undergoes an additive renormalization due to the mixing with other lattice operators with different chirality and the mixing pattern would be more complicated than that in~\cref{eq:H_W_renormalized} (see e.g.~refs.~\cite{Carrasco:2015xwa,DiCarlo:2019thl}). In the lattice calculation presented in this work, however, chiral fermions are employed and therefore in the following we will consider the operator $O_W$ renormalizing multiplicatively as in~\cref{eq:H_W_renormalized}, with $\mathcal{Z}_0=Z_V=Z_A$. Moreover, if a mass-independent scheme is adopted to renormalize the four-fermion operator, then the quantities $\mathcal{Z}_0$ and $\delta \mathcal{Z}$ will be the same regardless of the masses of the particles involved in the process. As a consequence, the contribution of the electromagnetic corrections proportional to $\delta \mathcal{Z}$ will cancel in the calculation of our quantity of interest, $\delta R_{K\pi}=\delta R_K-\delta R_\pi$, entering~\cref{eq:deltaRP}. 

In the full theory the (IR regulated) virtual decay rate can be written as 
\begin{equation}
  \Gamma_0(L) = 
  \mathcal{K}\,
  |{\Mcal}_{P}|^2~,
  \label{eq:DecayRate_full}
\end{equation}
where $\mathcal{K}$ is a factor containing the electro-weak coupling, the CKM matrix elements and the integration over the two-body phase space, while ${|\Mcal_P|^2 = \sum_{r,s} |\Mcal_P^{rs}|^2}$ is the magnitude squared of the QCD+QED virtual amplitude summed over the lepton and neutrino polarisations $r$ and $s$. In the rest frame of the decaying meson $P$ ($\pvec_P=\boldsymbol{0}$) the on-shell lepton and neutrino (Euclidean) momenta, $p_\ell=(\ii \omega_\ell,\pvec_\ell)$ and $p_\nu=(\ii \omega_\nu,\pvec_\nu)$, are such that $\pvec_\ell+\pvec_\nu=\boldsymbol{0}$ and the decay rate can be written in terms of 
\begin{equation}
  \mathcal K = \frac{G_F^2}{16\pi}\left|V_\mathrm{q_1q_2}\right|^2 \frac{1}{2 m_P} \left(1 - \frac{m_{\ell}^2}{m_P^2} \right)\,,
\end{equation}
and the renormalized QCD+QED matrix element
\begin{equation}
  \Mcal_P^{rs}(\pvec_\ell)= \mathcal{Z} \,\widebar{\Mcal}_P^{rs}(\pvec_\ell) = 
  \mathcal{Z}\, \braket{\ell^+,r,\pvec_\ell;\,\nu_\ell,s,\pvec_\nu|{O}_W|P^+,\boldsymbol{0}}^\phi\,.
  \label{eq:MatrixElement}
\end{equation}
Here $\widebar{\Mcal}_P^{rs}(\pvec_\ell)$ is the bare matrix element computed in the full QCD+QED theory, as indicated by the superscript $\phi$, while the factor $\mathcal{Z}=\mathcal{Z}_0+\tfrac{\alpha}{4\pi}\delta \mathcal{Z}$ denotes the renormalization constant of the weak operator $O_W$ entering the effective Hamiltonian of~\cref{eq:H_W_renormalized}. Note that ${\Mcal}_P^{rs}(\pvec_\ell)={\Mcal}_P^{rs}(|\pvec_\ell|)$ is a rotationally symmetric function of the lepton momentum~$\pvec_\ell$.
Energy conservation ($m_P=\omega_\ell+\omega_\nu$) has also been employed to rewrite
\begin{equation}
  |\pvec_\ell| = \omega_\nu = \frac{m_P}{2}\left(1-\frac{m_\ell^2}{m_P^2}\right) \ \ \text{and} \ \ \omega_\ell =\frac{m_P}{2}\left(1+\frac{m_\ell^2}{m_P^2}\right)~.
\end{equation}

We may expand now the QCD+QED squared matrix element $|\Mcal_P(\pvec_\ell)|^2$ in~\cref{eq:DecayRate_full} around the iso-symmetric QCD point keeping the lepton momentum to its physical value as
\begin{equation}
\label{eq:MatrixElement_expansion}
    |\Mcal_P(\pvec_\ell)|^2 = |\Mcal_P^\iso(\pvec_\ell)|^2 + \delta |\Mcal_P(\pvec_\ell)|^2 + \order(\varepsilon^2)\,.
\end{equation}
In iso-QCD the matrix element factorizes into a hadronic and leptonic part, namely
\begin{align}
    |\Mcal_P^{\iso}(\pvec_\ell)|^2 &= \mathcal{Z}_0^2\,|\braket{\ell^+,r,{\pvec}_{\ell};\,\nu_\ell,s,{\pvec}_{\nu}|{O}_W|P,\boldsymbol{0}}^\iso|^2 \nn\\
    & = \mathcal{Z}_0^2 \,|\mathcal{A}_P^\iso|^2 \,\left|\Lcal(\pvec_\ell)\right|^2\,,
\end{align}
where
\begin{equation}
  \label{eq:AP0}
\mathcal{Z}_0\,\mathcal{A}_P^\iso\equiv -\mathcal{Z}_0 \braket{0|J_H^0|P,\boldsymbol{0}}^\iso = \mathcal{Z}_0 \braket{0|\bar{q}_2\,\gamma^0\gamma_5\,q_1|P,\boldsymbol{0}}^\iso = \ii \, m_{P}^\iso f_P
\end{equation}
is the iso-QCD renormalized axial matrix element expressed in terms of the mass $m_{P}^\iso$ of the meson state $|P,\boldsymbol{0}\rangle^\iso$ and the decay constant $f_P$, while
\begin{equation}
  \label{eq:LeptonicTensor}
  \Lcal^{rs}(\pvec_\ell) 
  = \braket{ \ell^+,r,{\pvec}_{\ell}; \, \nu_\ell,s,{\pvec}_{\nu} | J_L^0 | 0 }^\iso  = \bar{u}^r_\nu({\pvec}_{\nu}) \, \gamma^0(1-\gamma_5) \,v^s_\ell({\pvec}_{\ell}) 
\end{equation} 
is the tree-level leptonic tensor with $v_\ell^r({\pvec}_{\ell})=\braket{\ell^+,r,{\pvec}_{\ell}|\ell|0}^\iso$ and $\bar{u}_\nu^r({\pvec}_{\nu})=\braket{\nu_\ell,s,{\pvec}_{\nu}|\bar{\nu}_\ell|0}^\iso$  the free Dirac spinors. We have considered here only the $\rho=0$ component of $O_W$~(\cref{eq:OW}) since this is the only one contributing to the axial matrix element in the meson rest frame. Using the completeness relations for spinors in Euclidean space
\begin{equation}
    \sum_{r,r'} v^{r'}_\ell(\pvec_\ell)\bar{v}^r_\ell(\pvec_\ell) = - \ii \pslash_\ell-\mell~,\qquad \sum_{s,s'} u^{s}_\nu(\pvec_\nu)\bar{u}^{s'}_\nu(\pvec_\nu) = - \ii \pslash_\nu~,
\end{equation}
one gets
\begin{equation}
|\Lcal({\pvec}_{\ell})|^2=\sum_{r,s}|\Lcal^{rs}({\pvec}_{\ell})|^2 = 8|{\pvec}_{\ell}|({\omega}_{\ell}-|{\pvec}_{\ell}|) = 4 m_\ell^2\left(1-\frac{m_\ell^2}{m_{P}^2}\right)
  \label{eq:LeptonicTrace}
\end{equation}
and hence
\begin{equation}
  |\Mcal_P^\iso(\pvec_\ell)|^2 = \mathcal{Z}_0^2 \,|\mathcal{A}_P^\iso|^2 \,\left|\Lcal(\pvec_\ell)\right|^2  = 4 m_\ell^2\, \left(1-\frac{m_\ell^2}{m_{P}^2}\right)^2m_{P}^{\iso\,2}f_{P}^2~.
\end{equation}

Following the convention of the PDG~\cite{Workman:2022ynf}, we define the ``tree-level'' decay rate as
\begin{equation}
  \Gamma_P^\mathrm{tree} = 
  \mathcal{K}\,
  |\Mcal_P^\mathrm{tree}|^2 = 
  \mathcal{K}\,  \bigg(\frac{m_P}{m_P^\iso}\bigg)^2 |\Mcal_P^\iso|^2 = \frac{G_F^2}{8 \pi} \left|V_\mathrm{q_1q_2}\right|^2 m_{\ell}^2 \bigg ( 1 - \frac{m_{\ell}^2}{m_{P}^2}\bigg )^2 m_P f_{P}^2 ~,
  \label{eq:DecayRate_tree}
\end{equation}
i.e. with all masses defined in the full theory and only the decay constant $f_{P}$ defined in iso-QCD. Combining the above~\cref{eq:MatrixElement,eq:MatrixElement_expansion,eq:DecayRate_tree} with~\cref{eq:deltaRP_virt} we obtain
\begin{align}
    \delta R_P^\mathrm{latt} 
    = \frac{\delta |\Mcal_P(\pvec_\ell)|^2}{|\Mcal_P^\iso(\pvec_\ell)|^2} - 2\,\frac{\delta m_P}{m_P^\iso}=2\left(\frac{\delta \mathcal{A}_P}{\mathcal{A}_{P}^\iso}-\frac{\delta {m}_P}{{m}_{P}^\iso} + \frac{\delta \mathcal{Z}}{\mathcal{Z}_{0}}\right)\,,
    \label{eq:deltaRP_latt_def}
\end{align}
where we have defined the leading IB corrections to the meson mass $\delta m_P$,
\begin{equation}
    m_P=m_P^\iso +\delta m_P + \order(\varepsilon^2)\,,
\end{equation}
and those to the bare matrix element as
\begin{equation}
\label{eq:deltaAP/AP}
    \frac{\delta \Acal_P}{\Acal_P^\iso} \equiv   \mathrm{Re}\,\Bigg\{ 
    -\frac{\sum_{r,s} \delta \widebar\Mcal_P^{rs}(\pvec_\ell) \big[\mathcal{L}^{rs}(\pvec_\ell)\big]^\dagger}{\Acal_P^\iso\,|\mathcal{L}(\pvec_\ell)|^2}
    \Bigg\}
    \,.
\end{equation}
As discussed above, the quantity ${\delta \mathcal{Z}}/{\mathcal{Z}_{0}}$ does not depend on the masses of the decaying meson and hence our target quantity $\delta R_{K\pi}^\mathrm{latt}$ is given by
\begin{equation}
  \label{eq:deltaRKPi_latt}
  \delta R_{K\pi}^\mathrm{latt} = 2\left(\frac{\delta \Acal_K}{\Acal_{K}^\iso}-\frac{\delta {m}_K}{{m}_{K}^\iso} \right) - 2\left(\frac{\delta \Acal_\pi}{\Acal_{\pi}^\iso}-\frac{\delta {m}_\pi}{{m}_{\pi}^\iso} \right)\,.
\end{equation}

\begin{figure}[b]
  \centering
  \includegraphics[width=0.9\textwidth]{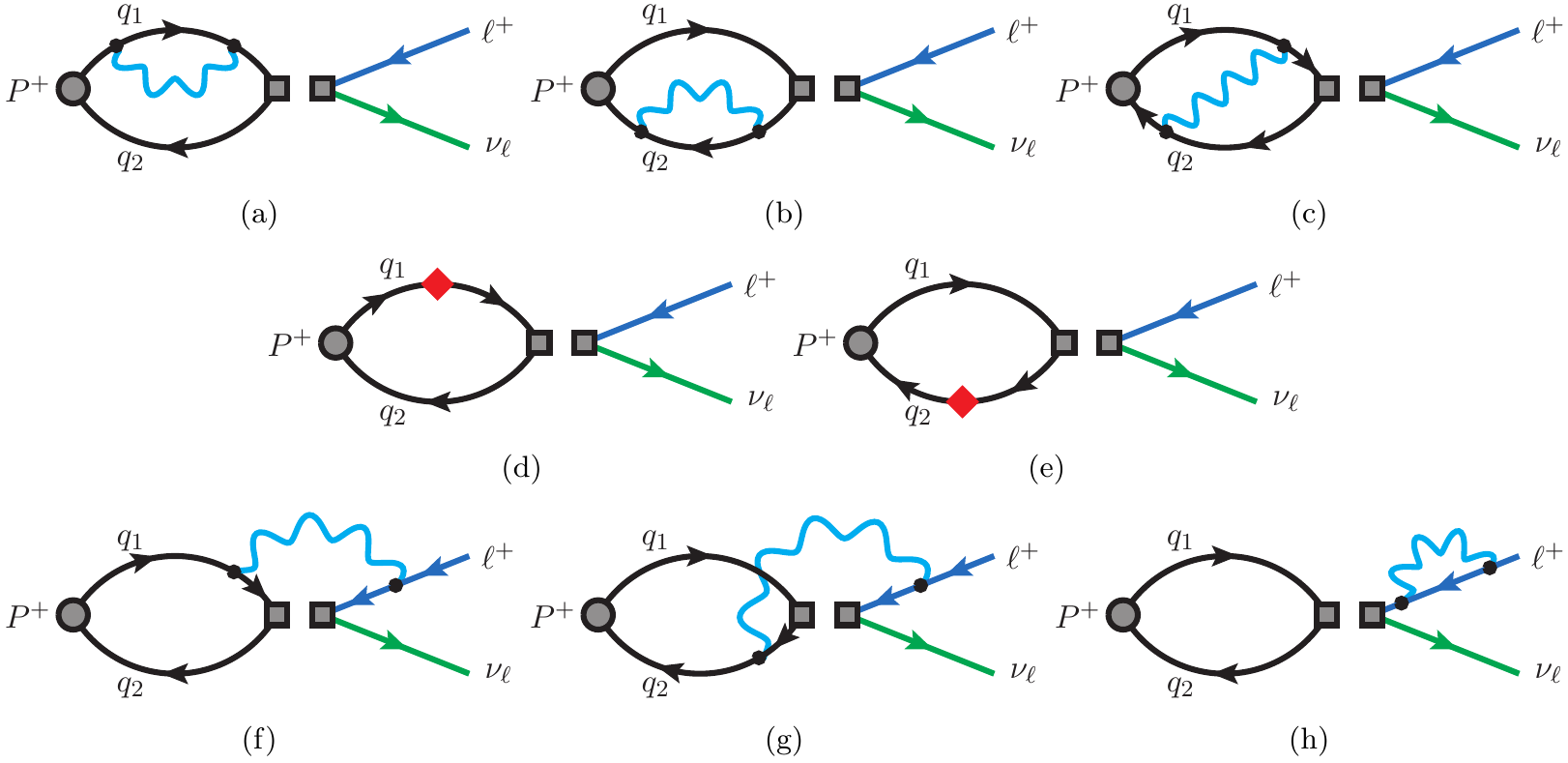}
  \caption{Quark-connected Feynman diagrams contributing to the leading IB corrections to the weak decay. The wiggly lines correspond to photons, and the diamond-shaped vertices are scalar insertions. }
  \label{fig:alldiags}
\end{figure}

\begin{figure}[t]
  \centering
  \includegraphics[width=0.9\textwidth]{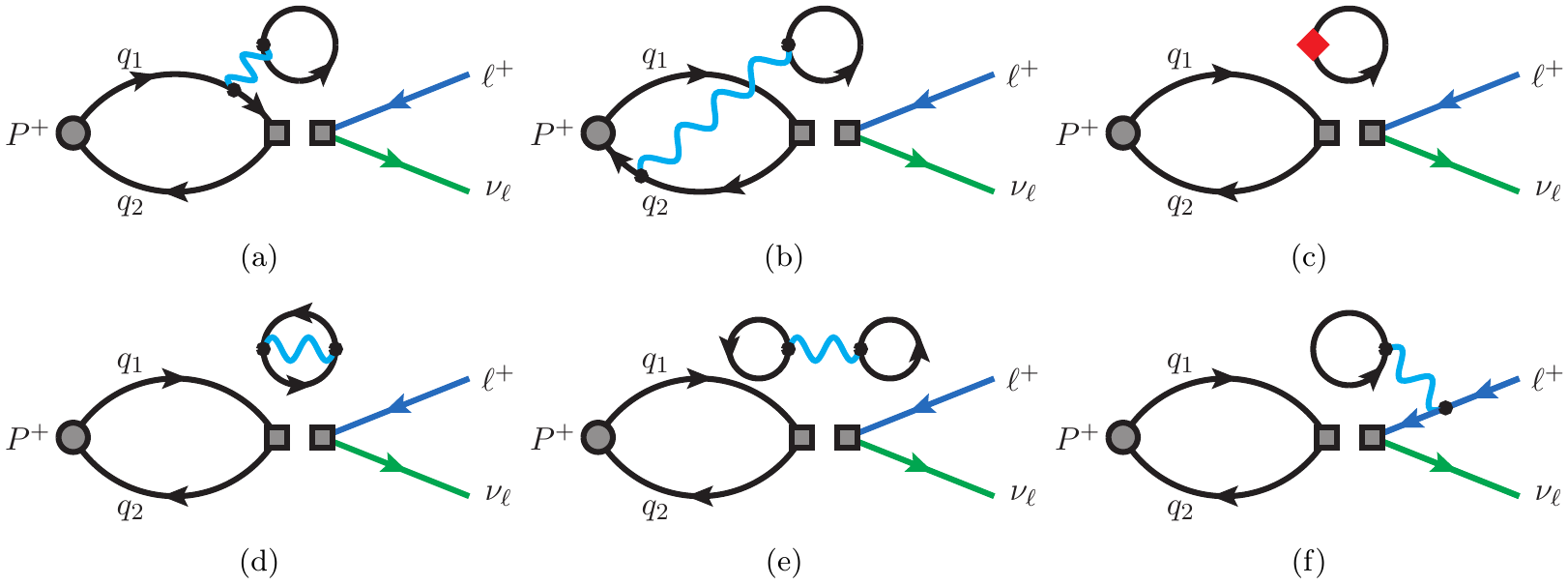}
  \caption{Quark-disconnected Feynman diagrams contributing to the leading IB corrections to the weak decay.}
  \label{fig:alldiags_disc}
\end{figure}

We can distinguish three kinds of corrections to the matrix element, that we denote as
\begin{equation}
\label{eq:dAP_contributions}
    \delta \Acal_P = \delta \Acal_P^\mathrm{f} + \delta \Acal_P^\mathrm{nf} + \delta \Acal_P^\mathrm{\ell}\,.
\end{equation}
The first term contains corrections involving only the quarks and these are proportional to either the quark fractional charges or to the bare quark mass splittings. These are obtained from the corrections to the bare matrix element
\begin{align}
    \big[\delta \widebar\Mcal_P^{rs}(\pvec_\ell)\big]^\mathrm{f} 
    &= \bigg[
    \frac{1}{2} \sum_{\mathrm{q},\mathrm{q}^\prime} e_{\mathrm{q}}e_{\mathrm{q}^\prime} \, \frac{\partial^2}{\partial e_{\mathrm{q}} \partial e_{\mathrm{q}\prime}}
    +\sum_{\mathrm{q}}\,  (\hat{m}^\phi_\mathrm{q}-\hat{m}_{\mathrm{q}}^\iso) \, \frac{\partial }{\partial\hat{m}_\mathrm{q}}\bigg]\,  \widebar\Mcal_P^{rs}(\pvec_\ell) \bigg|_{\sigmavec^\iso} \\
    &= -\,\Lcal^{rs}(\pvec_\ell)\,\bigg[
    \frac{1}{2} \sum_{\mathrm{q},\mathrm{q}^\prime} e_{\mathrm{q}}e_{\mathrm{q}^\prime} \, \frac{\partial^2}{\partial e_{\mathrm{q}} \partial e_{\mathrm{q}\prime}}
    +\sum_{\mathrm{q}}\,  (\hat{m}^\phi_\mathrm{q}-\hat{m}_{\mathrm{q}}^\iso) \, \frac{\partial }{\partial\hat{m}_\mathrm{q}}\bigg] \, \Acal_P^\phi \bigg|_{\sigmavec^\iso} \,,\nn 
\end{align}
with $e_{\mathrm{q}_1}=+2/3|e|$ and $e_{\mathrm{q}_2}=-1/3|e|$, and 
$\Acal_P^\phi$ the axial matrix element evaluated in the full theory. $\sigmavec^\iso$ indicates that the quantities are evaluated in the target iso-QCD theory, $\sigmavec^\iso=(g,0,\hat{\mvec}^\iso)$ as discussed in \cref{sec:isoqcd}. Since in this case the decay amplitude factorizes into a hadronic and a leptonic part, we refer to these contributions as ``factorizable''. The relevant diagrams contributing to these corrections are depicted in figures~\ref{fig:alldiags}(a)--\ref{fig:alldiags}(e) and \ref{fig:alldiags_disc}(a)--\ref{fig:alldiags_disc}(e).
The second term in \cref{eq:dAP_contributions} corresponds instead to the ``non-factorizable'' corrections to the matrix element where a photon is exchanged between a quark and the charged lepton ($e_\ell = -|e|$). These are given by
\begin{equation}
\label{eq:deltaM_nf}
    \big[\delta \widebar{\Mcal}_P^{rs}(\pvec_\ell)\big]^\mathrm{nf} 
    =\bigg[\frac{1}{2} e_\ell \sum_{\mathrm{q}} e_{\mathrm{q}} \, \frac{\partial^2}{\partial e_{\mathrm{q}} \partial e_{\ell}}  \bigg] \, \widebar{\Mcal}_P^{rs}(\pvec_\ell) \bigg|_{\sigmavec^\iso}\,,
\end{equation}
and the corresponding diagrams are shown in figures~\ref{fig:alldiags}(f)--\ref{fig:alldiags}(g) and~\ref{fig:alldiags_disc}(f).
Finally, the third term in \cref{eq:dAP_contributions} consists in the $\order(e_\ell^2)$ contribution of the lepton self-energy in~\cref{fig:alldiags}(h), which is proportional to $\Acal_P^\iso$ with a factor that can be computed analytically in perturbation theory,
\begin{align}
    \big[\delta \widebar\Mcal_P^{rs}(\pvec_\ell)\big]^\mathrm{\ell} &= \bigg[ \frac{1}{2} e_\ell^2  \, \frac{\partial^2 }{\partial e_{\ell}^2} \bigg]\, \widebar\Mcal_P^{rs}(\pvec_\ell) \bigg|_{\sigmavec^\iso} = - \Acal_P^\iso \,\bigg[ \frac{1}{2} e_\ell^2  \, \frac{\partial^2 }{\partial e_{\ell}^2} \bigg]\, \Lcal_\phi^{rs}(\pvec_\ell)\bigg|_{\sigmavec^\iso} \,,
\end{align}
with $\Lcal^{rs}_\phi(\pvec_\ell) 
  = \braket{ \ell^+,r,{\pvec}_{\ell}; \, \nu_\ell,s,{\pvec}_{\nu} | J_L^0 | 0 }^\phi$.
This perturbative correction, however, cancels in the difference $[\Gamma_0(L)-\Gamma_0^{(2)}(L)]$ in~\cref{eq:strategy} and therefore can be neglected in practice in the calculation. Of course, the lepton self-energy must be included in $\Gamma_P^\mathrm{pert}(m_\gamma)$.


Due to the numerical difficulty of evaluating the quark disconnected diagrams in~\cref{fig:alldiags_disc} on the lattice, in this work we employ the electro-quenched approximation. This consists in treating the sea quarks as if they were electrically neutral and hence, in practice, neglecting the diagrams in~\cref{fig:alldiags_disc}. The deviations from this approximation are expected to be small, and we assign an associated systematic uncertainty in our final prediction. 
We are currently working on overcoming this approximation and the progress of our preliminary study has been reported in ref.~\cite{Harris22}.

\subsection{Extracting matrix elements from Euclidean correlation functions}
\label{sec:mel_from_corr}

The IB corrections to meson masses, $\delta m_P$, and to the decay amplitude, $\delta \mathcal{A}_P$, which are needed to compute $\delta R_P^\mathrm{latt}$ in~\cref{eq:deltaRP_latt_def}, can be obtained from the study of the large time behaviour of suitably defined Euclidean correlation functions. Here the correlation functions are studied in the continuum and in a volume with infinite temporal extent. The subtraction of the effects due to the finite spatial extent of the lattice, $L$, are discussed later in~\cref{sec:FVE}, while finite-time corrections to these quantities will be addressed in~\cref{sec:latt_correlators}, together with the details on the lattice implementation of the correlation functions.

\paragraph{Tree-level correlation function:}

We start by defining the tree-level correlation function for the decay $P^+\to\ell^+\nu_\ell$, with the aim of extracting the tree-level matrix element $\mathcal{A}_{P}^\iso$ defined in~\cref{eq:AP0}. As discussed  in~\cref{sec:virtual}, in the absence of QED the matrix element for the operator $O_W$ is factorisable into a hadronic and a leptonic part. As a consequence, we can extract the hadronic matrix element $\mathcal{A}_{P}^\iso$ from a pure QCD two-point correlation function without the need of including leptons in the calculation.
Let ${\phi_P^\dagger(x)=\bar{q}_1(x)\gamma_5q_2(x)}$ be the interpolating operator for the pseudoscalar meson~$P^+$ 
and define the Euclidean correlation functions
\begin{equation}
\label{eq:C_PA_PP}
\resizebox{.9\hsize}{!}{
$\displaystyle
C_\mathrm{PA}(t)  \equiv  \int \dd{3}{\xvec} \, \braket{0|\mathrm{T}\big[A^0(t,\xvec)\phi^\dag_P(0)\big]|0}\,,\ 
  C_\mathrm{PP}(t) \equiv  \int \dd{3}{\xvec} \, \braket{0|\mathrm{T}\big[\phi_P(t,\xvec)\phi^\dag_P(0)\big]|0}\,,
  $}
\end{equation}
with $A^0(x)=\bar{q}_2(x)\gamma^0\gamma_5q_1(x)$ the temporal component of the hadronic axial current and the meson being projected on zero spatial momentum. For simplicity, we use translational invariance to create the meson at the origin. In practice, lattice correlators have been computed for several positions $x_P=(t_P,\xvec_P)$ and then shifted and averaged over all the volume to improve the statistical precision (see~\cref{sec:latt_correlators}). Note that these are generic correlation functions evaluated at a given point $\sigmavec$.
Fixing $t>0$, the correlation functions in~\cref{eq:C_PA_PP} have the following spectral decomposition
\begin{equation}
    \label{eq:C_PA_PP_spectr}
    C_\mathrm{PA}(t) = \frac{\mathcal{A}_P\,Z_P}{2\, m_P}\,\e^{-m_P t} \ + \  \dots \,,\quad
    C_\mathrm{PP}(t) =  \  \frac{|Z_P|^2}{2\, m_P}\ \,  \e^{-m_P t} \  + \  \dots\,,
\end{equation}
where $Z_P=\braket{P,\boldsymbol{0}|\phi^\dagger(0)|0}$ and the ellipses stand for contributions of heavier states that decay exponentially faster than the leading terms. The combined study of the two correlation functions evaluated in iso-QCD allows one to extract the meson mass $m_P^\iso$ and the matrix elements $Z_P^\iso$ and~$\mathcal{A}_P^\iso$. 

\paragraph{Factorizable correlators:}

When IB corrections only involve the constituent quarks of the decaying meson, the matrix element is still factorizable into a hadronic and a leptonic part. Also in this case we can make use of the correlation functions in~\cref{eq:C_PA_PP}.
Defining the leading factorizable corrections to the correlators as 
\begin{equation}
\label{eq:deltaC_fact}
    \delta C^\mathrm{f}_\mathrm{PA}(t) = \bigg[
    \frac{1}{2} \sum_{\mathrm{q},\mathrm{q}^\prime} e_{\mathrm{q}}e_{\mathrm{q}^\prime} \, \frac{\partial^2}{\partial e_{\mathrm{q}} \partial e_{\mathrm{q}\prime}}
    +\sum_{\mathrm{q}}\,  (\hat{m}^\phi_\mathrm{q}-\hat{m}_{\mathrm{q}}^\iso) \, \frac{\partial }{\partial\hat{m}_\mathrm{q}}\bigg] C_\mathrm{PA}(t) \bigg|_{\sigmavec^\iso}
\end{equation}
and analogously for $\delta C^\mathrm{f}_\mathrm{PP}(t)$, one gets the following decomposition for their ratios with the corresponding tree-level correlators
\begin{align}
    \label{eq:deltaC_fact_PA_spectr}
    \mathcal{R}_\mathrm{PA}^\mathrm{f}(t) &\equiv \frac{\delta C^\mathrm{f}_\mathrm{PA}(t)}{C_\mathrm{PA}^\iso(t)} = \frac{\delta \Acal_P^\mathrm{f}}{\Acal_P^\iso} + \frac{\delta Z_P}{Z_P^\iso} - \frac{\delta m_P}{m_P^\iso}\,(1+m_P^\iso t)+\dots\,,\\
    \label{eq:deltaC_fact_PP_spectr}
    \mathcal{R}_\mathrm{PP}^\mathrm{f}(t) &= \frac{\delta C^\mathrm{f}_\mathrm{PP}(t)}{C_\mathrm{PP}^\iso(t)} = 2\,\frac{\delta Z_P}{Z_P^\iso} - \frac{\delta m_P}{m_P^\iso}\,(1+m_P^\iso t) +\dots\,,
\end{align}
from a Taylor expansion of the spectral decomposition of the form \cref{eq:C_PA_PP_spectr}.
The slope in $t$ of the above ratios corresponds to the mass shift $\delta m_P$, and by combining the constant coefficients we can obtain the correction $\delta \Acal_P^\mathrm{f}/\Acal_P^\iso$.

\paragraph{Non-factorizable correlators:}
In order to obtain the non-factorizable IB corrections to the decay amplitude, we start defining the following QCD+QED correlation function
\begin{equation}
  \resizebox{.9\hsize}{!}{$\displaystyle
    C_\mathrm{P\ell}(t,t_\ell) = \int\dd{3}{\xvec} \dd{3}{\xvec_\ell} \dd{3}{\xvec_\nu} \e^{-\ii \pvec_\nu\cdot\xvec_\nu-\ii \pvec_\ell\cdot\xvec_\ell}   \bra{0} \mathrm{T} \big[\nu(t_\ell,\xvec_\nu) \bar{\ell}(t_\ell,\xvec_\ell)  O_W(0)  \phi_P^\dagger(-t,-\xvec) \big]\ket{0},
    $}
\end{equation}
where for simplicity we have set the temporal coordinate of the neutrino and the lepton to be equal. Also in this case we have used translational invariance to insert the weak Hamiltonian at the origin. Fixing $t>0$ and $t_\ell>0$ we have that in iso-QCD the above correlator becomes
\begin{equation}
\label{eq:C_0}
     C_\mathrm{P\ell}^\iso(t,t_\ell) = 
    -   \bra{0}  J_H^\rho(0) \ \phi_P^\dagger(-t,\pvec_P=\boldsymbol{0}) \ket{0} \times S_\nu(t_\ell,\pvec_\nu|0) \, \gamma^\rho(1-\gamma_5) \, S_\ell(0|t_\ell,\pvec_\ell)~,
\end{equation}
with 
\begin{align}
    \label{eq:prop_neutrino}
    S_\nu(t_\ell,\pvec_\nu|0) &= \int\dd{3}{\xvec_\nu}\, \e^{-\ii \pvec_\nu\cdot\xvec_\nu} S_\nu(t_\ell,\xvec_\nu|0) = \frac{\e^{-\omega_\nu t_\ell }}{2 \omega_\nu} \, \sum_s u_\nu^s(\pvec_\nu)\bar{u}_\nu(\pvec_\nu)\,,\\
    \label{eq:prop_lepton}
    S_\ell(0|t_\ell,\pvec_\ell) &= \int\dd{3}{\xvec_\ell}\, \e^{-\ii \pvec_\ell\cdot\xvec_\ell} S_\ell(0|t_\ell,\xvec_\ell) = -\,\frac{\e^{-\omega_\ell t_\ell }}{2 \omega_\ell} \, \sum_r v_\ell^r(\pvec_\ell)\bar{v}_\ell^r(\pvec_\ell)\,.
\end{align}
Using~\cref{eq:C_PA_PP_spectr} we get the following spectral decomposition in iso-QCD
\begin{equation}
    C_\mathrm{P\ell}^\iso(t,t_\ell) = - \sum_{r,s} \frac{\e^{-m_P^\iso t}\e^{-\omega_\ell t_\ell}\e^{-\omega_\nu t_\ell}}{8 m_P^\iso \omega_\ell\omega_\nu }  Z_P^\iso\,u_\nu^r(\pvec_\nu)\left\{\Acal_P^\iso \Lcal^{rs}(\pvec_\ell)\right\}\bar{v}_\ell^s(\pvec_\ell)+\dots\,,
\end{equation}
with $\mathcal{L}^{rs}(\pvec_\ell)$ defined in~\cref{eq:LeptonicTensor}.
Note that $C_\mathrm{P\ell}^\iso(t,t_\ell)$ is a matrix in Dirac space and that tracing with $\gamma^0_L=\gamma^0(1-\gamma_5)$ gives
\begin{equation}
    \mathrm{Tr}\big[\gamma^0_L\,C_\mathrm{P\ell}^\iso(t,t_\ell)\big] = - \frac{\e^{-m_P^\iso t}\e^{-\omega_\ell t_\ell}\e^{-\omega_\nu t_\ell}}{8 m_P^\iso \omega_\ell\omega_\nu }  Z_P^\iso \Acal_P^\iso |\Lcal(\pvec_\ell)|^2+\dots\,.
\end{equation}
We now define the non-factorizable correlator as
\begin{align}
    \delta C_\mathrm{P\ell}^\mathrm{nf}(t,t_\ell) &=  \int \dd{4}{x} \dd{4}{y} \bra{0}  \mathrm{T}\big[J_H^\rho(0) \, V^\mu(x)\, \phi_P^\dagger(-t,\pvec_P=\boldsymbol{0}) \big]\ket{0} \times \Delta_{\mu\nu}(x-y) \nn\\
    &\hspace{3cm} \times S_\nu(t_\ell,\pvec_\nu|0) \, \gamma^\rho(1-\gamma_5) \, S_\ell(0|y)\,\gamma^\nu \,S_\ell(y|t_\ell,\pvec_\ell) \,,
    \label{eq:deltaC_nf}
\end{align}
where $V^\mu(x)=\sum_\mathrm{q=1}^{\Nf} V^\mu_\mathrm{q}(x)$ is the Euclidean quark electromagnetic current and $\Delta_{\mu\nu}(x-y)$ the photon propagator. Here we have used $V^\nu_\ell(y) = \bar{\ell}(y)\gamma^\nu \ell(y)$ for the leptonic electromagnetic current.
The correlator in~\cref{eq:deltaC_nf} is obtained by applying the derivatives of~\cref{eq:deltaM_nf} to the QCD+QED correlator $C_\mathrm{P\ell}(t,t_\ell)$. The asymptotic behaviour of the non-factorizable correlator is
\begin{equation}
        \delta C_\mathrm{P\ell}^\mathrm{nf}(t,t_\ell) = \sum_{r,s} \frac{\e^{-m_P^\iso t}\e^{-\omega_\ell t_\ell}\e^{-\omega_\nu t_\ell}}{8 m_P^\iso \omega_\ell\omega_\nu }  Z_P^\iso\,u_\nu^r(\pvec_\ell)\ \big[\delta \widebar\Mcal_P^{rs}(\pvec_\ell)\big]^\mathrm{nf}\ \bar{v}_\ell^s(\pvec_\ell)+\dots\,.
\end{equation}
Tracing the correlator with $\gamma^0_L$ and making use of~\cref{eq:deltaAP/AP} we can obtain the desired non-factorizable correction to the decay amplitude as
\begin{equation}
    \label{eq:deltaCnf/C0}
    \mathcal{R}_\mathrm{P\ell}^\mathrm{nf}=\mathrm{Re}\left[\frac{\mathrm{Tr}\big[\gamma^0_L\, \delta C_\mathrm{P\ell}^\mathrm{nf}(t,t_\ell)\big]}{\mathrm{Tr}\big[\gamma^0_L\, C_\mathrm{P\ell}^\iso(t,t_\ell)\big]}\right] = \frac{\delta \Acal_P^\mathrm{nf}}{\Acal_P^\iso} +\dots\,.
\end{equation}

\subsection{Subtraction of finite-volume effects}
\label{sec:FVE}
In the calculation presented in this work we adopt the $\QEDL$ prescription, first introduced in ref.~\cite{Hayakawa:2008an}. As discussed in a number of publications~\cite{Hayakawa:2008an,Davoudi:2014qua,Borsanyi:2014jba,Davoudi:2018qpl}, charged states are not well-defined in a naive implementation of finite-volume $\QED$ (in which periodic boundary conditions are applied to the photon fields).
The $\QEDL$ approach solves this by discarding the zero spatial-momentum mode of the photon on each energy slice. The resulting momentum-space photon propagator in Feynman gauge then takes the simple form
\begin{equation}
	\Delta_L^{\mu\nu}(k_0,\kvec)=\delta ^{\mu\nu}\frac{1-\delta_{\kvec,\boldsymbol{0}}}{k_0^2+\kvec^2}~.
\end{equation}
This prescription solves the issue of zero-mode singularities in a periodic volume, at the cost of violating locality in space at finite volume. Nevertheless, this theory has a well-defined and local limit if the infinite-volume extrapolation is performed before the continuum limit. Additionally, $\QEDL$ has been the dominant prescription so far in high-precision lattice QCD+QED calculations, including radiative corrections to leptonic decays~\citep{Giusti:2017dwk,DiCarlo:2019thl} and isospin-breaking corrections to the muon anomalous magnetic moment~\cite{RBC:2018dos,Borsanyi:2020mff}.
Alternative strategies exist which preserve locality, such as introducing a photon mass~\cite{Endres:2015gda} or using non-periodic boundary conditions~\cite{Lucini:2015hfa}. However, these approaches affect other fundamental symmetries (gauge invariance and charge conservation, respectively) and their finite-volume behaviour for processes such as weak decays is currently not as well studied as in the case of $\QEDL$.

As it is described in detail in ref.~\cite{DiCarlo:2021apt}, the Feynman rules of $\QEDL$ can be used to predict the $L$ dependence of any lattice quantity by representing the latter in terms of QCD vertex functions at fixed order in $\QED$. In particular, this allows one to analytically predict the power-like volume dependence, order by order in $1/L$, for the virtual-photon contribution to the leptonic decay rate targeted in this work.
This strategy is already implicit in \cref{eq:strategy}, where the subtracted quantity in the first term, denoted by $\Gamma^{(2)}_0(L)$, is defined as the analytic $\QEDL$ prediction through $\order(1/L^2)$.

An extension of $\Gamma^{(2)}_0(L)$ to $n$ orders in $1/L$ can be written as
\begin{align}
  \Gamma^{(n)}_0(L) & =  \Gamma _{0}^{\textrm{tree}} \bigg [1 + \delta R^{(n)}_P(L) \bigg ]   \,,
\end{align}
where
\begin{align}\label{eq:deltaRndef}
  \delta R^{(n)}_P(L) & =  2\, \frac{\aem }{4\pi} \bigg (   \widetilde Y_P (L)  + \sum _{i=0}^{n} \frac{Y_{P, \, i}}{L^i}  \bigg ) \,,
\end{align}
isolates the $\order(\aem)$ contribution of direct interest to us. The first term in parentheses, $\widetilde Y_P (L)$, combines the infinite-volume universal (point-like) contributions to the decay rate with those that are logarithmic in $L$.\footnote{In the notation of ref.~\cite{DiCarlo:2021apt} this quantity can be defined introducing a photon mass $\lambda$ as
\begin{equation*}
  \widetilde Y_P (L) = \lim_{\lambda \to 0} \big [Y_{P, \,  \text{IV}}^{\text{uni}}(\lambda) + Y_{P, \,  \text{log}} \log \frac{L \lambda}{2 \pi}  \big ] \,.
\end{equation*}} The functional form is given by \cite{DiCarlo:2021apt,PhysRevD.95.034504}
\begin{equation}
  \label{eq:YtildeL}
   \widetilde Y_P (L)  =  	- \frac{5}{4} + 2\log\left(\frac{m_\ell}{m_W}\right) + 2\log\left(\frac{m_\ell L}{2\pi}\right)  
   + \frac{2\log r_\ell}{|\vell|}  \left[\log \frac{m_PL}{2\pi} + \log \frac{m_\ell L}{2\pi} - 1\right]  \,,
\end{equation}
with $\vell = \mathbf{p}_{\ell}/\omega _\ell$ defining the 3-velocity of the lepton and $r_\ell=m_\ell/m_P$ the lepton-pseudoscalar mass ratio.

Equation~\eqref{eq:YtildeL} depends only on the masses of particles and is, in this sense, universal or structure-independent. In fact, one can show that the same is true for $Y_{P, \,  0}$ and $Y_{P, \,  1}$, while for $Y_{P, \,  n > 1}$ structure dependence enters through contributions from, e.g.,~form factors and their derivatives. For this reason, the point-like approximation can be used to calculate $Y_{P, \,  0}$ and $Y_{P, \,  1}$ and the full machinery introduced in ref.~\cite{DiCarlo:2021apt} is first required for the determination of $Y_{P, \,  2}$ and higher-order coefficients.

A summary of the knowledge to-date on these coefficients is given by the following:
\begin{align}\label{eq:ourresultyn}
	Y_{P, \,  0} & = \frac{c_3-2\, (c_3(\vell)-B_{1}(\vell))}{2\pi} + 2\, (1-\log 2) \, ,\nn
	\\ 
	Y_{P, \,  1} & = -\frac{(1+r_{\ell}^2)^2c_2-4\, r_{\ell}^2c_{2}(\vell)}{m_{P}(1-r_{\ell}^4)} \, ,
	\\
	Y_{P, \,  2} & = -\frac{F_A^P}{f_P}\frac{4\pi \left[ (1+r_{\ell}^2)^2-4 \, r_{\ell}^2c_1(\vell) \right]}{m_P (1-r_{\ell}^4)}+\frac{8\pi\left[ (1+r_{\ell}^2)c_1-2 \, c_1 (\vell)\right] }{m_P^2 (1-r_{\ell}^4)} \, ,\nn
	\\
	Y_{P, \,  3} & = \frac{32\pi^2 c_{0} \, (2+r_{\ell}^2)}{m_{P}^3 (1+r_{\ell}^2)^3 }+ 	Y_{P, \,  3}^{\textrm{sd}} \, ,\nn
\end{align}
where $c_j$ and $c_j(\vell)$ are known finite-volume coefficients, and $B_1(\vell)$ is a known special function. These quantities are all defined in ref.~\cite{DiCarlo:2021apt}.

In $Y_{P, \,  2}$, the structure-dependent ratio $F_A^P/f_P$ appears where $F_A^P$ is the on-shell zero-momentum axial form factor describing radiative leptonic decays and $f_P$ is the iso-QCD pseudoscalar decay constant. A key message is that, while the full result including structure dependence is known for $Y_{P, \,  2}$, the same is not true for $Y_{P, \,  3}$, for which the structure-dependent piece, denoted $Y_{P, \,  3}^{\textrm{sd}}$, has yet to be determined. As a result, the finite-volume subtractions available currently include $\delta R_P^{(2)}(L)$ and $\delta R_P^{(3), \text{pt}}(L)$, where $Y_{P, \,  3}^{\textrm{sd}}$ is set to zero in the latter. In this work we use $\delta R_P^{(2)}(L)$ to determine our central value and take the absolute difference $\vert \delta R_P^{(3), \text{pt}}(L) - \delta R_P^{(2)}(L) \vert $ to estimate a systematic uncertainty associated with neglected finite-volume effects.

Finally, we also need to consider finite-size effects from $\QED$ corrections to the meson mass $m_{P}$. For the finite-volume state with zero spatial momentum, these are given by~\cite{Davoudi:2014qua,Borsanyi:2014jba,Tantalo:2016vxk,Davoudi:2018qpl,DiCarlo:2021apt}
\begin{equation}
  \label{eq:massfves}
  \resizebox{.9\hsize}{!}{$\displaystyle
	\Delta m_{P}^2(L) = 
	e^2 m_P^2 \Bigg\{ 
	\frac{c_2}{4\pi^2 m_P L} + \frac{c_1}{2\pi (m_P L)^2} + \frac{m_P ^2 \langle r_P^2 \rangle }{3 (m_PL)^3}
	+ \frac{\mathcal{C}}{(m_P L)^3} + \order \bigg[ \frac{1}{(m_P L)^4}\bigg] 
	\Bigg\} \, ,
  $}
\end{equation}
where $\langle r_P^2 \rangle$ is the squared electromagnetic charge radius known from experiments, dispersion theory and lattice simulations~\cite{Workman:2022ynf,Aoki:2021kgd}, and $\mathcal{C}$ is an unknown contribution, arising from the branch-cut in the Compton amplitude evaluated with zero spatial momentum for both the photon and the pseudoscalar. Because $\mathcal{C} > 0$ \cite{DiCarlo:2021apt}, subtracting the charge-radius dependent piece is guaranteed to reduce the $O(1/L^3)$ finite-volume effects, though it does not fully remove the $1/L^3$ scaling. In this work we use the predicted volume dependence through $\order(1/L^2)$ to estimate the infinite-volume pseudoscalar mass. As with the decay rate, we take the difference between the $1/L^2$ and partial $1/L^3$ results as a systematic uncertainty.

\subsection{Inclusion of real photon emission}
\label{sec:real_photon}

Lastly, we need to include the contributions from a real photon emission, namely the quantity $\delta R_P^\mathrm{pert}(\omega_\gamma^\mathrm{max})$ in~\cref{eq:deltaRP_limits}. 
To this end, we adopt the formulation discussed in detail in ref.~\cite{Carrasco:2015xwa}. Most notably, if the photon energy threshold, $\omega_\gamma^\mathrm{cut}$, is small enough, one may treat the initial hadron as a point-like particle and compute the inner bremsstrahlung term analytically. However, since structure-dependent contributions are negligible for the decays studied in this work, we can set $\omega_\gamma^\mathrm{cut}$ to the maximum value allowed for the photon energy, namely $\omega_\gamma^\mathrm{max}=m_P(1-r_\ell^2)/2$.  We report here the result obtained in ref.~\cite{Carrasco:2015xwa}, 
\begin{align}
        \delta R_P^\mathrm{pert}(\omega_\gamma^\mathrm{max}) &= \lim_{m_\gamma \rightarrow 0} \ \delta R_P^\mathrm{pert}(\omega_\gamma^\mathrm{max},m_\gamma)\nn \\
        & = \frac{\aem}{4\pi} \bigg[
        3\log \left(\frac{m^2_P}{m^2_W}\right) - 8\log(1-\rl) - \frac{3\,\rl[4]}{(1-\rl)^2}\log\rl -8\frac{1+\rl}{1-\rl}\,\mathrm{Li}_2 (1-\rl)\nn \\
        &\qquad\qquad + \frac{13-19\,\rl}{2(1-\rl)} + \frac{6-14\,\rl-4(1+\rl)\log(1-\rl)}{1-\rl}\log\rl \bigg]\,.
        \label{eq:deltaRP_pert_max}
\end{align}

\section{Lattice methodology}
\label{sec:methodology}

In this Section we discuss the lattice implementation of the correlation functions relevant for the calculation of IB corrections to the leptonic decay rate and give the details of our lattice setup.  

\subsection{Lattice QCD+QED path integrals}
\label{sec:lattice_QCQQED_path_int}
As anticipated in~\cref{sec:IB_effects}, IB corrections are computed in this work using the RM123 perturbative method~\cite{deDivitiis:2013xla}, which consists in expanding the path integral for a given physical observable around the iso-QCD point. In practice, since our lattice setup has been tuned to an iso-symmetric point different from the target one described in~\cref{sec:IB_effects}, we follow a two-step procedure to get our perturbative corrections. This consists in expanding both the full QCD+QED and the iso-QCD path integral around the simulation point, and get the desired correction as the difference of the two.

Let  $\langle \hat{\mathcal{O}}\rangle_{\sigmavec^\phi}$ be the expectation value of an observable $\hat{\mathcal{O}}$ calculated (in lattice units) in terms of the discretized Euclidean path integral in the full QCD+QED theory with bare parameters $\sigmavec^\phi=(g,e^\phi,\hat{\mvec}^\phi)$,
\begin{equation}
    \langle \hat{\mathcal{O}}\rangle_{\sigmavec^\phi} = \frac{1}{Z^\phi} \int \mathcal{D}[U] \mathcal{D}[A] \mathcal{D}[\psi,\bar{\psi}] \, \hat{\mathcal{O}}[\psi,\bar{\psi},U,A;\sigmavec^\phi] \, \e^{-S_F[\psi,\bar{\psi},U,A;\sigmavec^\phi]}\, \e^{-S_\gamma[A]} \,  \e^{-\frac{1}{g^2}S_G[U]},
\end{equation}
with $S_F[\psi,\bar{\psi},U,A;\sigmavec^\phi]$ being the fermionic action, and $S_\gamma[A]$ and $S_G[U]$ the $\QEDL$ and QCD gauge actions, respectively. $Z^\phi$ denotes instead the QCD+QED partition function. Here we keep the discussion general and allow the observable $\hat{\mathcal{O}}$ to depend on the electromagnetic coupling $e^\phi$. Let $\langle \hat{\mathcal{O}}\rangle_{\sigmavec^\iso}$ be  the corresponding expectation values calculated in the target iso-QCD theory, $\sigmavec^\iso=(g,0,\hat{\mvec}^\iso)$, 
\begin{equation}
    \langle \hat{\mathcal{O}}\rangle_{\sigmavec^\iso} = \frac{1}{Z^\iso} \int \mathcal{D}[U] \mathcal{D}[\psi,\bar{\psi}] \ \hat{\mathcal{O}}[\psi,\bar{\psi},U] \ \e^{-S_F[\psi,\bar{\psi},U;\sigmavec^\iso]}\,  \e^{-\frac{1}{g^2}S_G[U]}\,,
\end{equation}
and $\langle \hat{\mathcal{O}}\rangle_{\sigmavec}$ that evaluated at the simulation point $\sigmavec=(g,0,\hat{\mvec})$, which is obtained from the previous equation by substituting $\sigmavec^\iso\to\sigmavec$.

The expansion of $\langle \hat{\mathcal{O}}\rangle_{\sigmavec^\phi}$ around the simulation point $\sigmavec$ is then given by
\begin{equation}
    \langle \hat{\mathcal{O}}\rangle_{\sigmavec^\phi} = \langle \hat{\mathcal{O}}\rangle_{\sigmavec} + \langle \delta \hat{\mathcal{O}}\rangle_{\sigmavec} - \langle \hat{\mathcal{O} } \, \delta S_F\rangle_{\sigmavec} + \frac{1}{2}\,\langle \hat{\mathcal{O}}\, (\delta S_F)^2 \rangle_{\sigmavec} +\order(\bar \varepsilon^2)\,.
    \label{eq:path_int_expansion}
\end{equation}
The correction $\langle \delta \hat{\mathcal{O}}\rangle_{\sigmavec}$ in~\cref{eq:path_int_expansion} only appears if the observable $\mathcal{O}$ itself depends on the electromagnetic coupling $e^\phi$, which is not the case for the correlation functions studied in this work. The quantity $\delta S_F$ is instead the IB correction to the lattice fermionic action, i.e. 
\begin{equation}
    \delta S_F =  \sum_x \sum_{\mathrm{f}} \Big[(\hat{m}_\mathrm{f}^\phi-\hat{m}_\mathrm{f})\, \hat{\Scal}_\mathrm{f}(x) + \ii\, e_\mathrm{f}\,  \hat{V}^{\mu}_\mathrm{c,f}(x) \hat{A}_\mu(x) - \frac{1}{2}\,e_\mathrm{f}^2 \, \hat{T}_\mathrm{f}^\mu(x) \hat{A}_\mu^2(x)\Big] + \order(\bar \varepsilon^2)\,.
\end{equation}
Here $\Scal_\mathrm{f}(x)=\bar{\psi}_\mathrm{f}(x)\psi_\mathrm{f}(x)$ is the scalar density, while ${V}^\mu_\mathrm{c,f}(x)$ and ${T}^\mu_\mathrm{f}(x)$ are the electromagnetic conserved current and the seagull (or tadpole) current, respectively, which depend on the lattice regularization adopted (see e.g. refs.~\cite{deDivitiis:2013xla,Boyle:2017gzv}). The hats denote that all quantities are expressed in lattice units. In this work, however, we employ a definition of the fermion-photon coupling similar to the continuum one, where we use the renormalized local vector current, $V^\mu_\mathrm{f}(x)=Z_V \,\bar{\psi}_\mathrm{f}(x)\gamma^\mu\psi_\mathrm{f}(x)$, instead of the electromagnetic conserved one and do not include the tadpole current\footnote{Note that for leptons  $Z_V=1$ and hence $V_\ell^\mu(x)=\bar\ell(x)\gamma^\mu\ell(x)$.}.  This results in
\begin{equation}
\label{eq:deltaSF}
    \delta S_F =  \sum_x \sum_{\mathrm{f}} \Big[(\hat{m}_\mathrm{f}^\phi-\hat{m}_\mathrm{f})\, \hat{\Scal}_\mathrm{f}(x) + \ii\, e_\mathrm{f}\,  \hat{V}^{\mu}_\mathrm{f}(x) \hat{A}_\mu(x) \Big] + \order(\bar \varepsilon^2)\,,
\end{equation}
and the two approaches are expected to differ just by cut-off effects. The comparison of the two approaches has been thoroughly investigated, and we report on that in~\cref{app:loc-vs-cons}. 
Since the simulation point and the iso-QCD point only differ by the choice of the quark masses, the expansion of the iso-QCD path integral around the simulation point is given by 
\begin{equation}
    \langle \hat{\mathcal{O}}\rangle_{\sigmavec^\iso} = \langle \hat{\mathcal{O}}\rangle_{\sigmavec} + 
    \sum_{\mathrm{f}} (\hat{m}_\mathrm{f}^\iso-\hat{m}_\mathrm{f})\, \langle \hat{\mathcal{O}}\,\hat{\Scal}_\mathrm{f}\rangle_{\sigmavec} +\order(\bar \varepsilon^2)\,.
    \label{eq:path_int_expansion_iso}
\end{equation}
From~\cref{eq:path_int_expansion,eq:path_int_expansion_iso} it is then clear that IB corrections are obtained by computing correlation functions at the simulation point with the insertion of the operators $\Scal_\mathrm{f}(x)$ and $V^\mu_\mathrm{f}(x)$. We repeat that throughout this paper we work in the electro-quenched approximation. In practice, the bare parameters of the sea quark are kept fixed to their simulated values, which amounts to neglecting all quark-line disconnected diagrams. 

In the perturbative approach adopted in this work the $\mathrm{U}(1)$ gauge fields $A_\mu(x)$ are generated as stochastic fields sampled according to the $\QEDL$ gauge action in Feynman gauge
 \cite{Boyle:2017gzv,Giusti:2017dmp}
 \begin{equation}
     S_\gamma[A] = \frac{1}{V}\sum_{k : \kvec\neq 0} \bar{k}^2 \sum_\mu|\tilde{A}_\mu(k)|^2\, \quad
     \text{with} \quad \bar{k}^\mu = \frac{2}{a}\sin\left(\frac{ak^\mu}{2}\right)\,,
 \end{equation}
 with $\tilde{A}_\mu(k)$ being the photon field in momentum space, so that the expectation value $\langle A_\mu(x)A_\nu(y)\rangle_\gamma$ reproduces the photon propagator $\Delta_{\mu\nu}(x-y)$.

\subsection{Lattice setup}
For this calculation, we generate correlators for a $(L/a)^3 \times (T/a) = 48^3\times 96$ lattice using Möbius Domain Wall Fermions (DWF)~\cite{Brower:2012vk} with close-to-physical masses. The Domain wall height and the length of the fifth dimension are $aM_5=1.8$ and $L_\mathrm{s}/a=24$, respectively. See ref.~\cite{RBC:2014ntl} for more details. The QCD gauge configurations are generated by the RBC/UKQCD collaboration using the Iwasaki gauge action~\cite{IWASAKI1985141} with bare coupling $\beta=2.13$. The sea quark masses are $\hat{m}_\mathrm{ud}^\mathrm{sea}=0.00078$ for the light quarks and $ \hat{m}_\mathrm{s}^\mathrm{sea}=0.0362$ for the strange quark. We work in a unitary setup where we choose the valence light-quark masses to have the same value as the sea, $\hat{m}_\mathrm{ud}=\hat{m}_\mathrm{ud}^\mathrm{sea}$ and similarly for the valence strange quarks, $\hat{m}_\mathrm{s}=\hat{m}_\mathrm{s}^\mathrm{sea}$. In this setup, that we refer to as our \emph{simulation point} $\sigmavec$, the lattice spacing has been determined without QED to be $1/a=1.7295\,(38)$~GeV and the simulated pion mass of this ensemble is $m_\pi=139.15\,(36)$~MeV, corresponding to $m_\pi L=3.863\,(6)$.

To reduce the computational cost of inverting the Dirac operator for near-physical light quarks, we employ zMöbius fermions, which are a rational approximation of the Möbius formalism (see ref.~\cite{Mcglynn:2015uwh} and references therein), together with the deflation eigenvectors generated by the RBC/UKQCD collaboration for this $48^3\times 96$ ensemble. Light-quark propagators can then be obtained with a smaller value of $L_\mathrm{s}$, thereby reducing the simulation cost. This rational approximation of the Möbius DWF action must be corrected for, and we defer this discussion to~\cref{app:moebius}. 

\subsection{Implementation of the  hadronic correlators}
\label{sec:latt_correlators}

We now turn to discuss the lattice implementation of the correlation functions introduced in~\cref{sec:mel_from_corr}, where the relations with the corresponding matrix elements were obtained in the continuum and infinite-volume limit. As explained in~\cref{sec:lattice_QCQQED_path_int}, IB corrections to the expectation value of a given observable can be obtained in the iso-QCD simulated theory by inserting additional operators in the correlation function~\cite{deDivitiis:2013xla}.
As we discuss in the following this is obtained, in practice, by iteratively inverting the Dirac operator using suitable sources to get the appropriate sequential propagators.
All the correlation functions used in this calculation are generated using a set of 60 statistically independent QCD configurations and are then resampled with the bootstrap method. The QED gauge fields $A_\mu$ are generated using one stochastic source on each QCD gauge configuration. In this way the averages over QED and QCD gauge configurations are simultaneous.  The inversions of the Dirac operator and the quark field contractions have been performed using the Grid/Hadrons software framework~\cite{Boyle:2016lbp,Boyle:2022nef,antonin_portelli_2022_6382460}. 

In this calculation we study the decay of the meson $P^+$ in its rest frame,  $\pvec_P=\boldsymbol{0}$. To create the meson we use gauge-fixed wall sources. This corresponds to defining a zero-momentum interpolating operator of the form
\begin{equation}
    \phi_P^\dagger(t) \equiv \phi_P^\dagger(t,\pvec_P=\boldsymbol{0})= {a^6} 
    \sum_{\xvec_1,\xvec_2} \bar{q}_1(t,\xvec_1)\gamma_5 q_2(t,\xvec_2)\,,
\end{equation}
and evaluating expectation values of this operator fixed to Coulomb gauge. Any gauge-fixed expectation value involving $\phi_P^\dagger(t)$ can be re-expressed as a gauge invariant correlator with an alternative operator that includes a Wilson line between the quark fields.
Crucially, the gauge-invariant equivalent is local in time so that we can perform spectral decompositions using the standard Hilbert space of lattice QCD with pseudoscalar quantum numbers.\footnote{The same would not be true for gauge fixings that affect temporal gauge links. The idea of equivalence between gauge-invariant and gauge-fixed formulations is discussed in the context of QED in a seminal paper by Dirac~\cite{Dirac:1955uv} and more recently in ref.~\cite{Hansen:2018zre}.}
Note that using such definition of the meson interpolating operator, the dimensions of the correlators are different from those described in~\cref{sec:mel_from_corr}  because of the additional integration over the spatial coordinates of the quark fields.

\paragraph{Tree-level correlation function:} The tree-level correlation functions in~\cref{eq:C_PA_PP} are implemented at the simulation point in terms of quark propagators as
\begin{align}
    C_\mathrm{PA}(t)   
    &= a^3 \, \sum_\xvec \, \big\langle\mathrm{Tr}\big[S_\mathrm{q_2}(t,\xvec|0)^\dagger\, \gamma^0 \,S_\mathrm{q_1}(t,\xvec|0) \big]\big\rangle\,,\\
    C_\mathrm{PP}(t)  
    &= a^6 \sum_{\xvec_1,\xvec_2} \big\langle\mathrm{Tr}\big[S_\mathrm{q_2}(t,\xvec_2|0)^\dagger \, S_\mathrm{q_1}(t,\xvec_1|0) \big]\big\rangle\,,
\end{align}
where we have used $\gamma_5$-hermiticity, $S_\mathrm{q}(t_2,\xvec_2|t_1,\xvec_1)=\gamma_5 S_\mathrm{q}(t_1,\xvec_1|t_2,\xvec_2)^\dagger \gamma_5$, and defined the quark propagator with one end projected on zero momentum as
\begin{equation}
\label{eq:quark_prop_timemom}
S_\mathrm{q}(t_1,\xvec_1|t_2) \equiv S_\mathrm{q}(t_1,\xvec_1|t_2,\pvec=\boldsymbol{0}) = a^3\sum_{\xvec_2} S_\mathrm{q}(t_1,\xvec_1|t_2,\xvec_2)\,,  
\end{equation}
while the symbol~$\langle\,\cdot\,\rangle$ denotes the average over the gauge configurations.
Note that here we have generated the pseudoscalar meson at $t_P=0$ for simplicity. In the lattice calculation we have instead evaluated the correlation functions on each gauge configuration inserting the source at every timeslice ${t_P/a=\{1,\dots,T/a=96\}}$ and then shifted and averaged over the source positions to improve the statistical uncertainty. 

By considering the asymptotic form for the pseudoscalar correlator in~\cref{eq:C_PA_PP_spectr} on a torus with period $T$ in the temporal direction and periodic boundary conditions we obtain 
\begin{align}
    \label{eq:C_PA_spectr_FV}
    C_\mathrm{PA}(t) &= \frac{\mathcal{A}_P\,Z_P}{2\, m_P}\,\big\{\e^{-m_P t} - \e^{-m_P(T-t)}\big\} \,,\\
    \label{eq:C_PP_spectr_FV}
    L^3\,C_\mathrm{PP}(t) &=  \  \frac{|Z_P|^2}{2\, m_P}\, \ \big\{\e^{-m_P t} + \e^{-m_P(T-t)}\big\}\,,
\end{align}
having neglected exponentially suppressed contributions of excited states.

\paragraph{Factorizable correlators:} Let us define the sequential propagators obtained by inserting the correction to the fermionic action $\delta S_F$ (see~\cref{eq:deltaSF}) along the quark line as
\begin{equation}
    S_\mathrm{q}^{(1)}(x|y) = e_\mathrm{q}\,  S_\mathrm{q}^{A}(x|y) + (\hat{m}_\mathrm{q}^\phi-\hat{m}_\mathrm{q}) \, S_\mathrm{q}^\mathcal{S}(x|y)\,,
\end{equation}
where 
\begin{equation}
    \resizebox{.9\hsize}{!}{$\displaystyle
    S_\mathrm{q}^{A}(x|y) = \ii  \,a^4 Z_V\sum_z S_\mathrm{q}(x|z) \gamma^\mu A_\mu(z) S_\mathrm{q}(z|y)
    \ \, \text{and}\ \,
    S_\mathrm{q}^\mathcal{S}(x|y) = a^3\sum_z S_\mathrm{q}(x|z) S_\mathrm{q}(z|y)\,.$}
    \label{eq:SqS_SqA}
\end{equation}
We can analogously define the sequential quark propagator with a double insertion of $\delta S_F$, which generates the quark self-energy, as
\begin{equation}
    S_\mathrm{q}^{(2)}(x|y) = e_\mathrm{q}^2 \, S_\mathrm{q}^\mathrm{self}(x|y)+\order(\bar \varepsilon^2)\,
\end{equation}
with
\begin{equation}
   S_\mathrm{q}^\mathrm{self}(x|y) = -a^8 Z_V^2\,\sum_{z,w} S_\mathrm{q}(x|z) \gamma^\mu A_\mu(z) S_\mathrm{q}(z|w) \gamma^\nu A_\nu(w) S_\mathrm{q}(w|y)\,.
\end{equation}
Note that the propagators $S_\mathrm{q}^{(1)}(x|y)$ and $S_\mathrm{q}^{(2)}(x|y)$ are both $\gamma_5$-hermitian.

The factorizable correlators $\delta C^\mathrm{f}_\mathrm{PA}(t)$ and $\delta C^\mathrm{f}_\mathrm{PP}(t)$ can then be evaluated in terms of such sequential propagators. We define
\begin{align}
\label{eq:deltaC_fact_PA_latt}
    \delta C^\mathrm{f}_\mathrm{PA}(t) &= 
    4\pi \aem \, \delta C_\mathrm{PA}^\mathrm{em}(t) + \sum_\mathrm{q}(\hat{m}_\mathrm{q}^\phi-\hat{m}_\mathrm{q}^\iso)\,\delta C^\mathrm{\mathcal{S},\mathrm{q}}_\mathrm{PA}(t)
    \\
    &= \sum_\mathrm{q} e_\mathrm{q}^2\,\delta C^\mathrm{self,q}_\mathrm{PA}(t) + e_\mathrm{q_1}e_\mathrm{q_2}\,\delta C^\mathrm{exch}_\mathrm{PA}(t) + \sum_\mathrm{q}(\hat{m}_\mathrm{q}^\phi-\hat{m}_\mathrm{q}^\iso)\,\delta C^\mathrm{\mathcal{S},\mathrm{q}}_\mathrm{PA}(t)\,,\nonumber
\end{align}
and analogously $\delta C^\mathrm{f}_\mathrm{PP}(t)$, where
\begin{align}
\delta C^\mathrm{self,q_1}_\mathrm{PA}(t) &= a^3\sum_\xvec \big\langle 
    \mathrm{Tr}\big[
        S_\mathrm{q_2}(t,\xvec|0)^\dagger  \gamma^0  S_\mathrm{q_1}^\mathrm{self}(t,\xvec|0)
        \big]
        \big\rangle\,,\nn\\
\delta C^\mathrm{self,q_2}_\mathrm{PA}(t) &= a^3\sum_\xvec \big\langle 
    \mathrm{Tr}\big[
        S_\mathrm{q_2}^\mathrm{self}(t,\xvec|0)^\dagger  \gamma^0  S_\mathrm{q_1}(t,\xvec|0) 
        \big]
        \big\rangle\,,\nn\\
\label{eq:deltaC_PA_traces}
\delta C^\mathrm{exch}_\mathrm{PA}(t) &= a^3\sum_\xvec \big\langle 
    \mathrm{Tr}\big[
        S_\mathrm{q_2}^{A}(t,\xvec|0)^\dagger  \gamma^0  S_\mathrm{q_1}^{A}(t,\xvec|0) 
        \big]
        \big\rangle\,,\\
\delta C^{\mathcal{S},\mathrm{q_1}}_\mathrm{PA}(t) &= a^3\sum_\xvec \big\langle 
    \mathrm{Tr}\big[
        S_\mathrm{q_2}(t,\xvec|0)^\dagger \gamma^0  S_\mathrm{q_1}^\mathcal{S}(t,\xvec|0) 
        \big]
        \big\rangle\,,\nn\\
    \delta C^{\mathcal{S},\mathrm{q_2}}_\mathrm{PA}(t) &= a^3\sum_\xvec \big\langle 
    \mathrm{Tr}\big[
        S_\mathrm{q_2}^\mathcal{S}(t,\xvec|0)^\dagger  \gamma^0  S_\mathrm{q_1}(t,\xvec|0)
        \big]
        \big\rangle\,,\nn
\end{align}
    and
\begin{align}
    \delta C^\mathrm{self,q_1}_\mathrm{PP}(t) &= a^6\sum_{\xvec_1,\xvec_2}\big\langle 
    \mathrm{Tr}\big[
        S_\mathrm{q_2}(t,\xvec_2|0)^\dagger   S_\mathrm{q_1}^\mathrm{self}(t,\xvec_1|0) 
        \big]
        \big\rangle\,,\nn\\
    \delta C^\mathrm{self,q_2}_\mathrm{PP}(t) &= a^6\sum_{\xvec_1,\xvec_2} \big\langle 
    \mathrm{Tr}\big[
        S_\mathrm{q_2}^\mathrm{self}(t,\xvec_2|0)^\dagger  S_\mathrm{q_1}(t,\xvec_1|0) 
        \big]
        \big\rangle\,,\nn\\
    \label{eq:deltaC_PP_traces}
    \delta C^\mathrm{exch}_\mathrm{PP}(t) &= a^6\sum_{\xvec_1,\xvec_2} \big\langle 
    \mathrm{Tr}\big[
        S_\mathrm{q_2}^{A}(t,\xvec_2|0)^\dagger  S_\mathrm{q_1}^{A}(t,\xvec_1|0) 
        \big]
        \big\rangle\,,\\
    \delta C^{\mathcal{S},\mathrm{q_1}}_\mathrm{PP}(t) &= a^6\sum_{\xvec_1,\xvec_2} \big\langle 
    \mathrm{Tr}\big[
        S_\mathrm{q_2}(t,\xvec_2|0)^\dagger  S_\mathrm{q_1}^\mathcal{S}(t,\xvec_1|0) 
        \big]
        \big\rangle\,,\nn\\
    \delta C^{\mathcal{S},\mathrm{q_2}}_\mathrm{PP}(t) &= a^6\sum_{\xvec_1,\xvec_2} \big\langle 
    \mathrm{Tr}\big[
        S_\mathrm{q_2}^\mathcal{S}(t,\xvec_2|0)^\dagger   S_\mathrm{q_1}(t,\xvec_1|0)
        \big]
        \big\rangle\,,\nn
\end{align}
having used again $\gamma_5$-hermiticity together with~\cref{eq:quark_prop_timemom}. Note that the symmetries of the correlators ensure that $\delta C_\mathrm{PP}^\mathrm{self,u}=\delta C_\mathrm{PP}^\mathrm{self,d}$ and $\delta C_\mathrm{PA}^\mathrm{self,u}=\delta C_\mathrm{PA}^\mathrm{self,d}$, as well as $\delta C_\mathrm{PP}^\mathrm{\mathcal{S},u}=\delta C_\mathrm{PP}^\mathrm{\mathcal{S},d}$ and $\delta C_\mathrm{PA}^\mathrm{\mathcal{S},u}=\delta C_\mathrm{PA}^\mathrm{\mathcal{S},d}$ when $m_\rmu=m_\rmd=m_\rmud$. The five correlators in \cref{eq:deltaC_PA_traces} correspond to the hadronic part of the Feynman diagrams shown in figures~\ref{fig:alldiags}(a)-(e), respectively.

For the factorizable correlators, correcting the asymptotic behaviour in~\cref{eq:deltaC_fact_PA_spectr,eq:deltaC_fact_PP_spectr} for finite-time $T$ effects with (anti-)periodic boundary conditions and neglecting the contribution of excited states results in
\begin{align}
    \label{eq:deltaC_fact_PA_spectr_FV}
    \mathcal{R}_\mathrm{PA}^\mathrm{f}(t) &= \frac{\delta \Acal_P^\mathrm{f}}{\Acal_P} + \frac{\delta Z_P}{Z_P} - \frac{\delta m_P}{m_P}\,f_\mathrm{PA}(t,T)\,,\\
    \label{eq:deltaC_fact_PP_spectr_FV}
    \mathcal{R}_\mathrm{PP}^\mathrm{f}(t) &= 2\,\frac{\delta Z_P}{Z_P} - \frac{\delta m_P}{m_P}\,f_\mathrm{PP}(t,T)\,,
\end{align}
with
\begin{align}
    \label{eq:fPA}
    f_\mathrm{PA}(t,T) &= 1+m_P\big\{ \tfrac{T}{2}-(t-\tfrac{T}{2})\coth\big[m_P(t-\tfrac{T}{2})\big] \big\}\,,\\
    \label{eq:fPP}
    f_\mathrm{PP}(t,T) &= 1+m_P\big\{ \tfrac{T}{2}-(t-\tfrac{T}{2})\tanh\big[m_P(t-\tfrac{T}{2})\big] \big\}\,,
\end{align}
and $f_\mathrm{PA}(t,T)=f_\mathrm{PP}(t,T)\approx 1+m_P t$ for $t\ll T/2$.

In the following we will make use of the notation $\mathcal{R}_\mathrm{PA}^\mathrm{x}(t)$ (and analogously for $\mathcal{R}_\mathrm{PP}^\mathrm{x}(t)$) with $\mathrm{x}=\{{ \textrm{self,q}\,;\,\textrm{exch}\,;\,\textrm{$\mathcal{S}$,q}}\}$. This has to be interpreted as the contributions to $\mathcal{R}_\mathrm{PA}^\mathrm{f}(t)$ coming from the corresponding corrections to the correlator $\delta C_\mathrm{PA}^\mathrm{f}(t)$ in~\cref{eq:deltaC_fact_PA_latt}. Equivalently, we can decompose the correction to the meson mass up to $\mathrm{O}(\epsilon^2)$ as follows
\begin{align}
    \label{eq:deltamP_decomposition}
    \delta m_P &= 4\pi\aem \, \delta m_P^\mathrm{em} + \sum_\mathrm{q}(\hat{m}_\mathrm{q}^\phi-\hat{m}_\mathrm{q}^\iso)\,\delta m_P^\mathrm{\mathcal{S},\mathrm{q}} \\
    &= \sum_\mathrm{q} e_\mathrm{q}^2\,\delta m_P^\mathrm{self,q} + e_\mathrm{q_1}e_\mathrm{q_2}\,\delta m_P^\mathrm{exch} + \sum_\mathrm{q}(\hat{m}_\mathrm{q}^\phi-\hat{m}_\mathrm{q}^\iso)\,\delta m_P^\mathrm{\mathcal{S},\mathrm{q}}\,. \nonumber
\end{align}
For the mesons studied in this work we have $\mathrm{q}=\{\rmu,\rmd\}$ for $\delta m_{\pi^+}$, $\mathrm{q}=\{\rmu,\rms\}$ for $\delta m_{K^+}$, $\mathrm{q}=\{\rmd,\rms\}$ for $\delta m_{K^0}$, $\mathrm{q}=\{\rmu,\rmu\}$ for $\delta M_{\rmu\rmu}$ and $\mathrm{q}=\{\rmd,\rmd\}$ for $\delta M_{\rmd\rmd}$\,.

\paragraph{Non-factorizable correlators:} The non-factorizable correlator introduced in~\cref{eq:deltaC_nf} can also be evaluated on the lattice by using the sequential propagators described above. Defining
\begin{equation}
    \delta C^\mathrm{nf}_\mathrm{P\ell}(t,t_\ell) = e_\mathrm{q_1}e_\ell \, \delta C_\mathrm{P\ell}^\mathrm{nf,q_1}(t,t_\ell) + e_\mathrm{q_2}e_\ell \, \delta C_\mathrm{P\ell}^\mathrm{nf,q_2}(t,t_\ell),
\end{equation}
and using~\cref{eq:quark_prop_timemom} one has
\begin{align}
    \delta C_\mathrm{P\ell}^\mathrm{nf,q_1}(t,t_\ell) &=  
      \big\langle\mathrm{Tr}\big[S_\mathrm{q_2}(0|\!-\!t)^\dagger\, \, \gamma^\rho_L \,\, S_\mathrm{q_1}^{A}(0|\!-\!t)\big] \times S_\nu(t_\ell,\pvec_\nu|0) \, \gamma^\rho_L \, S_\ell^A(0|t_\ell,\pvec_\ell) \big\rangle\,, \\
    \delta C_\mathrm{P\ell}^\mathrm{nf,q_2}(t,t_\ell) &=  
      \big\langle\mathrm{Tr}\big[S_\mathrm{q_2}^A(0|\!-\!t)^\dagger\, \, \gamma^\rho_L \,\, S_\mathrm{q_1}(0|\!-\!t)\big] \times S_\nu(t_\ell,\pvec_\nu|0) \, \gamma^\rho_L \, S_\ell^A(0|t_\ell,\pvec_\ell) \big\rangle\,,
\end{align}
which correspond to the Feynman diagrams in figures~\ref{fig:alldiags}(f) and \ref{fig:alldiags}(g), respectively.
Here we have defined the (sequential) propagator of an anti-lepton with the insertion of an electromagnetic current and projected on definite external momentum as
\begin{equation}
    S_\mathrm{\ell}^{A}(0|t_\ell,\pvec_\ell) = \ii\, a^7\sum_{z,\xvec_\ell} S_\mathrm{\ell}(0|z) \gamma^\mu A_\mu(z) S_\mathrm{\ell}(z|t_\ell,\xvec_\ell) \, \e^{-\ii\pvec_\ell\cdot\xvec_\ell}\,.
\end{equation}
The tree-level correlator of~\cref{eq:C_0} evaluated at the simulated iso-symmetric point takes the form 
\begin{equation}
     C_\mathrm{P\ell}(t,t_\ell) = \big\langle \mathrm{Tr}\big[ S_\mathrm{q_2}(0|\!-\!t)^\dagger\, \, \gamma^\rho_L \,\, S_\mathrm{q_1}(0|\!-\!t) \big] \big\rangle \times S_\nu(t_\ell,\pvec_\nu|0) \, \gamma^\rho_L \,S_\ell(0|t_\ell,\pvec_\ell)\,.
\end{equation}
Also in this case translational invariance has been used to simplify the notation such that the weak current is inserted in the origin. However, lattice correlators have been computed by inserting the weak current on all possible timeslices $t_H/a=\{1,\dots,T/a=96\}$ and at all positions $\xvec_H$, and then averaged over the volume. The lepton propagator has been computed for 8 different lepton source-sink separations $t_\ell/a=\{12,16,\dots,40\}$ and its momentum is chosen in such a way that energy and momentum are conserved in the process.
Some comments concerning lattice lepton propagators are in order. First, we note that when evaluated on a torus, the lepton propagator $S_\ell(0|t_\ell,\pvec_\ell)$ takes the form (neglecting possible contact terms)
\begin{equation}
    S_\ell(0|t_\ell,\pvec_\ell) =  -\sum_r\bigg[\frac{\e^{-\omega_\ell t_\ell}}{2\Omega_\ell}\,  v^r_\ell(\pvec_\ell)\bar{v}^r_\ell(\pvec_\ell)+ \,\frac{\e^{-\omega_\ell(T-t_\ell)}}{2\Omega_\ell}\,  u^r_\ell(-\pvec_\ell)\bar{u}^r_\ell(-\pvec_\ell)\bigg] \frac{1}{1+\e^{-\omega_\ell T}}\,.
\end{equation}
The backward signal has a different Dirac structure compared to the forward one and $(2\Omega_\ell)^{-1}$ appears in the residue at the pole, with $\lim_{a\to 0} \Omega_\ell = \omega_\ell$. Such a backward term would contribute to the traces in~\cref{eq:deltaCnf/C0}. However, this contribution is not related to the matrix element $\Mcal_P^{rs}(\pvec_\ell)$ of our interest and therefore it has to be subtracted. To this end it is possible to define a projector $\mathcal{P}_{v_\ell(\pvec_\ell)}$ only onto the forward-propagating part, namely
\begin{equation}
    S_\ell(0|t_\ell,\pvec_\ell)\cdot \mathcal{P}_{v_\ell(\pvec_\ell)} = -\frac{\e^{-\omega_\ell t_\ell}}{2\Omega_\ell}\,  \sum_r v^r_\ell(\pvec_\ell)\bar{v}^r_\ell(\pvec_\ell)\,  \frac{1}{1+\e^{-\omega_\ell T}}\,.
\end{equation}
The definition and derivation of the projector $\mathcal{P}_{v_\ell(\pvec_\ell)}$ is discussed in~\cref{app:projectors}. Note that the same feature would appear also in the lattice neutrino propagator. However, being electrically neutral, the neutrino does not couple to the photon and, in addition, the term $e^{-\omega_\nu t_\ell}/(2\omega_\nu)$ in its  time-momentum representation (see~\cref{eq:prop_neutrino}) cancels in the ratio of~\cref{eq:deltaCnf/C0}. Therefore we can amputate the neutrino propagator and substitute it with the (continuum) completeness relation $[\sum_s u_\nu^s(\pvec_\nu)\bar{u}_\nu(\pvec_\nu)]^\mathrm{cont}=-\ii \pslash_\nu$.

The lattice correlators employed in the numerical calculation are then defined as
\begin{align}
    \delta \widetilde{C}_\mathrm{P\ell}^\mathrm{nf,q_1}(t,t_\ell) &=  
    -\ii \,\big\langle\mathrm{Tr}\big[S_\mathrm{q_2}(0|\!-\!t)^\dagger\, \, \gamma^\rho_L \,\, S_\mathrm{q_1}^{A}(0|\!-\!t)\big] \times \pslash_\nu \, \gamma^\rho_L \, S_\ell^A(0|t_\ell,\pvec_\ell) \cdot \mathcal{P}_{v_\ell(\pvec_\ell)} \big\rangle\,, \nn \\
    \delta \widetilde{C}_\mathrm{P\ell}^\mathrm{nf,q_2}(t,t_\ell) &=  
    -\ii \,  \big\langle\mathrm{Tr}\big[S_\mathrm{q_2}^A(0|\!-\!t)^\dagger\, \, \gamma^\rho_L \,\, S_\mathrm{q_1}(0|\!-\!t)\big] \times \pslash_\nu \, \gamma^\rho_L \, S_\ell^A(0|t_\ell,\pvec_\ell) \cdot \mathcal{P}_{v_\ell(\pvec_\ell)} \big\rangle\,,\\
    \widetilde{C}_\mathrm{P\ell}(t,t_\ell) &=
    -\ii \, \big\langle \mathrm{Tr}\big[ S_\mathrm{q_2}(0|\!-\!t)^\dagger\, \, \gamma^\rho_L \,\, S_\mathrm{q_1}(0|\!-\!t) \big] \big\rangle \times \pslash_\nu\, \gamma^\rho_L S_\ell(0|t_\ell,\pvec_\ell)\cdot \mathcal{P}_{v_\ell(\pvec_\ell)}\,.\nn
\end{align}

The spectral decompositions of $\delta \widetilde{C}_\mathrm{P\ell}^\mathrm{nf}(t,t_\ell)$ and $ \widetilde{C}_\mathrm{P\ell}(t,t_\ell)$, taking into account also the backward propagation of the meson on the torus, become\footnote{
    Note that the spectral decomposition for $\delta \widetilde{C}_\mathrm{P\ell}^\mathrm{nf}(t,t_\ell)$ given in~\cref{eq:deltaCnf_spectr_decomp_FV} is valid only for $t<T-t_\ell$. In this work we restrict the analysis of non-factorizable correlators in the region $t<T/2$, where the condition $t<T-t_\ell$ is satisfied for all values of $t_\ell$ used.
}
\begin{equation}
    \label{eq:deltaCnf_spectr_decomp_FV}
    \resizebox{.9\hsize}{!}{
    $\displaystyle
    \delta \widetilde{\mathcal{C}}_\mathrm{P\ell}^\mathrm{nf}(t,t_\ell) = \frac{1}{L^3}  \sum_{r,s} \frac{\e^{-\omega_\ell t_\ell}}{4 m_P \overline{\Omega}_\ell} \big\{\e^{-m_P t} + \kappa_\mathrm{P\ell} \, \e^{-m_P(T-t)}\big\}\,Z_P\,u_\nu^r(\pvec_\ell)\, \big[\delta \widebar\Mcal_P^{rs}(\pvec_\ell)\big]^\mathrm{nf}\, \bar{v}_\ell^s(\pvec_\ell)\,,$}
\end{equation}
\begin{equation}
    \resizebox{.9\hsize}{!}{
    $\displaystyle
    \widetilde{C}_\mathrm{P\ell}(t,t_\ell) = \frac{1}{L^3}\sum_{r,s} \frac{\e^{-\omega_\ell t_\ell}}{4 m_P \overline{\Omega}_\ell }  \,\big\{\e^{-m_P t} - \e^{-m_P(T-t)}\big\}\, Z_P\,u_\nu^r(\pvec_\ell)\left\{-\Acal_P \Lcal^{rs}(\pvec_\ell)\right\}\bar{v}_\ell^s(\pvec_\ell)\,,$}
\end{equation}
where $\overline{\Omega}_\ell=\Omega_\ell(1+\e^{-\omega_\ell T})$ and $\kappa_\mathrm{P\ell}$ (which has a residual dependence on $t_\ell$) parametrizes the correction to the matrix element due to the interaction of the backward propagating meson and the lepton. It follows that~\cref{eq:deltaCnf/C0} becomes
\begin{equation}
    \mathcal{R}_\mathrm{P\ell}^\mathrm{nf}(t,t_\ell)=\mathrm{Re}\left[\frac{\mathrm{Tr}\big[\gamma^0_L\, \delta \widetilde{C}_\mathrm{P\ell}^\mathrm{nf}(t,t_\ell)\big]}{\mathrm{Tr}\big[\gamma^0_L\, \widetilde{C}_\mathrm{P\ell}(t,t_\ell)\big]}\right] = \frac{\delta \Acal_P^\mathrm{nf}}{\Acal_P} \, f_\mathrm{P\ell}(t,T)\,,
    \label{eq:non_fact_fitansatz}
\end{equation}
where
\begin{equation}
    f_\mathrm{P\ell}(t,T)=\tfrac{1}{2}\big\{(1+\kappa_\mathrm{P\ell})-(1-\kappa_\mathrm{P\ell})\coth\big[m_P(t-\tfrac{T}{2})\big]\big\} \approx 1 \quad \text{for}~t\ll\tfrac{T}{2}\,.
\end{equation}

For the lepton propagator we use the free Shamir DWF action~\cite{Furman:1994ky} with $aM_5 = 1.0$ and $L_\mathrm{s}/a=8$. The Feynman rules for the free DWF propagator have been derived in ref.~\cite{Aoki:1997xg} and we give details of the relevant Feynman rules in the conventions used in the Grid software framework~\cite{Boyle:2016lbp,Boyle:2022nef} in~\cref{app:freeDWFfeynmanrules}. We have determined the bare input mass for the lepton such that the pole mass of the free propagator corresponds to the physical muon mass $m^\phi_\mu=105.6583755~\mathrm{MeV}$~\cite{Workman:2022ynf}. This results in a bare input lepton mass of $am^\textrm{input}_\ell=0.06107$ when using a previous determination of the lattice spacing $a^{-1}=1.730~\mathrm{GeV}$~\cite{RBC:2014ntl}. Details on how to determine the input bare mass for a desired target pole mass of the free Shamir DWF propagator are given in~\cref{app:freeDWFpolemass}.

We use twisted boundary conditions~\cite{Boyle:2003ui,Bedaque:2004kc,deDivitiis:2004kq,Sachrajda:2004mi} for the lepton propagator in order to fix the momentum of the lepton such that energy and momentum are conserved at the weak Hamiltonian. This is the case when the momentum of the lepton is given by $|\pvec_\ell|=\frac{m_P}{2}(1-(m_\ell/m_P)^2)$ for the pseudoscalar meson at rest. For the determination of $|\pvec_\ell|$ we used the physical mass for the muon $m_\ell\equiv m^\phi_\mu$ and the simulation point masses $m_P$ for pion and kaon as determined previously in ref.~\cite{RBCUKQCD:2015joy} on this gauge ensemble. We find $a|\pvec_\ell|=0.017054$ for the pion and $a|\pvec_\ell|=0.13783$ for the kaon. We distribute the momentum of the lepton equally in all three spatial directions, such that $\pvec_\ell=-\frac{|\pvec_\ell|}{\sqrt{3}}\{1,1,1\}$.

\paragraph{Omega baryon correlators:} 
Before closing the section we give details about the correlators for the $\Omega^-$ baryon, which is employed in the renormalization conditions imposed in~\cref{sec:IB_effects} to fix the bare parameters of the QCD+QED, QCD and iso-QCD actions. We define the zero momentum two-point function as
\begin{equation}
\label{eq:Omega_isoQCD_2pt}
C_{\Omega\Omega}(t)= \frac{a^3}{2} \sum_i \sum_\xvec \braket{0|\mathrm{T}\big[\psi_\Omega^i(t,\xvec)\widebar{\psi}_\Omega^i(0)\big]|0}\,,
\end{equation}
where the operator $\widebar{\psi}_\Omega^\mu(x)={\psi_\Omega^\mu}^\dagger(x)\gamma^0$ denotes the spin-3/2 interpolating operator for the $\Omega^-$ and we have summed over the spatial directions $i$. One form of baryon interpolator is given by
\begin{equation}
\label{eq:Omega_interpolator}
\psi_\Omega^\mu(x) = \epsilon^{abc}\, P_+ \,{s}_a(x) \left[ {s}^T_b(x) \, C \gamma^\mu \, {s}_c(x) \right] 
\end{equation}
where the ${s}$ represent the strange quark fields, $C$ is the charge conjugation matrix $C=\ii \gamma_2 \gamma_0$, and Roman indices identify color components of the fields. The projector $P_+=(1+\gamma^0)/2$ ensures that the interpolating operator $\bar{\psi}_\Omega^\mu$ generates states with positive parity quantum number ($\mathcal{P}=+1$) and annihilates states with negative parity quantum number ($\mathcal{P}=-1$). In order to improve the signal for the correlation function, in this calculation we employ Gaussian smearing for the strange quark fields ${\tilde{s}(t,\xvec) = a^3\sum_\yvec \exp[-(\xvec-\yvec)^2/(2 \sigma^2)] s(t,\yvec)}$ with a width of $\sigma/a = 9$, which requires gauge fixing of the QCD gauge configurations. 

One feature of lattice baryon interpolating operators is that, on a torus, they couple to negative parity states propagating backward in time. As a consequence, assuming ground state dominance, the correlator has the form
\begin{align}
\label{eq:Omega_no_trace}
C_{\Omega\Omega}(t) = & \, 
\left( |Z_\Omega|^2 \,\e^{-m_{\Omega} t} + |\widebar{Z}_\Omega|^2 \,\e^{-\widebar{\omega}_{\Omega}(T-t)} \right) P_+ 
\end{align}
where $\widebar{\omega}_{\Omega}$ is the energy of the state with parity $\mathcal{P}=- 1$. The operator-state overlaps for a state with spin projection $s \in \{\pm \frac{3}{2}, \pm \frac{1}{2} \}$ are defined by $Z_\Omega \, u_s^\mu = \bra{0} \psi_\Omega^\mu(0) \ket{\Omega,s}$ and $\widebar{Z}_\Omega\,  \gamma_5 u_s^\mu= \bra{0} \psi_\Omega^\mu(0) \ket{\widebar{\Omega},s}$, where $u^\mu_s$ is the positive energy solution to the spin-${3}/{2}$ Rarita-Schwinger equation (see e.g.~\cite{Shi-Zhong:2003} for a recent review), and $\ket{\Omega,s}$ and  $\ket{\widebar{\Omega},s}$ are states with positive and negative parity respectively.
In addition, quarks with anti-periodic boundary conditions in time have been assumed. 
Since baryon correlators are significantly affected by an exponential signal-to-noise-ratio problem, we restrict our analysis of the correlator to the time region $t\ll T/2$. In this interval we can then neglect the backward propagating signal and take for $t\gg 0$,
\begin{equation}
\label{eq:Omega}
\widetilde{C}_{\Omega\Omega}(t) = \frac{1}{2}\, \mathrm{Tr}\big[ C_{\Omega\Omega}(t)] \approx |Z_\Omega|^2 \,\e^{-m_{\Omega}t}\,.
\end{equation}
In analogy with~\cref{eq:deltaC_fact_PA_latt}, we can define the IB corrections to the correlator as
\begin{align}
    \label{eq:Omega_corrections}
    \delta \widetilde{C}_{\Omega\Omega}(t) &=  
    4\pi\aem\, \delta \widetilde{C}_{\Omega\Omega}^\mathrm{em}(t)
    + (\hat{m}_\mathrm{s}^\phi-\hat{m}_\mathrm{s})\,\delta \widetilde{C}^\mathrm{\mathcal{S},\mathrm{s}}_{\Omega\Omega}(t) \\
    & = e_\mathrm{s}^2\,\big[\delta \widetilde{C}^\mathrm{self,s}_{\Omega\Omega}(t)+\delta \widetilde{C}^\mathrm{exch}_{\Omega\Omega}(t)\big] + (\hat{m}_\mathrm{s}^\phi-\hat{m}_\mathrm{s})\,\delta \widetilde{C}^\mathrm{\mathcal{S},\mathrm{s}}_{\Omega\Omega}(t)\,, \nonumber
\end{align}
where $\widetilde{C}^\mathrm{self,s}_{\Omega\Omega}(t)$ and $\widetilde{C}^\mathrm{exch}_{\Omega\Omega}(t)$ denote the corrections due to the photon exchange between the constituent-strange quarks and $\widetilde{C}^\mathrm{\mathcal{S},\mathrm{s}}_{\Omega\Omega}(t)$ the correction given by the insertion of the quark scalar density on the quark lines. The ratio with the iso-QCD correlator has then the following asymptotic behaviour
\begin{equation}
    \label{eq:Omega_ratio}
    \mathcal{R}_{\Omega\Omega}(t) = \frac{\delta \widetilde{C}_{\Omega\Omega}(t)}{\widetilde{C}_{\Omega\Omega}(t)} = 2\,\frac{\delta Z_\Omega}{Z_\Omega} - \delta m_{\Omega} \, t\,.
\end{equation}
Also in this case we can decompose the correction to the $\Omega^-$ mass as 
\begin{equation}
    \delta m_\Omega = 4\pi\aem \,\delta m_\Omega^\mathrm{em}  + (\hat{m}_\mathrm{s}^\phi-\hat{m}_\mathrm{s})\,\delta m^\mathrm{\mathcal{S},\mathrm{s}}_{\Omega}\,.
\end{equation}

Details on the quark contractions for the $\Omega^-$ correlator, as well as a discussion on the derivation of its spectral decomposition can be found in~\cref{app:omega}.

\section{Numerical analysis}
\label{sec:numeval}

The virtual IB corrections to the ratio of inclusive decay rates evaluated on the lattice, as defined in~\cref{eq:deltaRKPi_latt}, is built from the IB corrections to the kaon and pion decay amplitudes and to their masses. As discussed in the previous section, such quantities can be extracted from the large-time behaviour of suitably defined Euclidean lattice correlators. In this section, the strategy for extracting the relevant quantities from lattice correlators using a global-fit analysis is presented.  Due to the various classes of correlators involved in this calculation, we adopt a data-driven approach to standardize the fitting criteria, which we explain below.

\subsection{Strategy for correlator fits}
\label{sec:num_strategy}
Extracting physical quantities from lattice correlators using a fit procedure requires that optimal fit ranges are identified for each correlator. In our work, when multiple lattice correlators have fit parameters in common, e.g.~the meson mass $m_P$, these data are fitted simultaneously fully taking into account such a constraint including the statistical correlation between the data. In this way, all parameters can be extracted from 7 independent frequentist fits.

In the case of the analysis of factorizable corrections, there are 12 correlators to study for the kaon, while for the pion the flavour symmetries of the correlation functions reduce the number of independent ones to 8. These correlators are listed below in~\cref{eq:analyses_corr}.
The functional forms of the fit ansätze used for the correlators are based on the spectral decompositions~\cref{eq:C_PA_spectr_FV,eq:C_PP_spectr_FV,eq:deltaC_fact_PA_spectr_FV,eq:deltaC_fact_PP_spectr_FV}, where only the ground-state contribution is included.
For both mesons the tree-level correlators depend on two parameters, while all the factorizable correlators depend on 3 parameters each, namely a constant term containing the relative corrections to the matrix elements $\mathcal{A}_P$ and $Z_P$, the correction to the meson mass and the simulation point mass $m_P$ entering the $\tanh$/$\coth$ functions in~\cref{eq:fPA,eq:fPP}.
The exact relation between the fit parameters and the physical quantities of interest is given in~\cref{eq:analyses_pars,eq:analyses_pars2,eq:analyses_pars3}.
Since all the correlators for a given meson depend on the same simulation point mass, we combine the fits as described below.
For what concerns the non-factorizable pion and kaon correlators, we decide instead to fit the ratios $\mathcal{R}_\mathrm{P\ell}^\mathrm{nf}(t,t_\ell)$ using a constant fit ansatz, i.e. setting $f_{\mathrm{P}\ell}(t,t_\ell)=1$ in~\cref{eq:non_fact_fitansatz}. This approximation corresponds to neglecting the contribution of backward signals and excited states and does not have a significant effect on the $\chi^2$ for the range considered. In this case there is then only one parameter for each meson. 
The $\Omega^-$ correlators, due to the usual rapidly degrading signal-to-noise ratio in baryon correlators, are also fitted in a region of small $t$, where we can safely neglect the contribution of the backward propagating baryon and excited states. This simplifies the fit ansätze for the tree level correlator $\widetilde{C}_{\Omega\Omega}$, and the ratios $\mathcal{R}_{\Omega\Omega}^{\mathrm{em}}$ and $\mathcal{R}_{\Omega\Omega}^{\mathrm{\mathcal{S},s}}$ to those given in~\cref{eq:Omega,eq:Omega_ratio}, respectively. Both the ansätze have two free parameters.

In order to select the best fit ranges we choose those with the maximum value for the Akaike Information Criterion~(AIC)~\cite{Akaike,Akaike2} similarly to the strategy followed by refs.~\cite{Borsanyi:2014jba,Jay:2020jkz,Borsanyi:2020mff}
\begin{equation}
    w= \exp\bigg[-\frac12\,(\chi^2-2 n_\mathrm{dof})\bigg]\,,
\end{equation}
where $n_\mathrm{dof}=n_\mathrm{data}-n_\mathrm{par}$ is the number of degrees of freedom of the fit and the $\chi^2$ function is defined as 
\begin{equation}
    \chi^2 = (\mathbf{C}-\mathbf{C}_\mathrm{M}(\mathbf{a}))^T  \boldsymbol{\Sigma}^{-1}  (\mathbf{C}-\mathbf{C}_\mathrm{M}(\mathbf{a}))\,.
\end{equation} 
Here $\mathbf{C}$ is a vector containing the data (i.e. the time correlators), $\mathbf{C}_\mathrm{M}(\mathbf{a})$ the corresponding model as a function of the fit parameters $\mathbf{a}$ and $\boldsymbol{\Sigma}$ the covariance matrix 
\begin{equation}
    \boldsymbol{\Sigma} = \frac{1}{n_\mathrm{B}-1}\sum_{i=1}^{n_\mathrm{B}}(\mathbf{C}_{i}-\langle\mathbf{C}\rangle)(\mathbf{C}_{i}-\langle\mathbf{C}\rangle)^T\,,
\end{equation}
with $n_\mathrm{B}$ the number of bootstrap samples.
The AIC weight function favours fits that have minimal $\chi^2$ with the largest $n_\mathrm{dof}$ possible, which penalises fits with a low $\chi^2$ per degree of freedom resulting from over-fitting the data.

The datasets $\mathbf{C}$ used for the 7 analyses can be summarized as follows
\begin{align}
    \boldsymbol{1.}&\, \ \mathbf{C}_K^\mathrm{f}=(C_{\mathrm{K}x},\,
        \mathcal{R}_{\mathrm{K}x}^\mathrm{self,u},\,\mathcal{R}_{\mathrm{K}x}^\mathrm{self,s},\,\mathcal{R}_{\mathrm{K}x}^\mathrm{exch},\,\mathcal{R}_{\mathrm{K}x}^\mathrm{\mathcal{S},u},\,\mathcal{R}_{\mathrm{K}x}^\mathrm{\mathcal{S},s}    
    )\quad \text{with} \ x=\mathrm{(K,A)}\,,\nn\\[5pt]
    \label{eq:analyses_corr}
    \boldsymbol{2.}&\, \ \,  \mathbf{C}_\pi^\mathrm{f}=(
        C_{\mathrm{\pi}y},\,
        \mathcal{R}_{\mathrm{\pi}y}^\mathrm{self,u},\,
        \mathcal{R}_{\mathrm{\pi}y}^\mathrm{exch},\,
        \mathcal{R}_{\mathrm{\pi}y}^\mathrm{\mathcal{S},u}
    )\quad \text{with} \ y=\mathrm{(\pi,A)}\,,\qquad
     \boldsymbol{3.}\, \ \mathbf{C}_K^\mathrm{nf}\ =( \mathcal{R}_{K\ell}^\mathrm{nf})\,,\\[5pt]
     \boldsymbol{4.}&\, \ \mathbf{C}_\pi^\mathrm{nf}=(\mathcal{R}_{\pi\ell}^\mathrm{nf})\,, 
      \quad
     \boldsymbol{5.}\, \ \mathbf{C}_\Omega=({C}_{\Omega\Omega})\,,
     \quad
     \boldsymbol{6.}\, \ \mathbf{C}_\Omega^\mathrm{em}=(\mathcal{R}_{\Omega\Omega}^\mathrm{em})\,,\quad
     \boldsymbol{7.}\, \ \mathbf{C}_\Omega^\mathrm{\mathcal{S},s}=(\mathcal{R}_{\Omega\Omega}^\mathrm{\mathcal{S},s})\,.\nn
\end{align}
The corresponding sets of fit parameters $\mathbf{a}$ are
\begin{align}
    \label{eq:analyses_pars}
    \boldsymbol{1.}&\, \ \mathbf{a}_K^\mathrm{f}=(
        \mathbf{a}_K,
        \mathbf{a}_K^\mathrm{self,u},\, 
        \mathbf{a}_K^\mathrm{self,s},\,
        \mathbf{a}_K^\mathrm{exch},\,
        \mathbf{a}_K^\mathrm{\mathcal{S},u},\,
        \mathbf{a}_K^\mathrm{\mathcal{S},s}   
    )\,, \nn\\[5pt]
    \boldsymbol{2.}&\, \ \, \mathbf{a}_\pi^\mathrm{f}=(
        \mathbf{a}_\pi,
        \mathbf{a}_\pi^\mathrm{self,u},\, 
        \mathbf{a}_\pi^\mathrm{exch},\,
        \mathbf{a}_\pi^\mathrm{\mathcal{S},u}
    )\,, \quad \, \, 
    \boldsymbol{3.} \, \ \mathbf{a}_K^\mathrm{nf}\ =( {a}_{K\ell}^\mathrm{nf})\,,\ \  \\[5pt]
    \boldsymbol{4.}&\, \ \mathbf{a}_\pi^\mathrm{nf}=({a}_{\pi\ell}^\mathrm{nf})\,,\quad
    \boldsymbol{5.} \, \ \mathbf{a}_\Omega=(\mathbf{a}_{\Omega\Omega})\,,\quad
    \boldsymbol{6.} \, \ \mathbf{a}_\Omega^\mathrm{em}=(\mathbf{a}_{\Omega\Omega}^\mathrm{em})\,,\quad
    \boldsymbol{7.} \, \ \mathbf{a}_\Omega^\mathrm{\mathcal{S},s}=(\mathbf{a}_{\Omega\Omega}^\mathrm{\mathcal{S},s})\,,\nn
\end{align}
where we have defined 
\begin{equation}
    \label{eq:analyses_pars2}
    \mathbf{a}_P = (
        m_P,\, |Z_P|^2,\, \Acal_P Z_P )\,, \quad 
    \mathbf{a}_P^\mathrm{x} = \Big(
        \delta m_P^\mathrm{x},\,
        2\, \frac{\delta Z_P^\mathrm{x}}{Z_P},\, 
        \frac{\delta \Acal_P^\mathrm{x}}{\Acal_P}+\frac{\delta Z_P^\mathrm{x}}{Z_P}\Big)\,, \quad
    {a}_{P\ell}^\mathrm{nf} = 
        \frac{\delta \Acal_P^\mathrm{nf}}{\Acal_P}\,,
\end{equation}
\begin{equation}
    \label{eq:analyses_pars3}
    \mathbf{a}_{\Omega\Omega} = (m_\Omega,\, |Z_\Omega|^2)\,,\quad
    \mathbf{a}_{\Omega\Omega}^\mathrm{x} = \Big(\delta m_\Omega^\mathrm{x},\, 2\, \frac{\delta Z_\Omega^\mathrm{x}}{Z_\Omega}\Big)\,.
\end{equation}

In the case of the factorizable correlators, however, the bootstrap covariance is rank-deficient as the number of original samples $n_\mathrm{cfg}=60$ is smaller than the dimension of the covariance matrix.
Some form of regularisation is then required to make the $\chi^2$-problem well-conditioned. To this end we choose to neglect the covariance between the rows of $\mathbf{C}^\mathrm{f}_{\mathrm K}$ and $\mathbf{C}^\mathrm{f}_{\pi}$ with and without photon lines.
This choice is motivated by the fact that the correlation matrix is approximately block diagonal, and furthermore, we verified that the optimum parameters do not change significantly if correlation is also neglected between the correlation functions with different operator insertions.
Finally, to reduce the number of degrees of freedom further, only a subsequence of correlator data separated by the \emph{thinning} parameter $\Delta t$ are included in the fit, which are reported for each fit in~\cref{tab:fit_summary}.
The regulated $\chi^2$ thus defined, the best-fit parameters are determined by minimizing the~$\chi^2$ function for a given fit range.

The choice of the fit ranges for each correlator is made using two different approaches depending on the number of possibilities.
For non-combined fits, like those on the $\Omega$ correlators, the maximum number of fit ranges spanning the region $t\in[0,T/2-1]$ (with $T/a=96$) is 1128. In the case of non-factorizable diagrams, including also all possible ranges in the lepton-time variable $t_\ell$, the maximum number of fit ranges is of $\order(10^5)$.
In this case it is computationally feasible to do fits for all possible fit ranges and to compare the values of $w$.
However, applying the same strategy to the combined factorizable fits would be computationally unfeasible, as the maximum number of possible fits is $\order(10^{24})$ or $\order(10^{36})$ for pion and kaon, respectively.
To find good fit range(s) with large AIC we utilize a genetic algorithm as described in~\cref{app:GA} to perform the optimization.
The outcome of this procedure is a set of fit ranges and their associated AIC weights $w$ from each analysis.
There is, however, a large multiplicity of good fit results. In order to capture the variability in the resulting good fits, we consider the 5 fits from each analysis that correspond to the highest AIC. This is an arbitrary and seemingly small number, which however already leads to a large multiplicity of $n_\mathrm{fit}=5^7=78125$ alternative combinations for the fit parameters. The propagation of the variations due to these alternatives to the final results is discussed in~\cref{sec:model_uncertainties}.

\begin{table}[t]
    \centering
    \begin{tabular}{|c|c|c|c|c|c|c|}
    \hline
    & $n_\mathrm{corr}$ & $n_\mathrm{par}$ & $n_\mathrm{dof}$ & $\Delta t/a$ & $\chi^2$  & $p$-value  \\ \hline\hline
    $\mathbf{1}$ & 8  & 18 & 80 & 2 & 49.98 & 1.00  \\ \hline
    $\mathbf{2}$ & 12 & 12 & 95 & 2 & 65.00 & 0.99  \\ \hline
    $\mathbf{3}$ & 5  & 1 & 24 & 2 & 21.42 & 0.61  \\ \hline
    $\mathbf{4}$ & 3  & 1 & 32 & 2 & 29.41 & 0.60  \\ \hline
    $\mathbf{5}$ & 1  & 2 & 7  & 1 & 5.14  & 0.64  \\ \hline
    $\mathbf{6}$ & 1  & 2 & 9  & 1 & 5.32  & 0.81  \\ \hline
    $\mathbf{7}$ & 1  & 2 & 6  & 1 & 1.73  & 0.94  \\ \hline
    \end{tabular}
    \caption{Details of the best fits for the 7 analysis performed in this work and presented in~\cref{fig:fact-tree,fig:fact-exch,fig:fact-selfu,fig:fact-mud,fig:nonfact,fig:omega_treelevel_effmass,fig:omega_ratios}: number of correlators ($n_\mathrm{corr}$), number of parameters ($n_\mathrm{par}$), number of degrees of freedom ($n_\mathrm{dof}$), size of thinning interval ($\Delta t/a$), chi-squared ($\chi^2$) and one-sided $p$-value.}
    \label{tab:fit_summary}
\end{table}

\begin{figure}[b]
    \centering
    \begin{subfigure}{.49\textwidth}
      \centering
      \includegraphics[width=\textwidth]{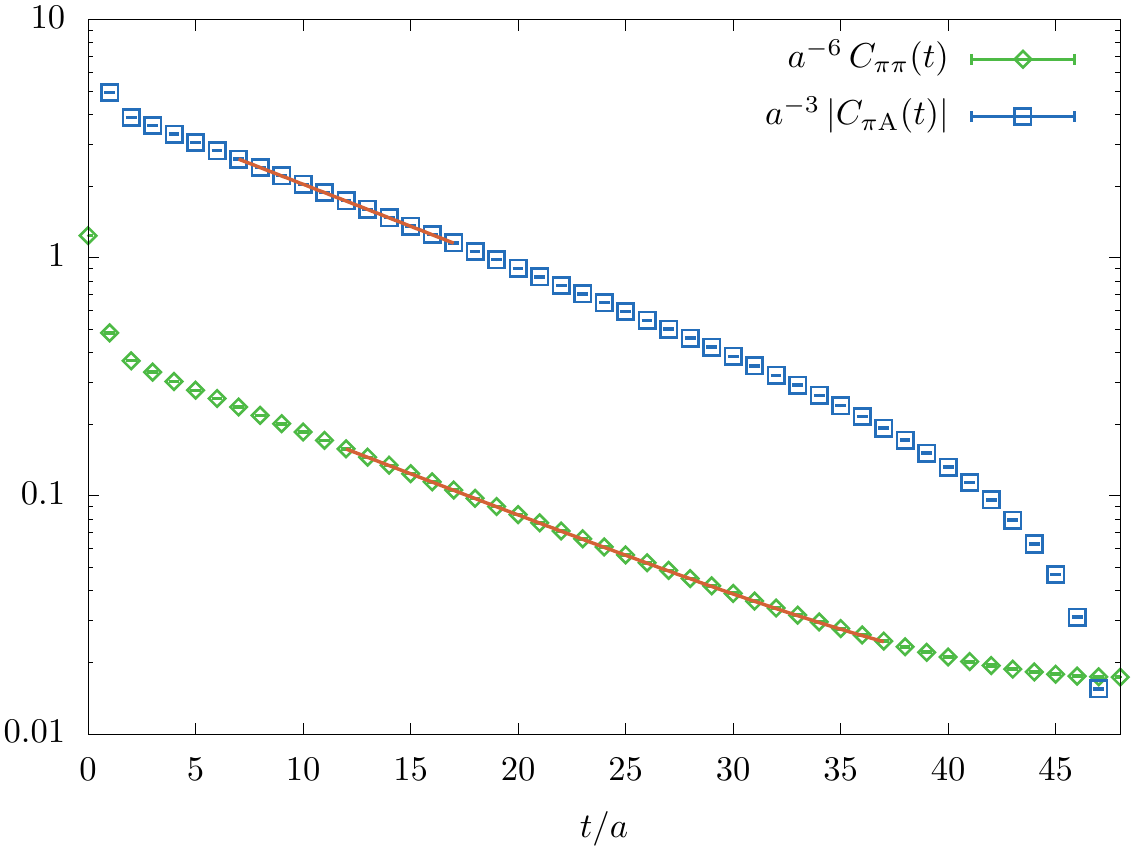}     
        \caption{pion
        }
        \label{fig:fact-pion-tree}
    \end{subfigure}%
    \hfill
    \begin{subfigure}{.49\textwidth}
        \includegraphics[width=\textwidth]{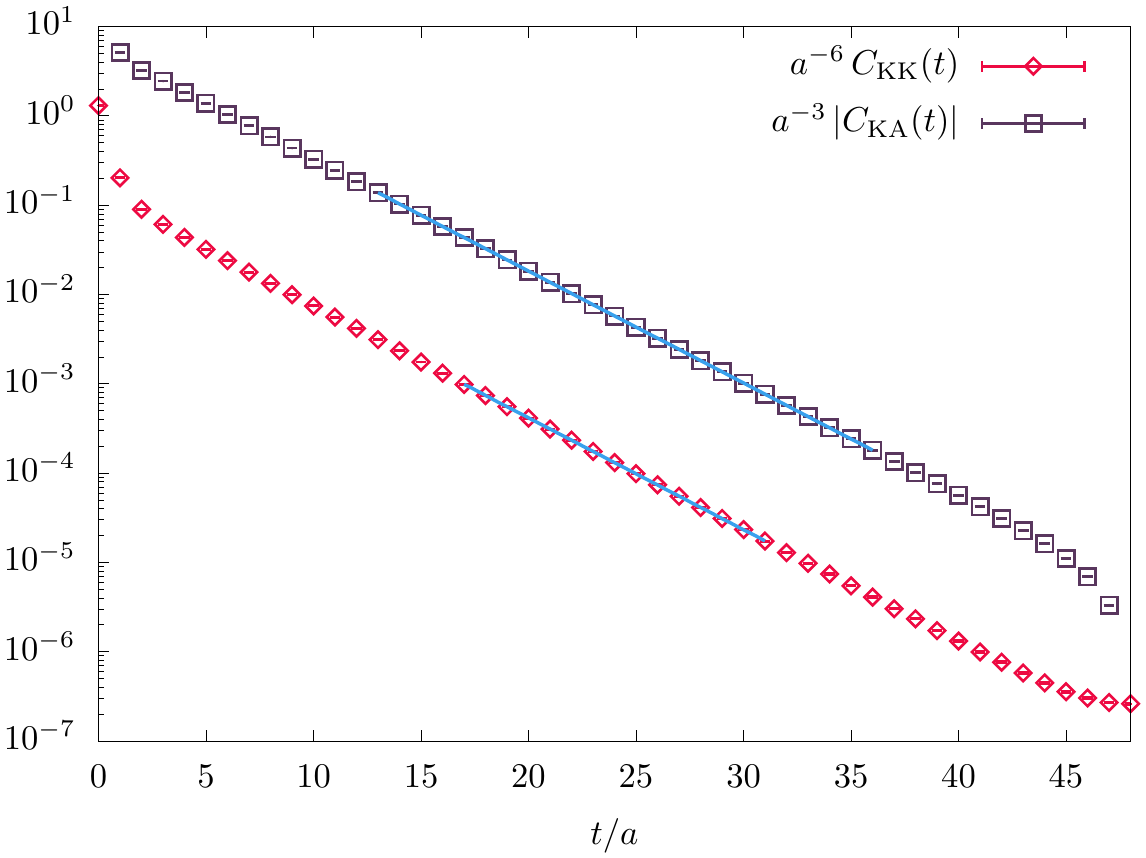}
        \caption{kaon
        }
        \label{fig:fact-kaon-tree}
    \end{subfigure}
    \caption{Tree-level correlators $C_{\mathrm{PP}}(t)$ and $|C_{\mathrm{PA}}(t)|$ for pion (a) and kaon (b). The solid lines with error band correspond to the best fits of the data.}
    \label{fig:fact-tree}
\end{figure}

\begin{figure}[t]
    \centering
    \begin{subfigure}{.49\textwidth}
      \centering
      \includegraphics[width=\textwidth]{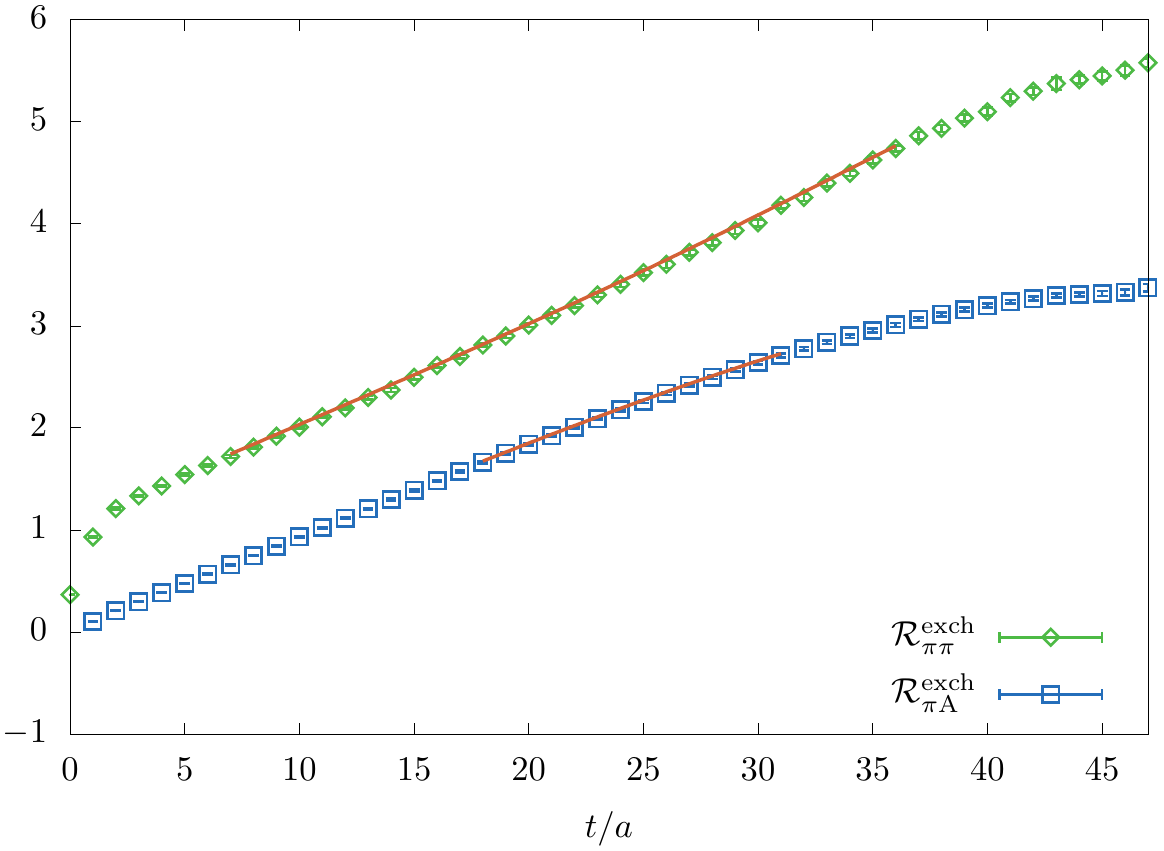}     
        \caption{pion
        }
        \label{fig:fact-pion-exch}
    \end{subfigure}%
    \hfill
    \begin{subfigure}{.49\textwidth}
        \includegraphics[width=\textwidth]{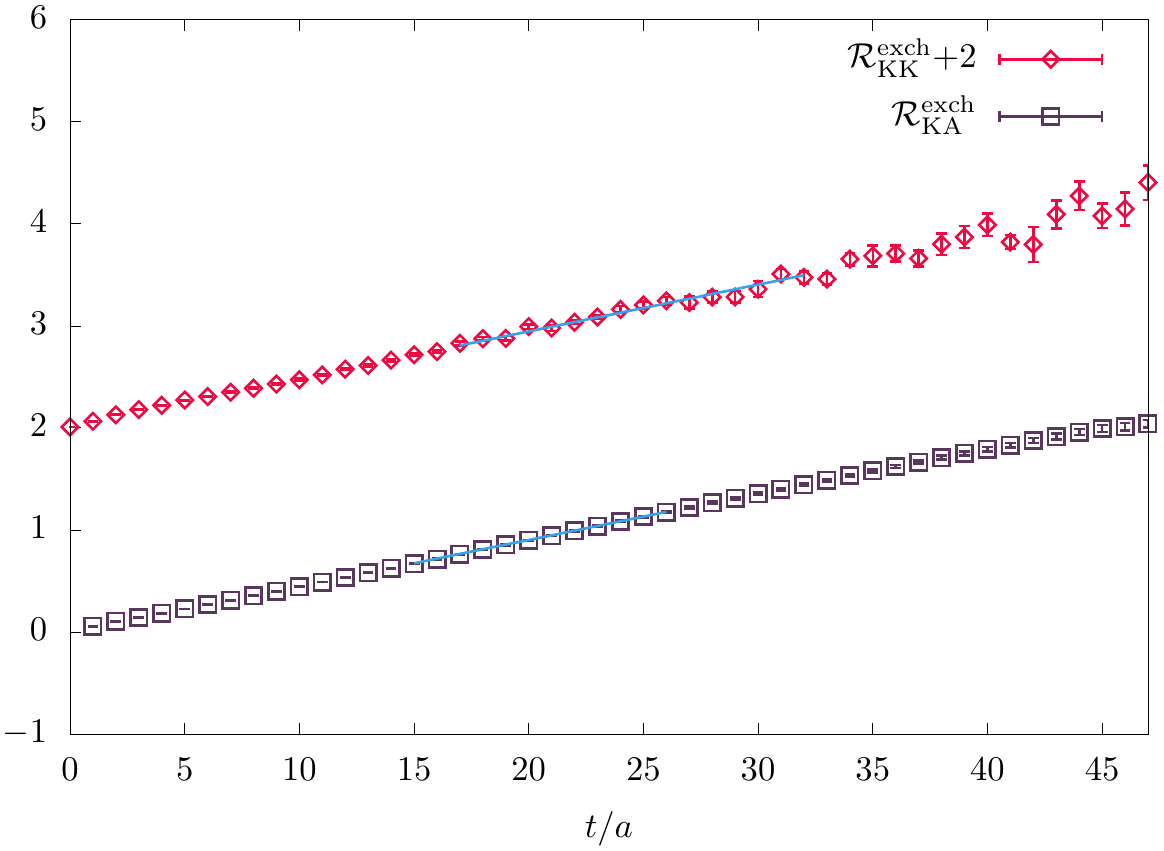}
        \caption{kaon
        }
        \label{fig:fact-kaon-exch}
    \end{subfigure}
    \caption{Factorizable diagram with a photon exchanged between the two constituent quarks, $\mathcal{R}^\mathrm{exch}_{\mathrm{PP}}(t)$ and $\mathcal{R}^\mathrm{exch}_{\mathrm{PA}}(t)$, for pion (a) and kaon (b). The solid lines with error band correspond to the best fits of the data.}
    \label{fig:fact-exch}
\end{figure}

\begin{figure}[H]
    \centering
    \begin{subfigure}{.49\textwidth}
      \centering
      \includegraphics[width=\textwidth]{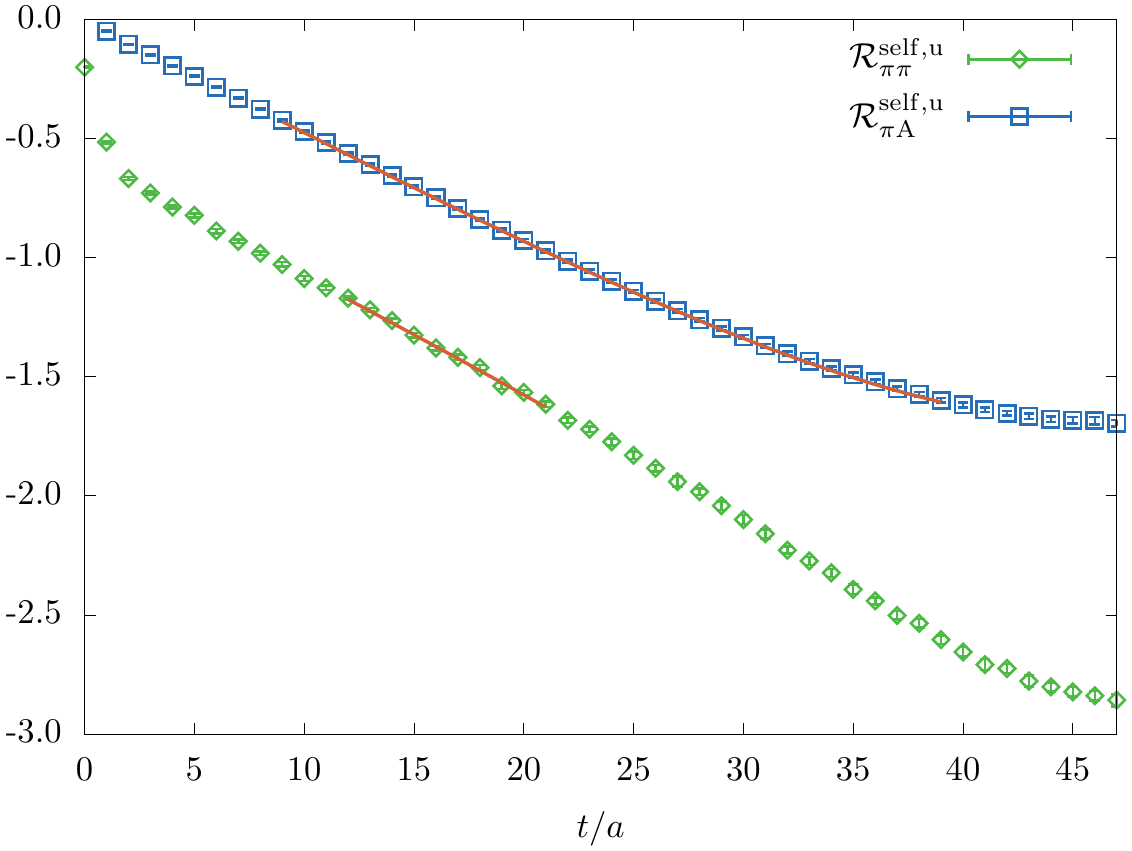}     
        \caption{pion
        }
        \label{fig:fact-pion-selfu}
    \end{subfigure}%
    \hfill
    \begin{subfigure}{.49\textwidth}
        \includegraphics[width=\textwidth]{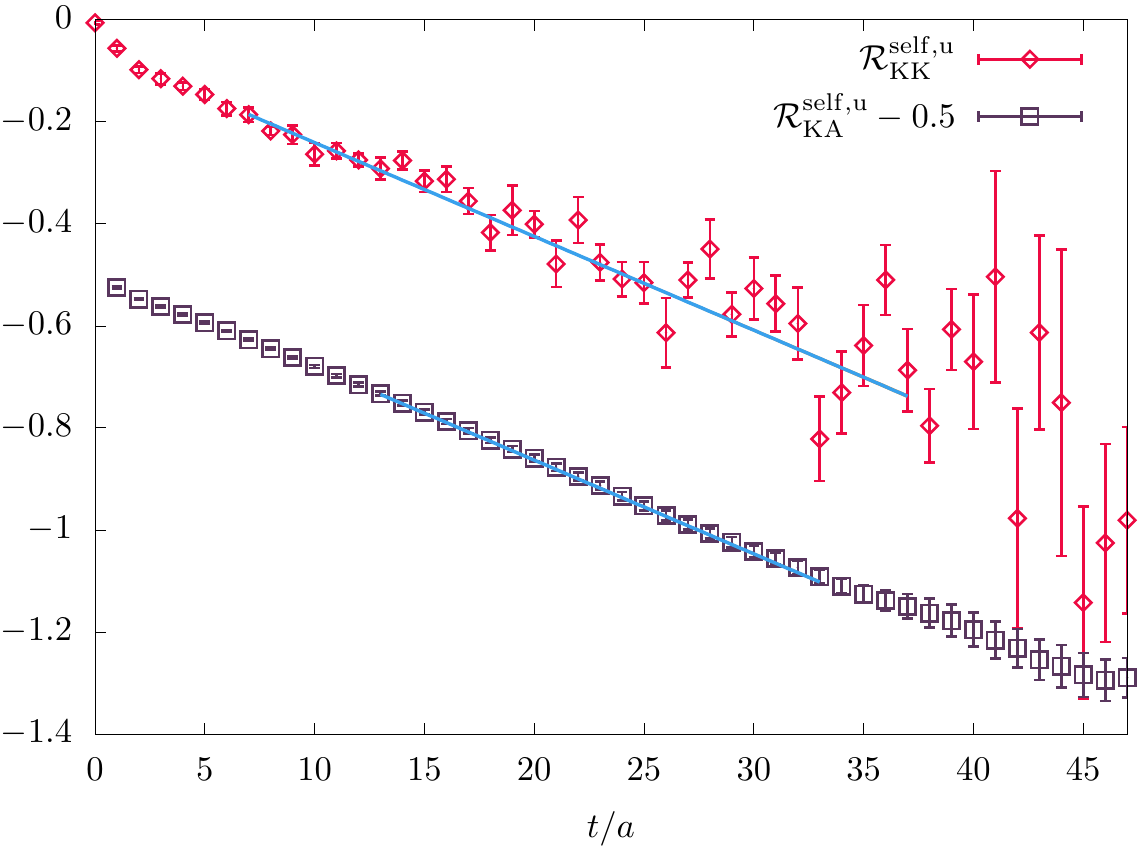}
        \caption{kaon
        }
        \label{fig:fact-kaon-selfu}
    \end{subfigure}
    \caption{Factorizable diagram with  $u$-quark self-energy correction, $\mathcal{R}^\mathrm{self,u}_{\mathrm{PP}}(t)$ and $\mathcal{R}^\mathrm{self,u}_{\mathrm{PA}}(t)$, for pion (a) and kaon (b). The solid lines correspond to the best fits of the data.}
    \label{fig:fact-selfu}
\end{figure}

\begin{figure}[H]
    \centering
    \begin{subfigure}{.49\textwidth}
      \centering
      \includegraphics[width=\textwidth]{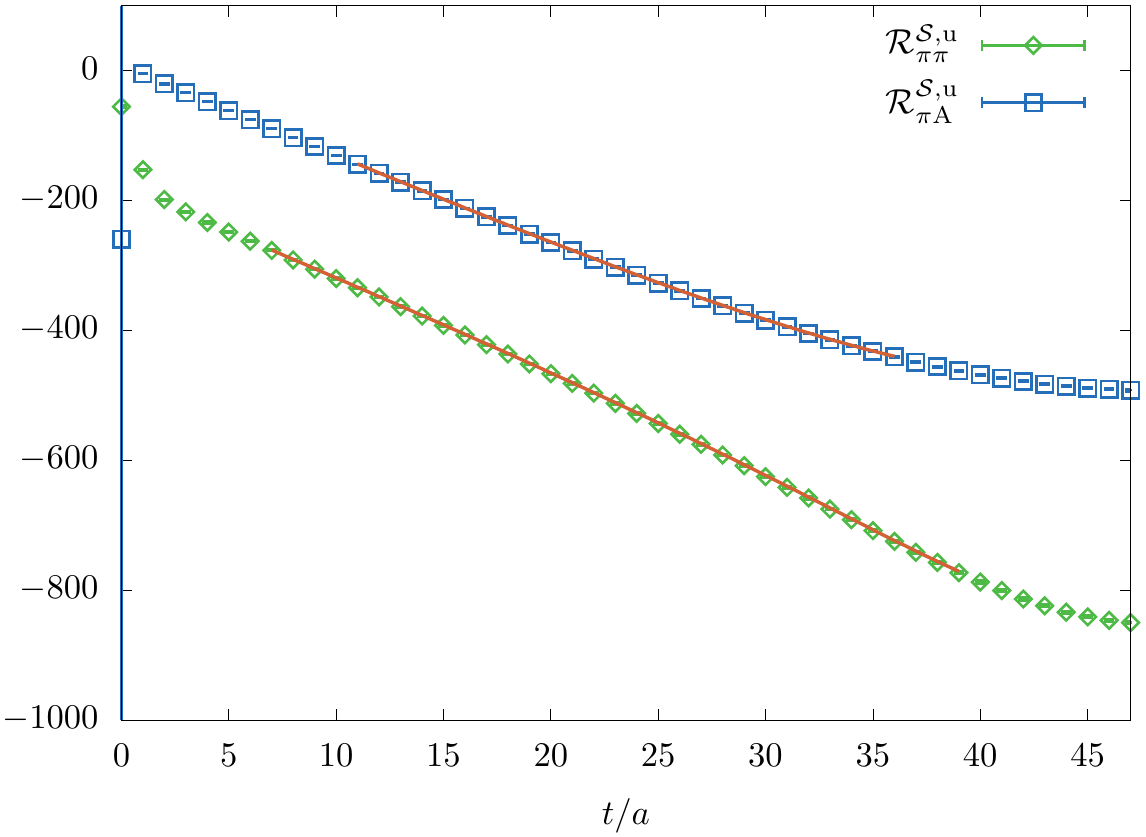}     
        \caption{pion
        }
        \label{fig:fact-pion-mud}
    \end{subfigure}%
    \hfill
    \begin{subfigure}{.49\textwidth}
        \includegraphics[width=\textwidth]{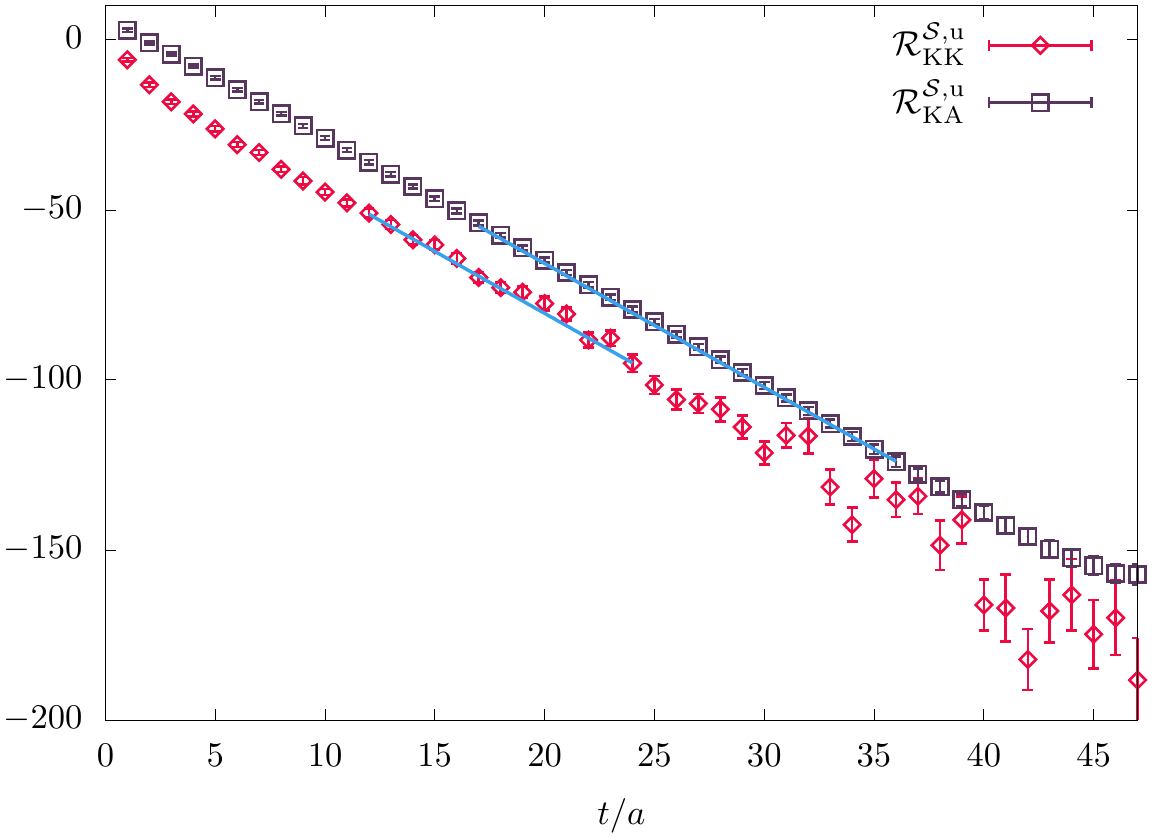}
        \caption{kaon
        }
        \label{fig:fact-kaon-mud}
    \end{subfigure}
    \caption{Factorizable diagram with a scalar insertion along the $u$-quark line, $\mathcal{R}^\mathrm{\mathcal{S},u}_{\mathrm{PP}}(t)$ and $\mathcal{R}^\mathrm{\mathcal{S},u}_{\mathrm{PA}}(t)$, for pion (a) and kaon (b). The solid lines correspond to the best fits of the data. }
    \label{fig:fact-mud}
\end{figure}

Here we only show the representative best fits of the correlators for each analysis, i.e. those corresponding to the highest AIC weight. In~\cref{fig:fact-tree} the tree-level pion and kaon correlators of~\cref{eq:C_PA_spectr_FV,eq:C_PP_spectr_FV} are shown on a logarithmic scale, their slope being related to the tree-level meson mass $m_P$. The electromagnetic corrections due to the exchange of photons between the two constituent quarks and to the $u$-quark self-energy are reported in~\cref{fig:fact-exch,fig:fact-selfu}, respectively, normalized by the tree-level diagrams. In this case the slopes of the correlators correspond to the corrections to the meson mass $\delta m_P^\mathrm{exch}$ and $\delta m_P^\mathrm{self,u}$ (see~\cref{eq:deltaC_fact_PA_spectr_FV,eq:deltaC_fact_PP_spectr_FV}). The correction due to the scalar insertion on the $u$-quark leg is shown instead in~\cref{fig:fact-mud}. 

\begin{figure}[H]
    \centering
    \begin{subfigure}{.49\textwidth}
      \centering
      \includegraphics[width=\textwidth]{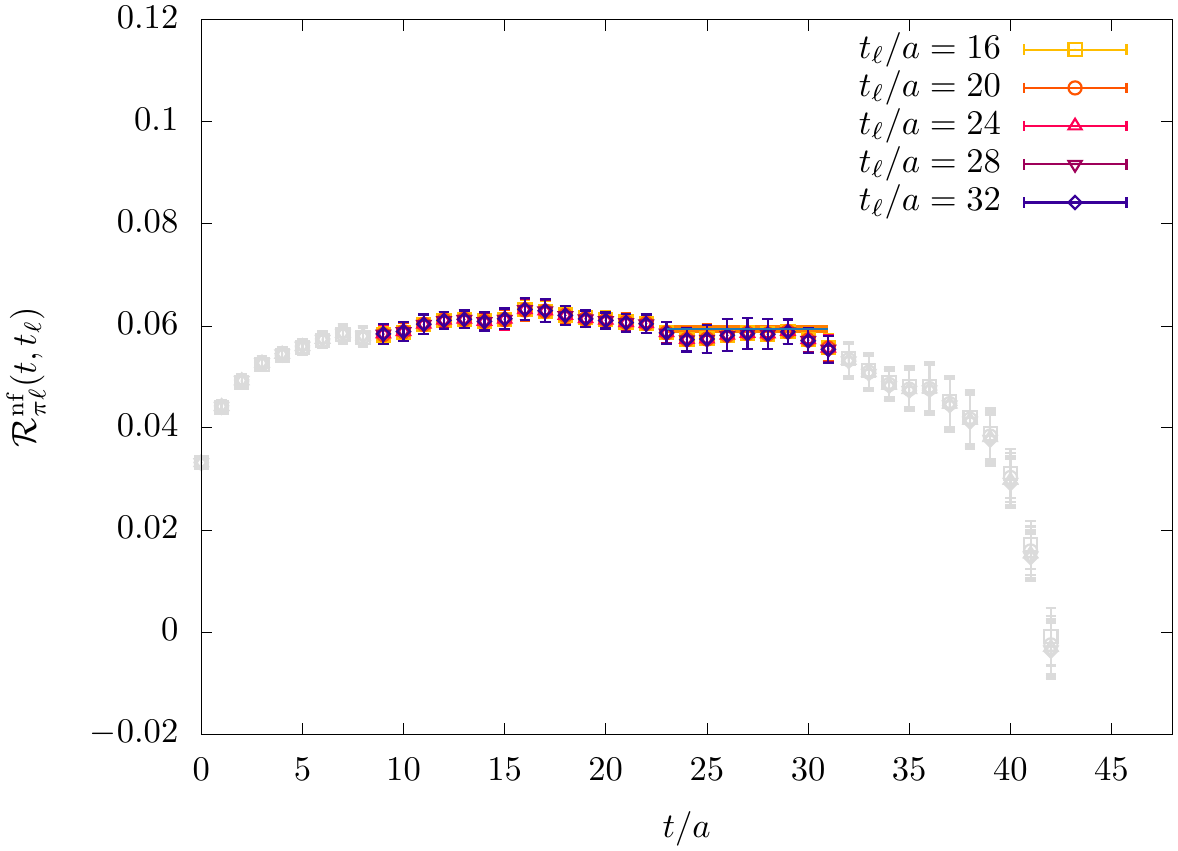}     
        \caption{pion
        }
        \label{fig:nonfact-pion}
    \end{subfigure}%
    \hfill
    \begin{subfigure}{.49\textwidth}
        \includegraphics[width=\textwidth]{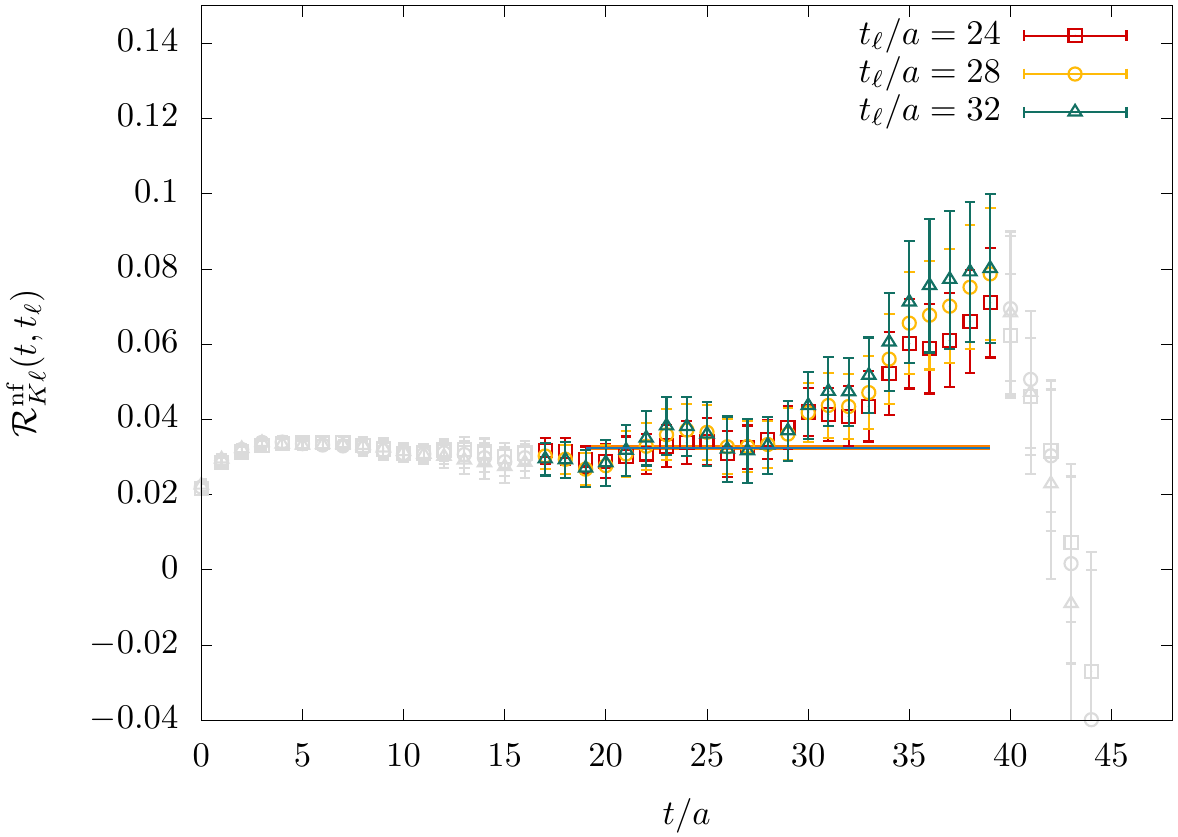}
        \caption{kaon
        }
        \label{fig:nonfact-kaon}
    \end{subfigure}
    \caption{Non-factorizable diagram $\mathcal{R}^\mathrm{nf}_{\mathrm{P}\ell}(t,t_\ell)$ for pion (a) and kaon (b), for multiple values of lepton source-sink separation $t_\ell/a$. The solid lines correspond to the best fits of the data. The grey points denote the data that are not included in any of the top 5 best fits.}
    \label{fig:nonfact}
\end{figure}

The non-factorizable correlators $\mathcal{R}_\mathrm{P\ell}^\mathrm{nf}$ defined in~\cref{eq:non_fact_fitansatz} are reported in~\cref{fig:nonfact} for both pions (left) and kaons (right). The expected time behaviour $f_\mathrm{P\ell}(t,T)$ is visible from the data, with plateaus in the region $t\ll T/2$. The dependence on the lepton source-sink time separation $t_\ell$ is suppressed by the use of the projector on the forward propagating signal (see~\cref{app:projectors} for more details). The constant fits to the data corresponding to the highest value of the AIC weight are reported in the figures, while the grey points identify the data which are not included in any of the top 5 best fits selected in our analysis.
The details for the best fits are reported in~\cref{tab:fit_summary} for the 7 analysis performed in this work.

\subsection{Tuning of the bare parameters}
\label{sec:tuning}

From each of the fits performed in the factorizable analyses (1) and (2) outlined in~\cref{eq:analyses_corr,eq:analyses_pars} we obtain an estimate of the masses of the charged pion, the charged and neutral kaon and the neutral BMW mesons at the simulation iso-QCD point, together with their leading IB corrections. Analogously, we obtain the mass of the $\Omega^-$ baryon and its corrections from the analyses (5), (6) and (7). 
Imposing the renormalization conditions in~\cref{sec:IB_effects}, we can then obtain the relevant mass shifts $(\hat{\mathbf{m}}^\phi-\hat{\mathbf{m}}^\iso)$, $(\hat{\mathbf{m}}^\qcd-\hat{\mathbf{m}}^\iso)$ and $(\hat{\mathbf{m}}^\phi-\hat{\mathbf{m}}^\qcd)$ that allow one to define the IB correction $\delta \hat{X}(\sigmavec^\iso)$ to a given observable $\hat{X}$, as well as its decomposition into strong isospin-breaking and electromagnetic effects~(see~\cref{eq:X_SIB,eq:X_gamma}). 

The mass shift $(\hat{\mvec}^\phi-\hat{\mvec})$ from the physical to the simulation point is obtained by imposing~\cref{eq:ren_cond_full} and simultaneously solving the following system of equations
\begin{equation}
    \frac{\mathrm{M}_j^2}{m_\Omega^2}\,\bigg[1+ 2\,\aem\bigg(\frac{\delta \mathrm{M}_j^\mathrm{em}}{\mathrm{M}_j}-\frac{\delta m_\Omega^\mathrm{em}}{m_\Omega}\bigg) + 2\sum_\mathrm{q}\bigg(\frac{\delta \mathrm{M}_j^{\mathcal{S},\mathrm{q}}}{\mathrm{M}_j}-\frac{\delta m_\Omega^{\mathcal{S},\mathrm{q}}}{m_\Omega}\bigg)(\hat{m}_\mathrm{q}^\phi-\hat{m}_\mathrm{q}) \bigg]=\bigg(\frac{\mathrm{M}_j^2}{m_\Omega^2}\bigg)^\pdg\,,
    \label{eq:ren_cond_phys_explicit}
\end{equation}
where $j=1,2,3$ and $\mathbf{M}=\{m_{\pi^+},m_{K^+},m_{K^0}\}$. Finite-volume effects are applied to the meson masses on the right-hand side of~\cref{eq:ren_cond_phys_explicit} making use of the formula in~\cref{eq:massfves}.
Once the vector $(\hat{\mvec}^\phi-\hat{\mvec})$ is known, the QCD mass shifts $(\hat{\mvec}^\qcd-\hat{\mvec})$ are obtained from~\cref{eq:ren_cond_QCD} using the BMW mesons $\mathbf{N}=\{M_{\rmu\rmd},\Delta M,M_{K\chi}\}$ and solving the system 
\begin{align}
    &\frac{\mathrm{N}_j^2}{m_\Omega^2}\,\bigg[1 + 2\sum_\mathrm{q}\bigg(\frac{\delta \mathrm{N}_j^{\mathcal{S},\mathrm{q}}}{\mathrm{N}_j}-\frac{\delta m_\Omega^{\mathcal{S},\mathrm{q}}}{m_\Omega}\bigg)(\hat{m}_\mathrm{q}^\qcd-\hat{m}_\mathrm{q}) \bigg]=\\
    &\qquad
    \frac{\mathrm{N}_j^2}{m_\Omega^2}\,\bigg[1+ 2\,\aem\bigg(\frac{\delta \mathrm{N}_j^\mathrm{em}}{\mathrm{N}_j}-\frac{\delta m_\Omega^\mathrm{em}}{m_\Omega}\bigg) + 2\sum_\mathrm{q}\bigg(\frac{\delta \mathrm{N}_j^{\mathcal{S},\mathrm{q}}}{\mathrm{N}_j}-\frac{\delta m_\Omega^{\mathcal{S},\mathrm{q}}}{m_\Omega}\bigg)(\hat{m}_\mathrm{q}^\phi-\hat{m}_\mathrm{q}) \bigg]\,,\nn
\end{align}
with $j=1,2,3$.
Finally, the iso-QCD point is determined solving the system in~\cref{eq:ren_cond_iso} for $(\hat{\mvec}^\iso-\hat{\mvec})$, namely for $j=1,2,3$
\begin{align}
    &\frac{\mathrm{N}_j^2}{m_\Omega^2}\,\bigg[1 + 2\sum_\mathrm{q}\bigg(\frac{\delta \mathrm{N}_j^{\mathcal{S},\mathrm{q}}}{\mathrm{N}_j}-\frac{\delta m_\Omega^{\mathcal{S},\mathrm{q}}}{m_\Omega}\bigg)(\hat{m}_\mathrm{q}^\iso-\hat{m}_\mathrm{q}) \bigg]=\\
    &\qquad
    \frac{\mathrm{N}_j^2}{m_\Omega^2}\,\bigg[1+ 2\,\aem\bigg(\frac{\delta \mathrm{N}_j^\mathrm{em}}{\mathrm{N}_j}-\frac{\delta m_\Omega^\mathrm{em}}{m_\Omega}\bigg) + 2\sum_\mathrm{q}\bigg(\frac{\delta \mathrm{N}_j^{\mathcal{S},\mathrm{q}}}{\mathrm{N}_j}-\frac{\delta m_\Omega^{\mathcal{S},\mathrm{q}}}{m_\Omega}\bigg)(\hat{m}_\mathrm{q}^\phi-\hat{m}_\mathrm{q}) \bigg](1-\delta_{j,2})\,.\nn
\end{align}

Using only the best fit from each of the analyses (i.e. the one corresponding to the highest AIC weight), we obtain the following bare quark masses in lattice units
\begin{equation}
    \resizebox{.9\hsize}{!}{$
    \begin{pmatrix}
        \hat{m}_{\rmud}^\iso\\ \delta \hat{m}^\iso\\ \hat{m}_{\rms}^\iso
    \end{pmatrix} = 
    \begin{pmatrix}
        0.00068\,(2)\\ 
        0\\
        0.0353\,(4)
    \end{pmatrix}\,, \
    \begin{pmatrix}
        \hat{m}_{\rmud}^\qcd\\ \delta \hat{m}^\qcd\\ \hat{m}_{\rms}^\qcd
    \end{pmatrix} = 
    \begin{pmatrix}
        0.00068\,(2)\\
        -0.0010\,(4)\\
        0.0353\,(4)
    \end{pmatrix}\,, \
    \begin{pmatrix}
        \hat{m}_{\rmud}^\phi\\ \delta \hat{m}^\phi\\ \hat{m}_{\rms}^\phi
    \end{pmatrix} = 
    \begin{pmatrix}
        0.00068\,(2)\\
        -0.0010\,(4)\\
        0.0352\,(4)
    \end{pmatrix} 
    $}\,.
\end{equation}
The difference between the simulation point and the physical point is given by
\begin{equation}
    \begin{pmatrix}
        \hat{m}_{\rmud} - \hat{m}_{\rmud}^\phi\\ 
        \delta\hat{m} - \delta \hat{m}^\phi\\ 
        \hat{m}_{\rms}-\hat{m}_{\rms}^\phi
    \end{pmatrix} = 
    \begin{pmatrix}
        0.00010\,(2)\\
        0.0010\,(4)\\
        0.0010\,(4)
    \end{pmatrix}\,,
\end{equation}
and an important feature to notice is the similar size between the deviations in $\hat{m}_{\rmud}$, $\hat{m}_{\rms}$, and 
$\delta \hat{m}^\phi$. This justifies the linearity assumption made in~\cref{sec:IBeffects_lattice}, where we assumed that the $\hat{m}_{\rmud}$ and $\hat{m}_{\rms}$  corrections to match with the physical point were of the same size as the isospin-breaking effects.

Finally, we also obtain the following ratios 
\begin{align}
    \bigg[\frac{\mathbf{M}^\iso}{m_\Omega^\iso}\bigg]^2 &= \big(0.006530\,(4),\,
    0.08761\,(3),\,
    0.08761\,(3) \big)\,,\label{eq:isoratios}\\
    \bigg[\frac{\mathbf{M}^\qcd}{m_\Omega^\qcd}\bigg]^2 &= \big(0.006530\,(4),\,
    0.08653\,(2),\,
    0.08869\,(3) \big)\,,\\
    \bigg[\frac{\mathbf{N}^\phi}{m_\Omega^\phi}\bigg]^2 &= \big(0.006530\,(4),\,
    -0.00464\,(2),\,
    0.08434\,(2)\big)\,.
\end{align}
Assuming $m_{\Omega}^\iso=m_\Omega^\phi$, we can form the ratio in~\cref{eq:isoratios} using the iso-QCD meson masses in the GRS scheme quoted in ref.~\cite{DiCarlo:2019thl}, 
\begin{equation}
    \bigg[\frac{\mathbf{M}^\iso}{m_\Omega^\iso}\bigg]^2_{\mathrm{GRS}}=\big(0.00652\,(2),\, 0.08746\,(4),\, 0.08746\,(4)\big)\,.
\end{equation}
The pion component agrees between the two schemes, the difference in the kaon part is more significant, but represents only
a per-mille relative difference, which as we will see in~\cref{sec:results} is well covered by our systematic errors.

\subsection{Estimation of model uncertainties}
\label{sec:model_uncertainties}
\begin{figure}[b]
    \centering
    \includegraphics[width=0.75\textwidth]{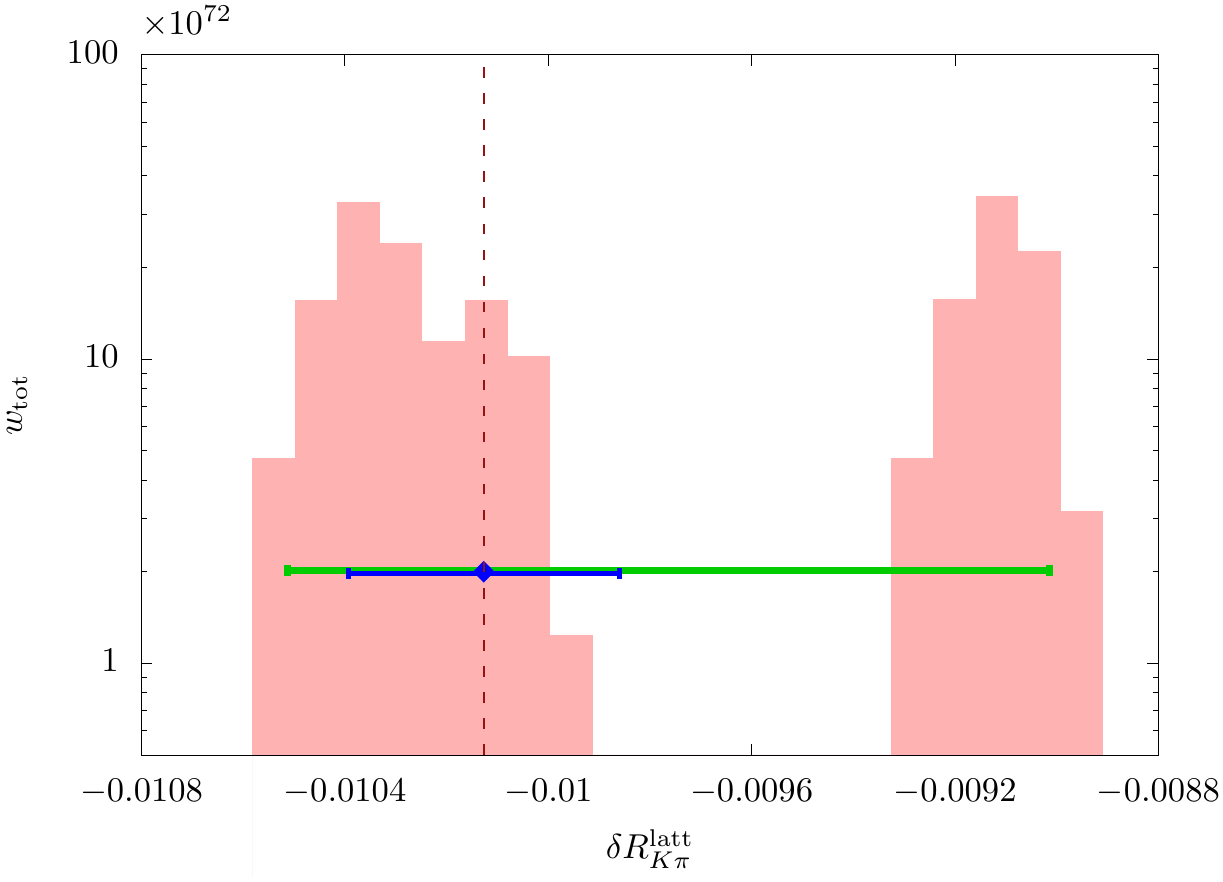}
    \caption{Histogram displaying the distribution of $\delta R^\text{latt}_{K\pi}$. The blue and green error bands are the statistical and fit systematic errors, respectively.}
    \label{fig:histogram_deltaRKPi_latt}
\end{figure}
As described in~\cref{sec:num_strategy}, given a fit-scan procedure we obtain a set of fit ranges and their associated AIC weights from each analysis. In this calculation we choose to consider the 5 best fits from each analysis, thus obtaining a total of $n_\mathrm{fit}=78125$ determinations of the fit parameters for each bootstrap sample. 
We can then combine the fit parameters, tune the bare-quark masses and use~\cref{eq:deltaRKPi_latt} to get $n_\mathrm{fit}$ estimates of $\delta R_{K\pi}^\mathrm{latt}$ for each bootstrap.
In order to extract a value from this set, we build a histogram of the $n_\mathrm{fit}$ values of $\delta R_{K\pi}^\mathrm{latt}$ reweighting each entry with the total AIC weight for that choice of analyses, namely
\begin{equation}
    w_\mathrm{tot} = \prod_{i=1}^7 w_i = \exp\bigg[-\frac12\,\sum_{i=1}^7 (\chi^2_i-2 n_{\mathrm{dof},i})\bigg]\,.
\end{equation}
Here the summation applies because the 7 analyses are independent. The relative size between the $n_\mathrm{fit}$ different weights informs us which prediction is preferable to the others. The choice of limiting our study to only the fit ranges associated to the top 5 AIC weights in each analysis is motivated by the fact that, with this reweighting procedure, the exponential suppresses the relatively inferior fit results. Given the reweighted histogram built from the $n_\mathrm{fit}$ values of $\delta R_{K\pi}^\mathrm{latt}$, which is shown in~\cref{fig:histogram_deltaRKPi_latt}, we determine the central value for this quantity as the median of the histogram. Choosing the median instead of the mean makes the result not subject to drastic variations due to outlier predictions. In~\cref{fig:histogram_deltaRKPi_latt} the median is indicated in blue together with its statistical error, while the green error bar is the fit systematics. The statistical error is estimated from the variance of the bootstrap samples of the medians, while the systematic error is determined from the distribution of $\delta R_{K\pi}^\mathrm{latt}$ as the $2\sigma$ interval around the central value (i.e.~the central 95\% band). The distribution of $\delta R_{K\pi}^\mathrm{latt}$ in ~\cref{fig:histogram_deltaRKPi_latt} shows two peaks. They suggest that there are two sets of fit intervals with statistically distinct fit results but with comparably good AIC weights. However, we note that both peaks are covered by our systematic error. Alternative strategies were attempted to stress the stability of our result, including different assumptions about correlation and different weight functions\footnote{We tried the flat distribution, the two-sided $p$-value, and ad-hoc functions favouring high number of degrees of freedom with small $\chi^2$.}, all leading to results within the quoted systematic uncertainty. The value obtained for $\delta R_{K\pi}^\mathrm{latt}$ is then
\begin{equation}
    \delta R_{K\pi}^\mathrm{latt} = -0.0101\,(3)_\mathrm{stat.}(^{+11}_{-4})_\mathrm{fit}\,.
\end{equation}

\section{Results}
\label{sec:results}

The finite-volume lattice estimate of $\delta R_{K\pi}^\mathrm{latt}$ obtained in the previous section can be combined with the function $\delta R^{(n)}_P(L)$ discussed in~\cref{sec:FVE} in order to subtract the logarithmic divergence and power-like electromagnetic finite-volume effects up to $\order(1/L^2)$. The prediction of $\delta R_{K\pi}$ is then obtained according to~\cref{eq:deltaRP_limits} by adding the contribution of the real-photon emission $\delta R_{K\pi}^\mathrm{pert}(\omega_\gamma^\mathrm{max})$, which is computed in perturbation theory~\cite{Carrasco:2015xwa} and reported in~\cref{eq:deltaRP_pert_max}. 
To evaluate the finite-volume correction, we compute~\cref{eq:deltaRndef,eq:ourresultyn} using the finite-volume coefficients determined in ref.~\cite{DiCarlo:2021apt} and the simulation point meson masses and decay constants, together with $F_A^{\pi}$ and $F_A^{K}$ from $\chi$PT at $\order(p^4)$ and $\order(p^6)$, respectively~\cite{Bijnens:1992en,Cirigliano2012,Desiderio:2020oej}. For our lattice of size $L_{48}\equiv 48a$ we get
\begin{equation}
    \delta R^{(2)}_{K\pi}(L_{48}) = 
    \delta R^{(2)}_{K}(L_{48})-\delta R^{(2)}_{\pi}(L_{48}) = -0.00730 
    \,.
\end{equation}
Evaluating~\cref{eq:deltaRP_pert_max} for the physical values of the meson masses $m_\pi$ and $m_K$~\cite{Workman:2022ynf} we obtain instead
\begin{equation}
    \delta R_{K\pi}^\mathrm{pert}(\omega_\gamma^\mathrm{max}) = \delta R_{K}^\mathrm{pert}(\omega_\gamma^\mathrm{max})-\delta R_{\pi}^\mathrm{pert}(\omega_\gamma^\mathrm{max}) = -0.00583\,.
\end{equation}

Combining the previous results and including all sources of systematic uncertainties, which we are going to discuss in the rest of the section, our result for $\delta R_{K\pi}$ obtained at $L=L_{48}$ amounts to 
\begin{equation}
	\delta R_{K\pi} = -
    0.0086\,(3)_{\mathrm{stat.}}({}^{+11}_{-4})_\mathrm{fit}(5)_{\mathrm{disc.}}(5)_{\mathrm{quench.}}(39)_\mathrm{vol.}
    \,.
	\label{eq:deltaRKPi_result}
\end{equation}
The first error is statistical, and it is obtained from the variance of the bootstrap distribution of $\delta R_{K\pi}^\mathrm{latt}$. The second error is the systematic uncertainty associated with our fit strategy and estimated as the 2$\sigma$ interval around the median of the distribution of $\delta R_{K\pi}^\mathrm{latt}$ (see~\cref{fig:histogram_deltaRKPi_latt}), as discussed in~\cref{sec:model_uncertainties}.

The calculation presented in this work has been performed on a single lattice spacing and, as a consequence, we are not able to extrapolate $\delta R_{K\pi}^\mathrm{latt}$ to the continuum limit. Thus, we quote a systematic uncertainty associated with the residual $\order(a^2)$ discretization effects. This is estimated as $(a\Lambda)^2$ with $a^{-1}=1730\,\mathrm{MeV}$ and $\Lambda=400\,\mathrm{MeV}$~\cite{RBC:2018dos}. This gives $(a\Lambda)^2\sim 5\%$, which is applied to the central value of $\delta R_{K\pi}^\mathrm{latt}$ before the finite-volume subtraction and results in $-0.0086\,(5)_\mathrm{disc.}$.

Electromagnetic interactions involving sea quarks have been neglected in this work. 
Such electro-quenching effects are SU(3) and $1/N_\mathrm{c}$ suppressed for $\order(\aem)$ contributions and expected to be of $\sim\order(10\%)$~\cite{Budapest-Marseille-Wuppertal:2013rtp} of the QED correction to the rate.
Separating $\delta R_{K\pi}^\mathrm{latt}$ into its electromagnetic and strong-isospin breaking contributions $(\delta R_{K\pi}^\mathrm{latt})^\gamma$ and $(\delta R_{K\pi}^\mathrm{latt})^\mathrm{SIB}$ (according to the separation scheme outlined in~\cref{sec:IB_effects}) we take the 10\% of the e.m. part $(\delta R_{K\pi}^\mathrm{latt})^\gamma$ as our electro-quenching error. Using the median of the $(\delta R_{K\pi}^\mathrm{latt})^\gamma$ distribution, $(\delta R_{K\pi}^\mathrm{latt})^\gamma=-0.0047$, we get $-0.0086\,(5)_\mathrm{quench.}$.

As discussed above, we use the finite-volume correction including the full $1/L^2$ scaling (denoted by $\delta R_{P}^{(2)}(L)$) in order to determine our central value for the infinite-volume observable $\delta R_P$. We then estimate the systematic uncertainty, associated with the truncation of the finite-volume expansion, by forming the difference between $\delta R_{P}^{(2)}(L)$ and the correction including the point-like $1/L^3$ contribution (denoted by $\delta R_{P}^{(3), \text{pt}}(L)$). These quantities are given explicitly by combining~\cref{eq:deltaRndef,eq:YtildeL,eq:ourresultyn} from~\cref{sec:FVE}.

Since we are only targeting the difference between pion and kaon decay rates, the finite-volume correction we actually require is the difference
\begin{equation}
\delta R^{(n)}_{K\pi}(L) = \delta R^{(n)}_K(L) - \delta R^{(n)}_\pi(L)\,.
\end{equation}
The systematic uncertainty on this is then estimated via
\begin{align}
\sigma_L & \equiv	 \delta R_{K\pi}^{(3),\mathrm{pt}}(L)-\delta R_{K \pi}^{(2)}(L) \,, \\
& = \frac{\alpha_{\text{em}}}{2 \pi} \frac{32\pi^2}{(m_\pi L)^3} \bigg [\frac{ 2+(m_{\ell}/m_\pi)^2}{ [1+ (m_{\ell}/m_\pi)^2]^3 } - \frac{ 2+(m_{\ell}/m_K)^2 }{ [1+ (m_{\ell}/m_K)^2]^3 } \frac{m_\pi^3}{ m_K^3} \bigg ] \,,
\end{align}
where we have given the explicit expression as it will play a crucial role in our error budget.
We stress that $\sigma_L$ is positive. As we will see below, both $\delta R_{K \pi}^{(2)}(L)$ and the final observable $\delta R_{K \pi}$ are negative. This implies that, if one were to estimate $\delta R_{K \pi}$ using $\delta R_{K\pi}^{(3),\mathrm{pt}}(L)$, the result would be reduced (a negative number with increased magnitude) as compared to the central value we report using $\delta R_{K \pi}^{(2)}(L)$.

To give numerical results for $\delta R_{K\pi}^{(n)}$, we require values for the meson masses and decay constants, the muon and $W$-boson mass, and also values for the form factors $F_A^K$ and $F_A^\pi$. As above, we take $F_A^K$ and $F_A^\pi$ from $\chipt$ at $\order(p^6)$ and $\order(p^4)$, respectively~\cite{Bijnens:1992en,Cirigliano2012,Desiderio:2020oej}, and meson masses and decay constants from our simulation. The full set of inputs is then
\begin{equation}
\begin{alignedat}{3}
m_{\pi} & = 0.1395 \, \textrm{GeV} \, ,
&
m_{K} & = 0.4992 \, \mathrm{GeV} \, ,
\\[5pt]
m_{\mu} & = 0.1057\, \mathrm{GeV} \, ,
&
m_W & = 80.38 \, \mathrm{GeV}\,,
\\[5pt]
f_{\pi} & = 0.1310 \, \mathrm{GeV} \, ,
&
f_{K} & = 0.1564 \, \mathrm{GeV} \, ,
\\[5pt]
F_{A}^{\pi , \, \chipt} & = 0.0119 \, ,
&
F_{A}^{K , \, \chipt} & = 0.0340 \, ,
\end{alignedat}
\end{equation}
where results for $F_{A}$ are reported to three digits, all other numbers to four digits and uncertainties are neglected, since these are completely subdominant in our determination.

Substituting these values into the expressions for $\delta R_{K\pi}^{(n)}(L)$ and evaluating at the lattice volume used in this calculation, $L_{48}$, one finds
\begin{equation}
\delta R_{K\pi}^{(1)}(L_{48}) = - 0.00468
\,, \qquad
\delta R_{K\pi}^{(2)}(L_{48}) = -0.00730
\,, \qquad
\delta R_{K\pi}^{(3),\text{pt}}(L_{48}) = -0.00337
\,. \nonumber
\end{equation}
From these numerical results it is clear that the convergence appears quite poor for the volume used. In particular the ratio
\begin{equation}\label{eq:L3relshift}
\frac{\sigma_{L_{48}}}{\delta R^{(2)}_{K\pi}(L_{48})}
\simeq -0.54 \,,
\end{equation}
implies that the finite-volume correction is assigned a $54 \%$ systematic error in our method. As emphasized above, this is due to the fact that we have only incomplete knowledge of the correction through $1/L^3$, since the structure-dependent piece has not been calculated. 
Propagating this through~\cref{eq:deltaRP_limits}, we obtain
\begin{equation}
\delta R_{K\pi} = -0.0086\,(39)_\mathrm{vol.} \,.
\end{equation}

\begin{figure}[h]
	\centering
	\begin{minipage}{0.49\textwidth}
	\centering
	\includegraphics[width=\textwidth]{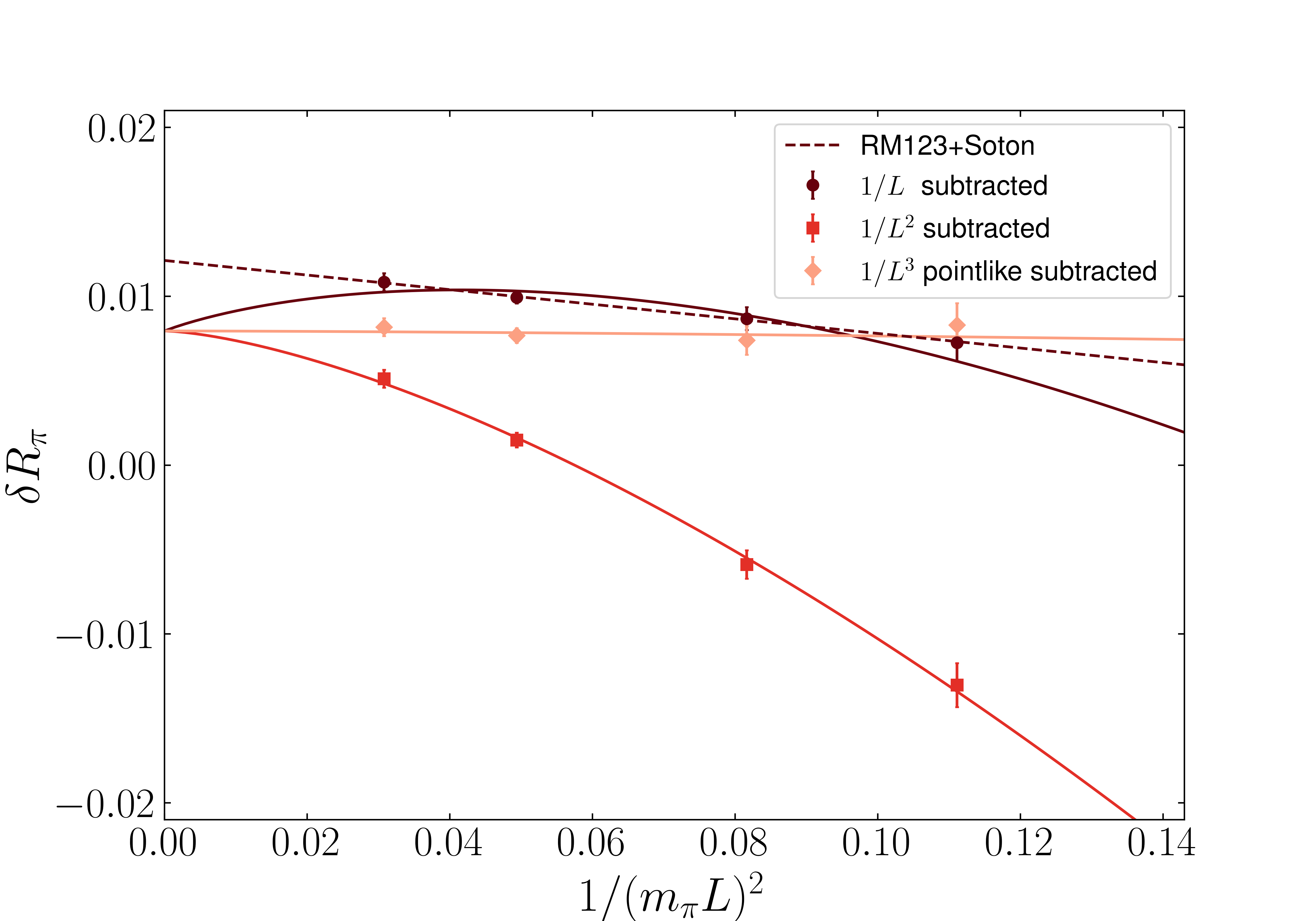}
	\caption*{(a)}
	\end{minipage}
	\hfill
	\begin{minipage}{0.49\textwidth}
	\centering
	\includegraphics[width=\textwidth]{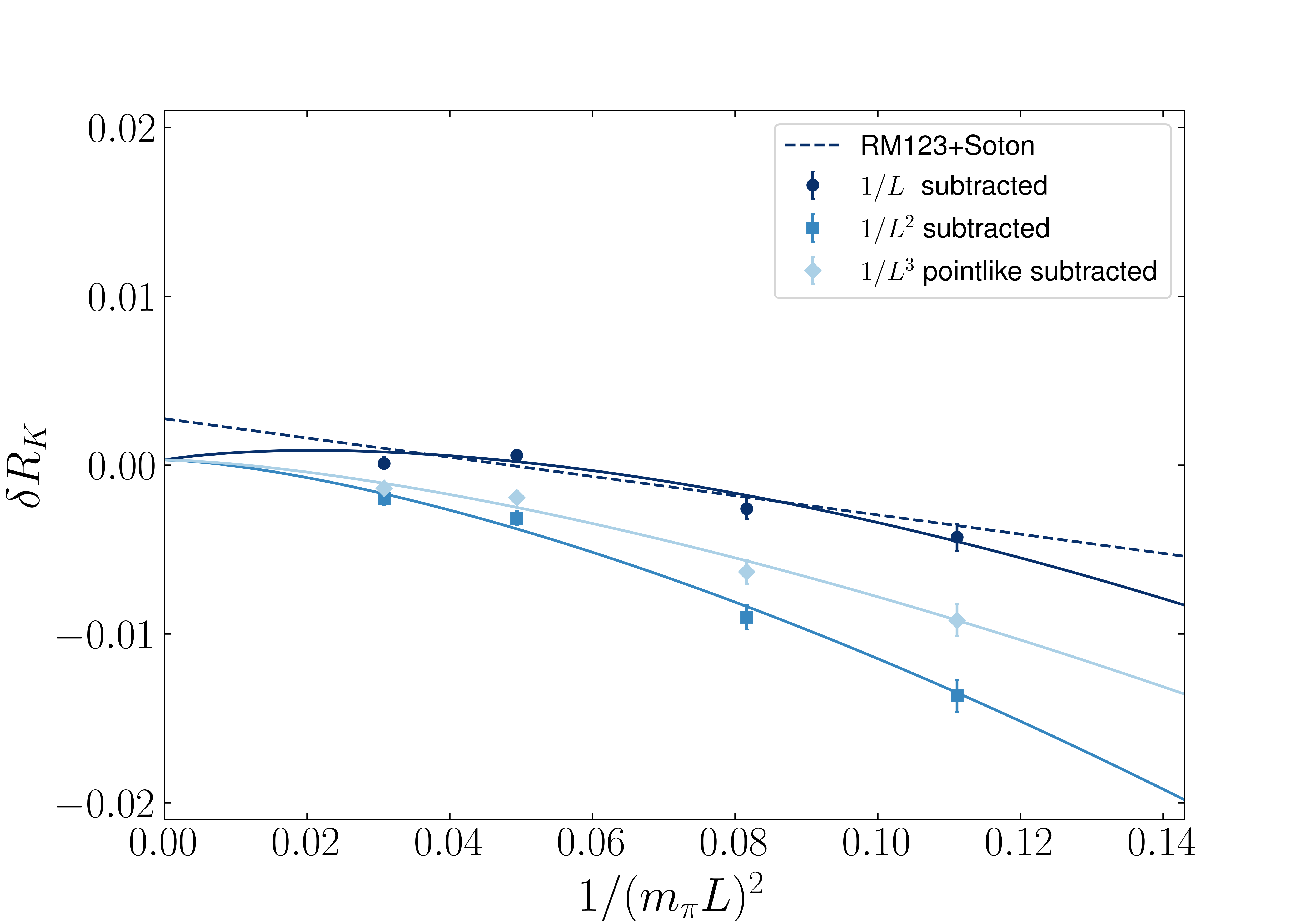}
	\caption*{(b)}
	\end{minipage}
	\caption{The volume-dependence of (a) $\delta R_{\pi}$ and (b) $\delta R_{K}$ based on data taken from ref.~\cite{DiCarlo:2019thl} supplemented with analytic knowledge from ref.~\cite{DiCarlo:2021apt}. The $L$ dependence is presented at fixed, heavier-than-physical quark masses corresponding to $m_\pi \approx 320$ MeV and $m_K \approx 580$ MeV. As indicated in the legend and described in detail in the text, the various points correspond to different subtractions and the curves to fits of residual $L$ dependence.}\label{fig:rm123sdata}
	\end{figure}

We close this section by presenting additional information on the finite-volume expansion, making use of the analytic results of ref.~\cite{DiCarlo:2021apt} as well as data from the previously published lattice calculation by the RM123S group~\cite{DiCarlo:2019thl}. This calculation uses a different lattice discretization and also extrapolates from heavier-than-physical pions. A key advantage relative to this work, however, is that it includes results at multiple volumes. The data are displayed in~\cref{fig:rm123sdata}, separately for $\delta R_\pi$ and $\delta R_K$. The results are for $m_\pi \approx 320$ MeV and $m_K \approx 580$ MeV, and four different volumes.

Our aim is to examine this data in light of a key conclusion of ref.~\cite{DiCarlo:2021apt}, namely that the structure-dependent part of $Y_{P,\,2}$ (the $1/L^2$ coefficient) is numerically negligible. As this was not known at the time, the approach of ref.~\cite{DiCarlo:2019thl} was to subtract the point-like $1/L$ prediction and to numerically investigate the residual volume dependence. The circular data points in~\cref{fig:rm123sdata}, labelled in the legend as ``$1/L$ subtracted'', show the result of this analysis and coincide to figure~9 of ref.~\cite{DiCarlo:2019thl}. 
As can be seen from the $1/L$-subtracted data and the dashed curves in~\cref{fig:rm123sdata}, a linear description vs.~$1/L^2$ for the residual volume dependence is realistic. This results in a numerical prediction for the structure dependence that is much larger than the analytic result of ref.~\cite{DiCarlo:2021apt}. Another way to reach this same conclusion is to examine the residual $L$ dependence in the pointlike $1/L^2$ subtracted data. The fact that this shows a clear residual slope was interpreted as the effect of the structure dependence at $1/L^2$. This can also be seen in~\cref{fig:rm123sdata} in the square data points labelled ``$1/L^2$ subtracted'' (strictly, here we subtract the full $1/L^2$ behaviour and ref.~\cite{DiCarlo:2019thl} the point-like part, but the distinction is numerically insignificant.)

We argue that the puzzle is resolved by the observation that the data is equally well described by $1/L^3$ behaviour. To explore this we first subtract the point-like $1/L^3$ prediction and find that the $L$ dependence is reduced. This is shown in~\cref{fig:rm123sdata} as the diamonds, labelled in the legend as ``$1/L^3$ pointlike subtracted''.
We then perform a fit of the form $a+b/L^3$ to the point-like $1/L^3$-subtracted data. We find this describes the data reasonably and can be interpreted as an estimate of the residual $1/L^3$ behaviour, again arising from structure dependence.\footnote{Given the discussion above, the reader might note that we are mimicking the approach of ref.~\cite{DiCarlo:2019thl} but one order higher in $1/L$. To this point we stress one key difference; the point-like $1/L^n$ contributions are known to vanish for $n>3$.} The three solid curves in~\cref{fig:rm123sdata} show the result of the $a+b/L^3$ fit for each of the three subtraction scenarios. We stress that the curves are related by analytic terms and that only one fit was performed.

From these considerations, we conclude that the $L \to \infty$ limit is challenging for $\QEDL$ and that analytical knowledge of the $L$ dependence can be of great importance in controlling the systematic error associated to these extrapolations. We are working on several directions to address this issues, including an analytic determination of the structure dependence at $1/L^3$. As is discussed in ref.~\cite{DiCarlo:2021apt}, this will require evaluating a branch-cut contribution (similar to that appearing for the pseudoscalar mass in~\cref{eq:massfves}).

We now turn to the determination of $|\vus|/|\vud|$. For this purpose, symmetrizing the fit systematic in~\cref{eq:deltaRKPi_result} and summing in quadrature all the errors but the ``vol.'' one, we get $\delta R_{K\pi}=-0.0086\,(13)(39)_{\mathrm{vol.}}$. Combining this result with the value of the iso-QCD ratio $f_K/f_\pi$ we can predict $|\vus|/|\vud|$ at leading order in IB corrections as 
\begin{equation}
    \frac{|\vus|}{|\vud|}
    =
	\bigg[\frac{\Gamma(K^+\to \mu^+\nu_\mu[\gamma])}{\Gamma(\pi^+\to \mu^+\nu_\mu[\gamma])}\frac{m_K}{m_\pi} \frac{(m_\pi^2-m_\mu^2)}{(m_K^2-m_\mu^2)}
    \bigg]^{1/2}\frac{f_\pi}{f_K} \bigg(1-\frac{1}{2}\,\delta R_{K\pi} \bigg)\,.
\end{equation}
Averaging\footnote{FLAG does not quote an average for $f_K/f_{\pi}$, but for the isospin-corrected ratio $f_{K^+}/f_{\pi^+}$. We produced the value $f_{K}/f_{\pi} =  1.1930\,(33)$ following exactly the averaging procedure described in the review. Although iso-QCD has been tuned in slightly different ways in the calculations entering this average, from the corresponding values of $m_\pi$ and $m_K$ we expect scheme ambiguities to be below the quoted uncertainty (see discussion in~\cref{sec:scheme_ambiguities,sec:tuning}).} the $\Nf=2+1$ lattice results reviewed in FLAG~\cite{Aoki:2021kgd,Follana:2007uv,Durr:2010hr,RBC:2010qam,RBC:2012cbl,RBC:2014ntl,Scholz:2016kcr,Durr:2016ulb,QCDSF-UKQCD:2016rau}, and using the PDG average for the ratio of experimental decay widths~\cite{Workman:2022ynf}, we obtain
\begin{equation}
    |\vus|/|\vud| = 0.23154 \ (28)_{\rm exp.} \, (15)_{\delta R_P} \, (45)_{\delta R_P,\mathrm{vol.}} \,(65)_{f_{P}},
    \label{eq:vusvud}
\end{equation}
where the first error comes from the experimental measurements, the second is our uncertainty on $\delta R_{K\pi}$ excluding the finite-volume systematics quoted separately, and the last error comes from the average of lattice determinations for $f_{K}/f_{\pi}$.  Interestingly, we find that the error from $f_K/f_\pi$ dominates the uncertainty on $|\vus|/|\vud|$.  The same conclusion is obtained using the RM123S result. In fact, taking $\delta R_{K\pi}^\mathrm{RM123S}=-0.0126\,(14)$~\cite{DiCarlo:2019thl} and the $\Nf=2+1+1$ FLAG average, $f_K/f_\pi=1.1966\,(18)$~\cite{Aoki:2021kgd}, one obtains
$(|\vus|/|\vud|)^\mathrm{RM123S} = 0.23131 \, (28)_{\rm exp} \, (17)_{\delta R_P} \, (35)_{f_{P}}$. This is a clear motivation for future new computations of $f_K/f_\pi$ on the lattice, with the aim of reducing the uncertainty by a factor 2 to 3 to bring it below the current experimental uncertainties on the decay width ratio. Finally, the second-largest uncertainty in~\cref{eq:vusvud} comes from the challenges with finite-volume QED as discussed above. It is foreseeable that this conservative uncertainty will be drastically reduced in the near future, which can be done through the addition of multiple volumes to compute the $1/L^3$ coefficient or the usage of a different QED formulation with smaller volume corrections. In conclusion, there are identified ways forward to reduce in the short-term future the two main systematic errors on $|\vus|/|\vud|$, and beyond those the precision reached on $\delta R_{K\pi}$ is sufficient and below the experimental input uncertainties.

\section{Conclusions}
\label{sec:conclusions}

The study of light-meson leptonic decays is of great relevance for the extraction of the CKM matrix elements $|\vus|$ and $|\vud|$, especially in light of current  outstanding $3\sigma$ tensions in the first-row unitarity~\cite{Workman:2022ynf,Cirigliano:2022yyo}. To either confirm or resolve such tensions, a combined effort of both theory and experiment is necessary. New experimental measurements and analyses are possible for some facilities (e.g.~at NA62, as suggested by the authors of ref.~\cite{Cirigliano:2022yyo}) and can help to clarify the situation. On the theoretical side, precise and controlled calculations of leptonic and semi-leptonic decay rates, including non-perturbative effects of strong interactions, as well as QED and strong isospin-breaking effects, would allow stringent tests of the SM. 

In this paper we have presented the first physical-quark-mass lattice calculation of the leading isospin-breaking effects on the ratio of the rates of kaon and pion decays into muons. This has been performed using chiral domain wall fermions with close-to-physical masses on a single gauge ensemble, i.e. at a fixed value of the lattice spacing and on a finite volume. Finite-volume QED interactions have been regulated according to the $\QEDL$ prescription by removing the spatial zero mode of the photon propagator and the electro-quenched approximation has been employed, thus assigning zero electric charge to the sea quarks.  Including all sources of systematic uncertainty, we obtain 
\begin{equation}
	\delta R_{K\pi} = -
    0.0086\,(3)_{\mathrm{stat.}}({}^{+11}_{-4})_\mathrm{fit}(5)_{\mathrm{disc.}}(5)_{\mathrm{quench.}}(39)_\mathrm{vol.}
    \,.
\end{equation}
This result is compatible with the lattice result obtained by the RM123S collaboration~\cite{DiCarlo:2019thl},
as well as with the $\chi$PT estimate of ref.~\cite{Cirigliano:2011tm}.

Although our statistical error is very competitive with e.g.~the RM123S calculation, the final precision of our estimate of $\delta R_{K\pi}$ is affected by a large systematic uncertainty. This is dominated by the error associated with residual finite-volume effects, which amounts to around $45\%$ of the central value of $\delta R_{K\pi}$. The origin of such a large uncertainty, as explained in~\cref{sec:results}, is due to the lack of knowledge of structure-dependent effects at $\order(1/L^3)$, which are specific to the $\QEDL$ prescription. The discussion in~\cref{sec:results} emphasises the crucial role of finite-volume effects in the extraction of $\delta R_{K\pi}$ and the need for a dedicated study of the $\order(1/L^3)$ contributions.
Two ways of reducing the finite-volume systematic error will be explored in future calculations. On the one hand, work is in progress to understand and determine the $1/L^3$ finite-volume $\QEDL$ contributions. On the other hand, performing the same calculation on multiple volumes can certainly help to extrapolate to the infinite-volume limit. Repeating the calculation on gauge ensembles with different lattice spacings would also allow to reduce the systematic uncertainties associated to discretization effects. For what concerns  electro-quenching, a plan is in place to overcome this approximation calculating quark-disconnected electromagnetic corrections. The progress of our preliminary study has been reported in ref.~\cite{Harris22}.

To conclude, our calculation provides an important step towards future flavour physics precision tests. 
The anticipated extensions of the calculation presented in this work, resulting in smaller systematic uncertainties, will allow for a new theoretical prediction for the ratio $|\vus|/|\vud|$.
However, as discussed at the end of~\cref{sec:results}, a real progress will only be possible if also the precision of the iso-QCD decay constants $f_K/f_\pi$ is improved. At this point, the uncertainties coming from theoretical predictions will no longer dominate over those from experimental inputs in the extraction of $|\vus|/|\vud|$.
We are currently also investigating the prospects for a non-perturbatve determination of the leading isospin-breaking corrections to semi-leptonic $K\rightarrow\pi\ell\nu$ decays, which are relevant for an independent determination of $|\vus|$. Together, these results will provide novel and stringent precision tests of the CKM matrix unitarity.

\acknowledgments

We warmly thank Luigi Del Debbio, Fabian Joswig and our colleagues in the RBC and UKQCD collaborations for many helpful discussions. We also thank the RM123-Soton collaboration for kindly providing lattice data (from ref.~\cite{DiCarlo:2019thl}) for our study of finite-volume effects. V.G. thanks M. Tomii and J. Flynn for useful discussions about free domain wall fermion propagators.
N.~H.-T.~wishes to thank the Higgs Centre for Theoretical Physics at the University of Edinburgh for hosting visits where part of this work was completed.
This work used the DiRAC Extreme Scaling service at the University of Edinburgh, operated by the Edinburgh Parallel Computing Centre on behalf of the STFC DiRAC HPC Facility (\url{www.dirac.ac.uk}). This equipment was funded by BEIS capital funding via STFC capital grant ST/R00238X/1 and STFC DiRAC Operations grant ST/R001006/1. DiRAC is part of the National e-Infrastructure. 
P.B. has been supported in part by the U.S. Department of Energy, Office of Science, Office of Nuclear Physics under the Contract No. DE-SC-0012704 (BNL).
M.D.C., F.E., T.H., V.G., M.T.H., and A.P. are supported in part by UK STFC grant ST/P000630/1. Additionally M.T.H. is supported by UKRI Future Leader Fellowship MR/T019956/1.   
F.E., V.G., R.H., F.\'Oh., A.P. and A.Z.N.Y. received funding from the European Research Council (ERC) under the European Union's Horizon 2020 research and innovation programme under grant agreement No 757646 and A.P. additionally under grant agreement No 813942.
N.~H.-T.~is funded in part by the Albert Einstein Center for Fundamental Physics at the University of Bern, and in part by the Swedish Research Council, project number 2021-06638.
A.J. and J.R have been supported in part by UK STFC grant ST/P000711/1 and ST/T000775/1.
J.R. is also supported in part by UK STFC DiRAC operational grants ST/S003762/1 and ST/W002701/1.

\section*{Appendices}

\appendix
\section{Comparison of local and conserved electromagnetic currents}
\label{app:loc-vs-cons}
\subsection{Theory}

A certain freedom always exists in the detailed choice of how to discretize local composite fields in a lattice calculation. In particular, various equally valid discretizations can be defined that differ in their renormalization and cut-off effects.
In this work, we use the (ultra-)local discretization of the electromagnetic
current on the lattice, defined as
\begin{align}
    V^\mu_{fg}(x) = Z_\mathrm{V}\bar\psi_{f}(x)\gamma^\mu\psi_{g}(x) \,.
\end{align}
This is an extension of the current appearing in~\cref{eq:deltaSF}, as here we allow the possibility of an
off-diagonal flavour current in order to simplify the discussion of particular
quark contractions in isolation.

The local current does not exactly satisfy the QED Ward-Takahashi identity. In
other words, the coupling to the photon field breaks QED gauge invariance
explicitly for non-zero lattice spacing.
In addition to introducing a finite renormalization of the electromagnetic
current at order $\aem$, the lack of gauge symmetry implies that new 
singularities may arise when the position of the current coincides with other local fields.
Such short-distance effects occur, for example, when the vertex is integrated
over the space-time volume and as a result coincides with the axial current
 as in the first correlation function defined in~\cref{eq:deltaC_PA_traces}.
By contrast, when gauge-invariance is preserved using a discretization of the
current which is exactly conserved, singularities associated with overlapping operators are highly constrained
by the Ward-Takahashi identities.

Nevertheless, by power counting one can show that in our set-up no such extra divergences
arise, nor is the automatic $\mathrm O(a)$ improvement of the chiral fermion
discretization spoiled. 
To see this, first consider diagram (a) of \cref{fig:alldiags}. We examine the limit in which 
both electromagnetic vertices approach the
position of the axial current. To identify this diagram in isolation we introduce fictitious valence-quark flavours denoted $1,2,3,4$ (discussed in more detail below) and write
\begin{equation}
    a^8\frac{1}{x^2}\Big[V^\mu_{12}(x) V^\mu_{23}(x) A^\nu_{34}(0)\Big]
    \stackrel{x\rightarrow a}{=}
     \delta Z_\mathrm{A} \, A^\nu_{14}(0) +  \mathrm O(a^2)  \,.
\end{equation}
Here the factor $x^{-2}$ on the left-hand side arises from the short-distance behaviour of the photon
propagator and the $a^8$ arises from the discretized space-time measure. The three key claims in this equation, all justified in the following paragraphs, are (i) that no power divergences (positive powers of $1/a$) arise, (ii) that the constant order simply defines a contribution to the renormalization of the axial current, and (iii) that the leading corrections that vanish as $a \to 0$ are $\mathrm O(a^2) $ rather than $\mathrm O(a) $.

We have introduced additional (degenerate) flavours $1,2,3,4$ in the paradigm of a
partially quenched theory to isolate the contribution from the diagram of
interest. This flavour structure ensures that only operators with energy dimension greater than or equal to three can contribute to the right-hand side, since all contributing operators must carry anti-$1$ and $4$ quantum numbers, and must therefore be built from at least two quarks. The difference between the lowest dimension of operators contributing (three) and the dimension on the left-hand side (nine) is therefore six, and this leads to a $1/a^6$ scaling accompanying the quark bilinear. This is however cancelled by the power of $a^8/x^2 \to a^6$ on the left-hand side, implying that no power divergences arise. This demonstrates point (i) above.

Without any additional symmetries, all rotationally covariant quark bilinears could contribute.
However, in our set-up we have an approximate chiral symmetry, broken due to an exponentially suppressed contribution from the finite extent of the fifth dimension. Taking this to be negligible, we need only catalogue dimension three operators with the correct chiral rotation properties, and the \{1,4\} axial-current is the unique choice with dimension three. 
Thus, the effect of breaking gauge invariance results in an additional
renormalization of the axial current at the next-to-leading order in the
electromagnetic coupling. This demonstrates point (ii) above.

Finally, the discrete lattice chiral symmetry~\cite{Niedermayer:1998bi} forbids mixing with dimension-four
operators with the appropriate definition~\cite{Capitani:2000xi}, which might otherwise introduce linear lattice artefacts in such
off-shell correlation functions.
This is our third and final point (iii) and a similar analysis of the remaining diagrams illustrates that the use of the local current poses no particular difficulties with our chosen discretization.
We now turn to a numerical demonstration that the discrepancy in $\delta R_{K\pi}$ between this current and the conserved vector current at fixed lattice spacing has a value consistent with our expectations for an $\mathrm O(a^2)$ effect.

\subsection{Numerical check}

We perform a numerical test on a smaller $24^3\times 64$ lattice using Shamir-Domain-Wall fermions~\cite{Shamir:1993zy}, with $aM_5=1.8$ and $L_\mathrm{s}/a=16$. We limit the statistics to 10 QCD configurations, with interpolating operator inserted on every other timeslice (32 in total). The pion mass for this ensemble is ${m}_\pi \approx 339.789$~MeV.

\begin{figure}[t]
    \centering
    \begin{minipage}{0.5\textwidth}
        \centering
        \includegraphics[width=1\textwidth]{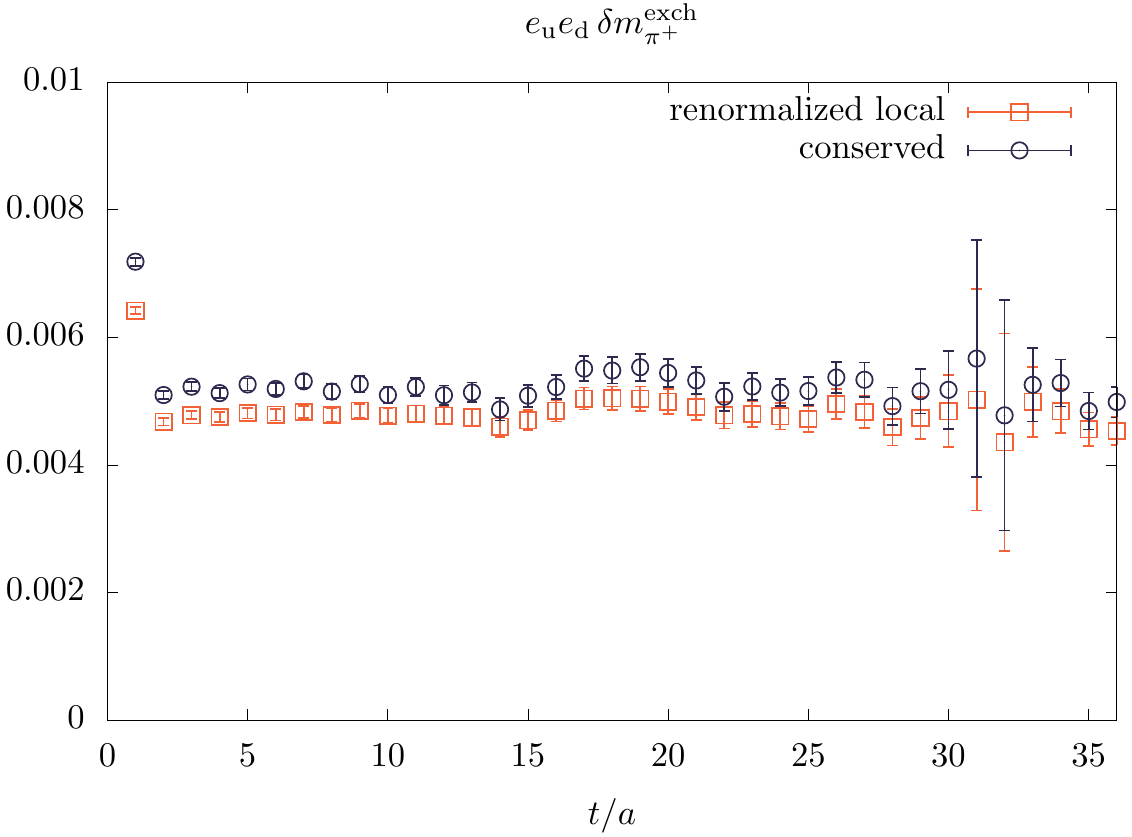}
    \end{minipage}\hfill
    \begin{minipage}{0.5\textwidth}
        \centering
        \includegraphics[width=1\textwidth]{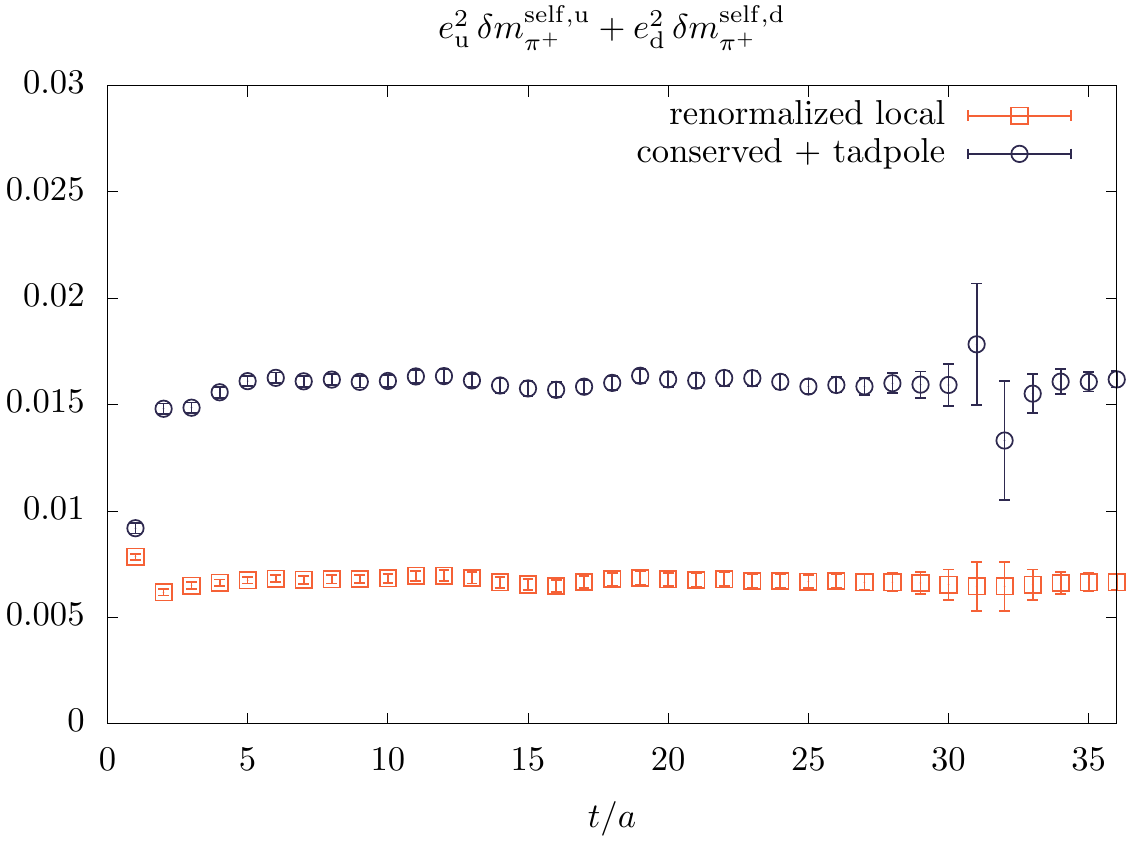}
    \end{minipage}
    \caption{A comparison of the QED correction to the pion mass from the exchange ({left}) and the self energy+tadpole ({right}) diagrams between the local and conserved currents.}
    \label{fig:local-vs-cons_QED_corr}
\end{figure}

The difference between the formulation of local and conserved electromagnetic current is the presence of a tadpole diagram in the latter, which arises from the second derivative of the Dirac operator with respect to the electric charge.
We may extract the contributions to the QED mass corrections from correlator ratios via
\begin{equation}
    \delta m_P^\mathrm{x} = -\frac{\mathcal{R}_\mathrm{PP}^\mathrm{x}(t+1)-\mathcal{R}_\mathrm{PP}^\mathrm{x}(t)}{f_\mathrm{PP}(t+1,T)-f_\mathrm{PP}(t,T)}\,,
\end{equation}
where $\mathrm{x}=\{ \mathrm{self,{q_1}} \,;\, \mathrm{self,{q_2}} \,;\, \mathrm{exch}\}$ while the ratios $\mathcal{R}_\mathrm{PP}^\mathrm{x}(t)$ and the function $f_\mathrm{PP}(t,T)$ are defined in~\cref{eq:deltaC_fact_PP_spectr_FV,eq:fPP}, respectively.
The result for pions is shown in~\cref{fig:local-vs-cons_QED_corr}. For the exchange diagram where the electromagnetic current is inserted on both propagators, we notice that the use of the renormalized local current or the conserved current give very similar results. However, in the case where there are two current insertions on the same quark propagator, the presence of the tadpole contributes additionally to the mass correction, as expected. 
This discrepancy will manifest in the results obtained from the combined fits performed on the tree-level and factorizable correlators. 

Since, for this numerical check, we are working on a gauge ensemble away from the physical point we simply define the iso-QCD point to be equal to the simulation one, i.e.~$\sigmavec^\iso=\sigmavec$, such that $m_P^\iso=m_P$. Furthermore, the  (fictitious) physical point $\sigmavec^\phi$ is defined imposing the following conditions
\begin{align}
    \big(m_{\pi^+}^\phi\big)^2 &= m_\pi^2 +  \left(m_{\pi^+}^{\textrm{\tiny PDG}}\right)^2 - \left(m_{\pi^0}^{\textrm{\tiny PDG}}\right)^2\,,\\
    \big(m_{K^+}^\phi\big)^2 &= m_K^2 + \frac{1}{2}\Big[\big(m_{K^+}^{\textrm{\tiny PDG}}\big)^2 - \big(m_{K^0}^{\textrm{\tiny PDG}}\big)^2\Big]\,,\\
    \big(m_{K^0}^\phi\big)^2 &= m_K^2 - \frac{1}{2}\Big[\big(m_{K^+}^{\textrm{\tiny PDG}}\big)^2 - \big(m_{K^0}^{\textrm{\tiny PDG}}\big)^2\Big]\,,
\end{align}
i.e.~we keep the pion and kaon mass splittings at their experimentally measured values (taken from PDG~\cite{Workman:2022ynf}). These conditions allow us to obtain the quark mass shifts $(\hat{\mvec}^\phi-\hat{\mvec})$ needed to compute IB corrections. The physical value of $\aem$ is tuned instead to its Thomson limit, as done in~\cref{sec:IB_effects}. 

In~\cref{tab:local-vs-conserved_QED_derivatives} we report the photon corrections to the meson masses, as well as the factorizable and non-factorizable contributions to the decay amplitude, obtained using either the conserved electromagnetic current or the renormalized local vector current. We see that all diagrams except the non-factorizable correction give as expected significantly different results using the two different approaches. However, when combining these corrections with those obtained from the insertion of the scalar density, $\sum_x(\hat{m}_\mathrm{q}^\phi-\hat{m}_\mathrm{q})\hat{\mathcal{S}}_\mathrm{q}(x)$ (see~\cref{eq:deltaSF}), the estimates for a physical observable obtained with the two approaches become comparable. In fact, this is the case for $\delta R_{K\pi}^\mathrm{latt}$,
\begin{equation}
    \big(\delta R_{K\pi}^\mathrm{latt}\big)_\mathrm{loc} =  -7.04\,(20)\times 10^{-3}\,, \qquad \big(\delta R_{K\pi}^\mathrm{latt}\big)_\mathrm{cons}=-6.91\,(20)\times 10^{-3}\,.
\end{equation}
We can see that the two results are compatible with each other, the difference $(\delta R_{K\pi}^\mathrm{latt})_\mathrm{loc}- (\delta R_{K\pi}^\mathrm{latt})_\mathrm{cons}$ being consistent with zero within errors. The slightly larger value ($\sim 2\%$) of $(\delta R_{K\pi}^\mathrm{latt})_\mathrm{loc}$ can be associated to $\order(a^2)$ cut-off effects, which as explained in the previous subsection are expected to contribute.
\begin{table}[h]
    \centering
    \begin{tabular}{| c | c | c |} 
    \hline
     & local & conserved \\
    \hline \hline
    $ (\delta \hat{m}_{\pi^+}^\mathrm{2})^\mathrm{e.m.}$ & $0.005476\,(38)$ & $0.009160\,(30)$\\
    \hline
    $ (\delta \hat{M}_{\rmu\rmu}^\mathrm{2})^\mathrm{e.m.}$ & $0.000434\,(39)$ & $0.005782\,(29)$\\
    \hline
    $ (\delta \hat{m}_{K^+}^\mathrm{2})^\mathrm{e.m.}$ & $0.008148\,(47)$ & $0.012260\,(54)$\\
    \hline
    $ (\delta \hat{m}_{K^0}^\mathrm{2})^\mathrm{e.m.}$ & $0.0005142\,(98)$ & $0.002250\,(13)$\\
    \hline\hline
    $ (\delta \Acal_{\pi^+}/\Acal_\pi)^\mathrm{f}$ & $5.388\,(49)\times 10^{-2}$ & $1.1743\,(64)\times 10^{-1}$\\
    \hline
    $ (\delta \Acal_{K^+}/\Acal_K)^\mathrm{f}$ & $2.218\,(48)\times 10^{-2}$ & $5.265\,(57)\times 10^{-2}$ \\
    \hline\hline
    $ (\delta \Acal_{\pi^+}/\Acal_\pi)^\mathrm{nf}$ & $5.374\,(59)\times 10^{-2}$  & $5.287\,(43)\times 10^{-2}$\\
    \hline
    $ (\delta \Acal_{K^+}/\Acal_K)^\mathrm{nf}$ & $4.493\,(41)\times 10^{-2}$ & $4.494\,(48)\times 10^{-2}$ \\
    \hline
    \end{tabular}
    \caption{Comparison of photon corrections to meson masses and to the decay amplitude computed using local and conserved current. }
    \label{tab:local-vs-conserved_QED_derivatives}
\end{table}
\section{zM\"obius to M\"obius correction}
\label{app:moebius}

The zM\"obius DWF action~\cite{Mcglynn:2015uwh} is an approximation of the M\"obius DWF action~\cite{Brower:2012vk} and is used in this work due to faster numerical convergence. The real parameters of the M\"obius DWF action are matched to complex ones in the zM\"obius DWF action, using the Remez algorithm, leading to a reduced $L_\mathrm{s}$ dimension. On the ensemble used in this work, an $L_\mathrm{s}/a=24$ is used for M\"obius and $L_\mathrm{s}/a=10$ for zM\"obius. 

A further drastic improvement in the iterations needed for a light-quark inversion is achieved via deflation: we compute the lowest $N_\mathrm{vec}=2000$ eigenvectors of the Dirac operator to obtain a starting guess, reducing computational cost of light-quark inversions substantially. 

For the light-quark inversions using the M\"obius action, for which we do not have eigenvectors available on disk, we employ the M\"obius accelerated DWF (MADWF) algorithm~\cite{Yin:2011np}. This algorithm constructs a guess for the final solve by transforming the 5D Domain-Wall Dirac operator $D^\mathrm{5D}_\mathrm{DW}$ via Pauli-Villars solves into a 4D approximation of the overlap operator $D^\mathrm{4D}_\mathrm{ov}$. The solution of the $D^\mathrm{4D}_\mathrm{ov}$ inversion is then used to reconstruct an approximated solution for $D^\mathrm{5D}_\mathrm{DW}$. Using this solution as a guess for the final solve on $D^\mathrm{5D}_\mathrm{DW}$ leads to an overall reduction in computational cost. One key insight used in this work is that the Domain-Wall Dirac operator $D^\mathrm{5D}_\mathrm{DW}$ does not have to be the same in the first and last step of this algorithm. For our light-quark solves with the M\"obius action, we therefore produced the guess of the MADWF algorithm using a zM\"obius Dirac operator, allowing us to benefit from deflation. We found that the zM\"obius MADWF guess was able to significantly speed up the final M\"obius solve. Compared to an undeflated light-quark M\"obius solve, the deflated zM\"obius solve has an iteration count reduced by a factor 20 and the MADWF M\"obius solve with a deflated zM\"obius MADWF guess is faster by a factor 10.

To correct for the bias introduced by the zM\"obius approximation, we perform an all-mode averaging (AMA) \cite{Blum:2012uh} correction step. Within AMA, for each observable $O$ we compute the estimator $\langle \tilde{O} \rangle_\mathrm{M}$ using the M\"obius action from two source times ($t_\text{src}=0,T/2$). On the same source times, we compute the cheaper estimator $\langle \tilde{O} \rangle_\mathrm{zM}$ using the zM\"obius action. Finally, we compute another zM\"obius estimator $\langle O \rangle_\mathrm{zM}$ from all  $96$ source times available on the ensemble used in this work. The final bias-corrected estimator is then given by
\begin{equation}
    \langle O \rangle = \langle O \rangle_\mathrm{zM} + \langle \tilde{O} \rangle_\mathrm{M} - \langle \tilde{O} \rangle_\mathrm{zM}\,.
\end{equation}

\begin{figure}[H]
  \centering
  \begin{minipage}{0.5\textwidth}
      \centering
      \includegraphics[width=1\textwidth]{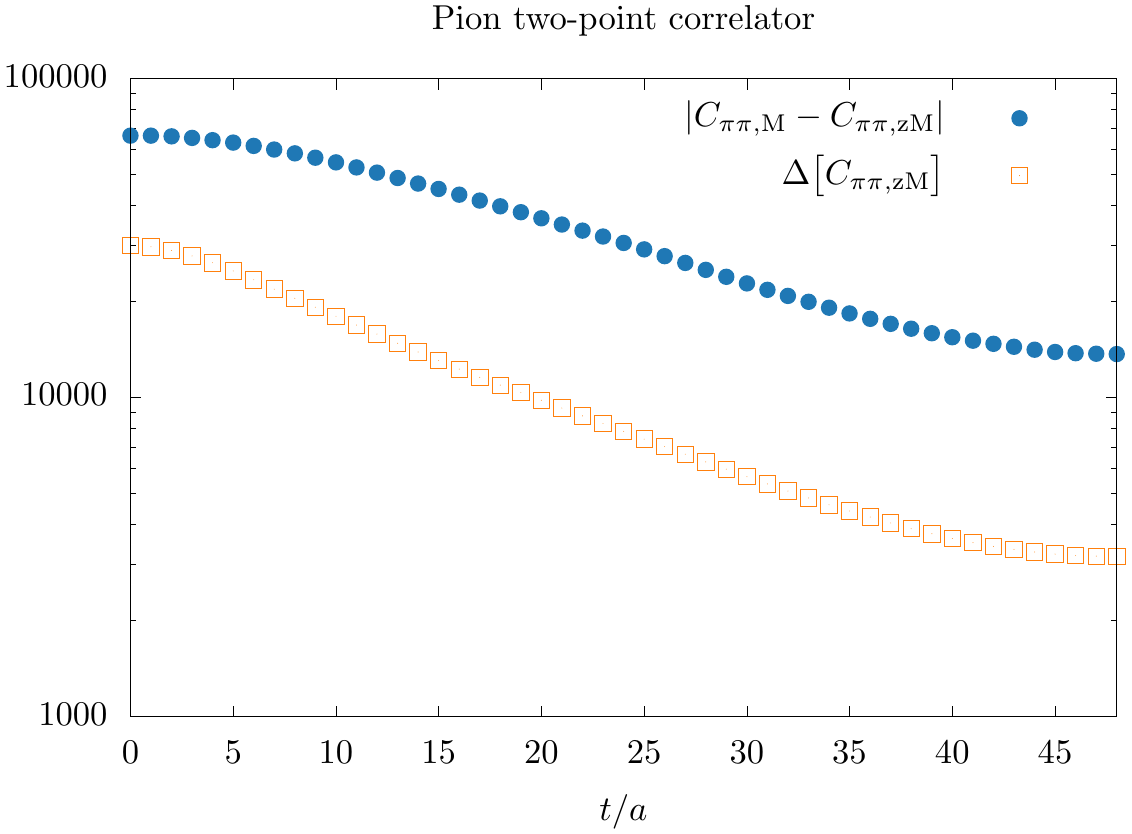}
  \end{minipage}\hfill
  \begin{minipage}{0.5\textwidth}
      \centering
      \includegraphics[width=1\textwidth]{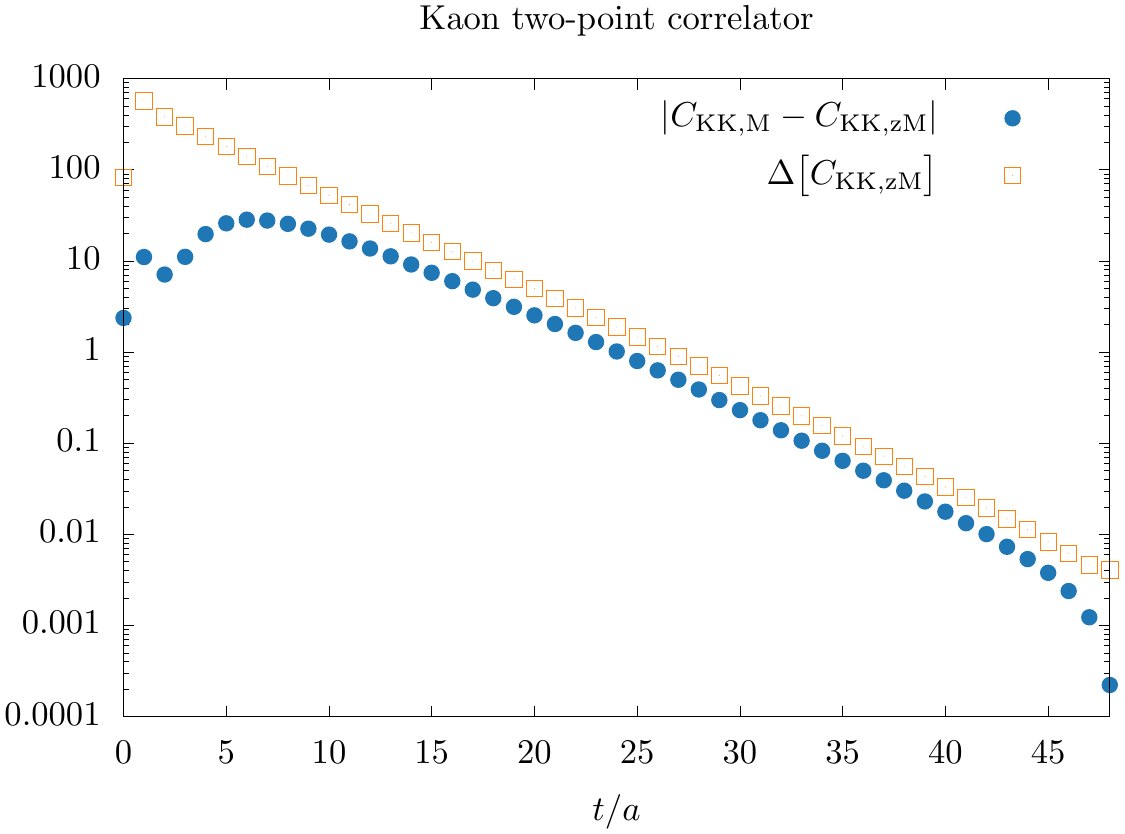}
  \end{minipage}\\
  \begin{minipage}{0.5\textwidth}
    \centering
    \includegraphics[width=1\textwidth]{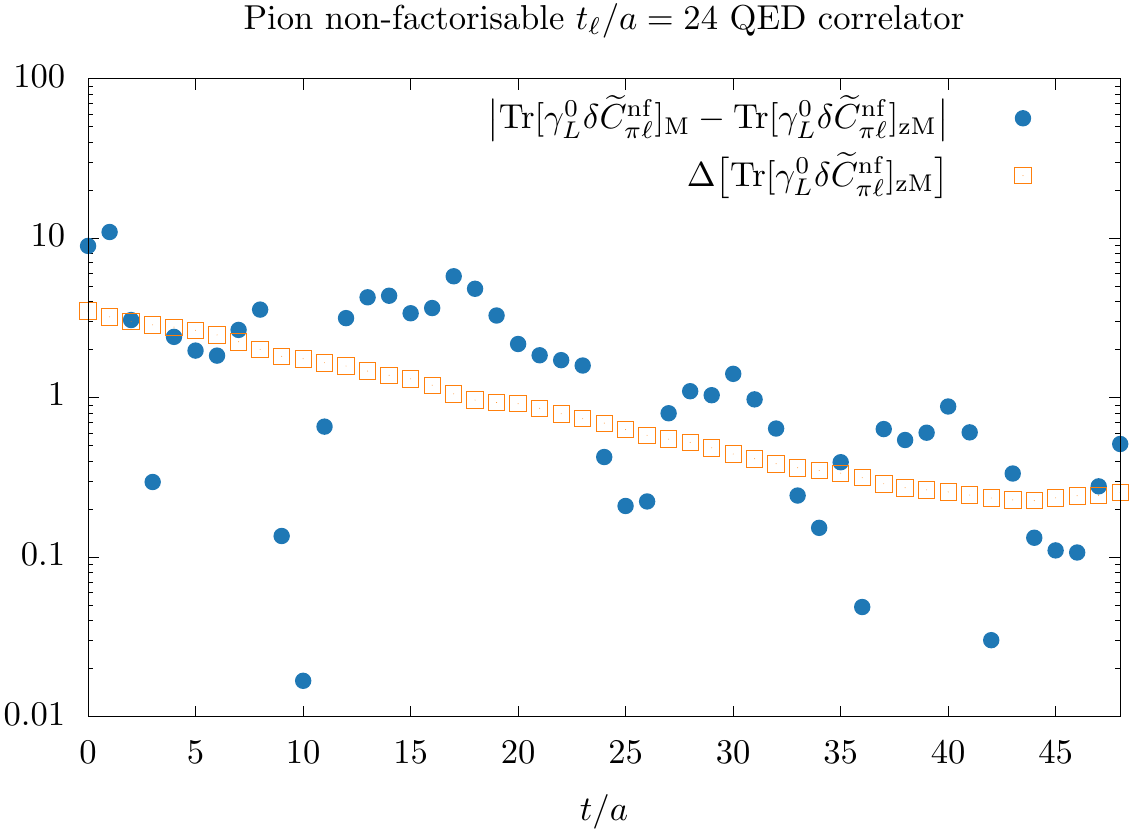}
  \end{minipage}\hfill
  \begin{minipage}{0.5\textwidth}
    \centering
    \includegraphics[width=1\textwidth]{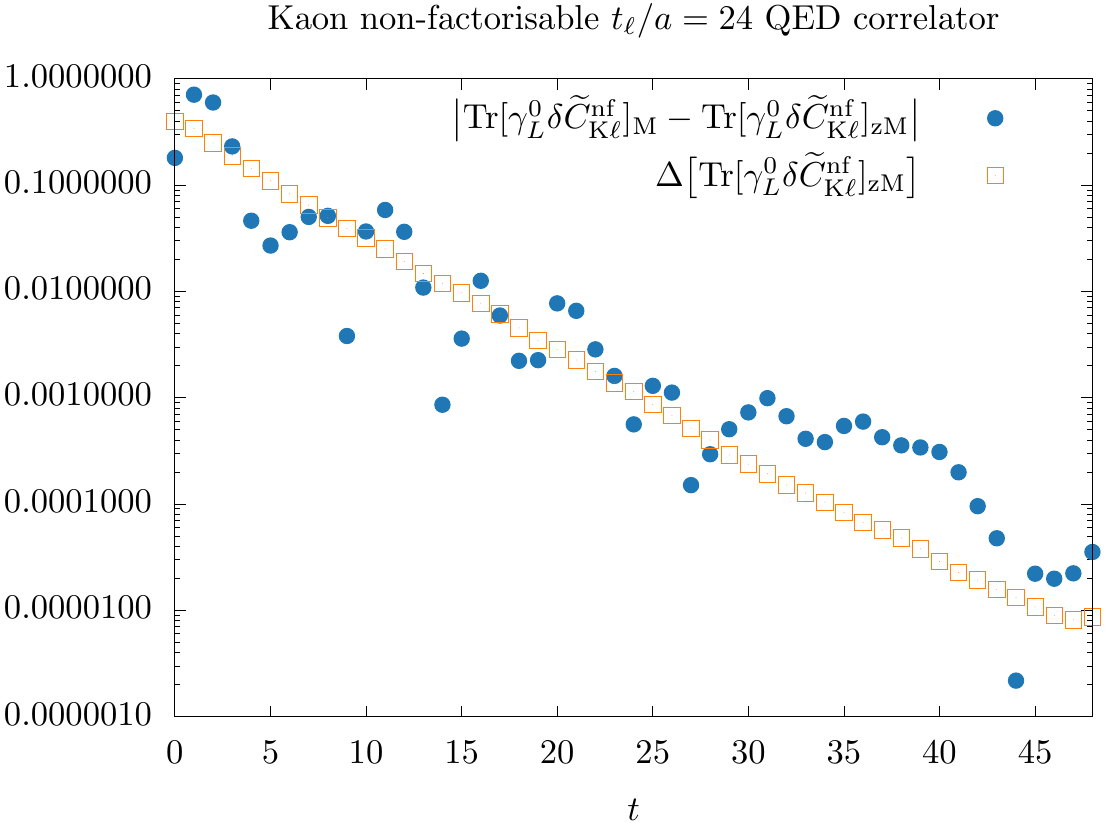} 
  \end{minipage}\hfill
  \caption{Comparison of the magnitude of the zM\"obius-to-M\"obius bias correction (blue circles) to the statistical error of the zM\"obius estimators (orange squares) for the $\pi$ and $K$ meson. All plots show the correlations functions, in lattice units, computed using point sources and wall sinks. In the pion two-point correlation function (top left) the correction is larger than the noise, while for the kaon two-point function (top right) the correction is smaller than our statistical precision. In the non-factorisable correlation functions (bottom plots) signal and correction are of compatible magnitude.}
  \label{fig:mobius-correction-noise}
\end{figure}

A comparison of the magnitude of bias correction $\langle \tilde{O} \rangle_\mathrm{M} - \langle \tilde{O} \rangle_\mathrm{zM}$ to the statistical error of the estimator $\langle O \rangle_\mathrm{zM}$ is shown in~\cref{fig:mobius-correction-noise}. We find that the correction is negligible on most observables, with the exception of the non-factorisable correlation functions and the pion two-point correlation function. 

\section{Free domain wall fermion propagators}
\label{app:freeDWF}

In this appendix we discuss the free domain wall fermion propagators, which have been used for the implementation of the lepton in the non-factorizable correlation functions. Throughout this section all quantities are expressed in lattice units.

In this work we use the following convention for the five-dimensional Shamir-Domain-Wall-Fermion-Dirac operator~\cite{Boyle:2016lbp,Boyle:2022nef}
\begin{align}
 D_{s,t}(x,y) &= -\delta_{s,t} \frac{1}{2} \sum_\mu 
\left[(1-\gamma_\mu)U_\mu\delta_{y,x+\mu} + 
(1+\gamma_\mu)U^\dagger_\mu\delta_{y,x-\mu}\right] - 
\delta_{s,t}(M_5-1-4)\delta_{x,y} \nn\\
&\quad  -\delta_{t,s+1}P_-\delta_{x,y}-\delta_{t,s-1}P_+\delta_{x,y} + 
m\delta_{x,y}\delta_{s,L_\mathrm{s}}\delta_{t,1}P_- + 
m\delta_{x,y}\delta_{s,1}\delta_{t,L_\mathrm{s}}P_+
\label{eq:DWFoperator}
\end{align}
where $s,t\in\{1,L_\mathrm{s}\}$ label the slices in the fifth dimension and $M_5$ is the Domain Wall height. The Domain Wall Fermion action is given by
\begin{equation}
 S[\Psi,\overline{\Psi},U]=-\sum_{x,y}\sum\limits_{s,t=1}^{L_\mathrm{s}} \overline{\Psi}_s(x)\, D_{s,t}(x,y)\,\Psi_t(y)\,.
\end{equation}
The physical quark fields are given by
\begin{equation}
 q(x)=P_-\Psi_1(x)+P_+\Psi_{L_\mathrm{s}}(x)\qquad\qquad\text{and}\qquad
  \overline{q}(x)=\overline{\Psi}_1(x)P_++\overline{\Psi}_{L_\mathrm{s}}(x)P_-\,,
\end{equation}
with $P_{\pm}=(1\pm\gamma_5)/2$.
\subsection{Feynman rules for free propagator}
\label{app:freeDWFfeynmanrules}
A derivation of the free Domain-Wall-Fermion propagator in momentum space can be found in ref.~\cite{Aoki:1997xg}. However, the conventions used for the five dimensional Dirac operator in ref.~\cite{Aoki:1997xg} differ from the ones given in~\cref{eq:DWFoperator} and thus, in the following, we give results for the convention used in our work. These can be obtained by following the same steps as the derivation in ref.~\cite{Aoki:1997xg}. \par
The free momentum-space action is given by
\begin{align}
 \tilde{D}_{st}(p) = \ii\sum_\mu\gamma_\mu\sin p_\mu \delta_{st} &+ 
\big(W(p)\delta_{s,t}-\delta_{s-1,t}+m\delta_{s,1}\delta_{t,L_\mathrm{s}}\big)P_+ \nn\\
&+\big(W(p)\delta_{s,t}-\delta_{s+1,t}+m\delta_{L_\mathrm{s},1}\delta_{t,1}\big)P_-\,,
\label{eq:DWFoperatormom}
\end{align}
with 
\begin{equation}
 W(p)  = 1-M_5+2\sum_\mu\sin^2\frac{p_\mu}{2}\,.
\end{equation}
The inverse of the operator in~\cref{eq:DWFoperatormom} (i.e.\ the propagator) can be written as
\begin{equation}
\resizebox{.9\hsize}{!}{$
    S_{s,t}(p) = -\Big(\ii\sum_\mu\gamma_\mu\sin p_\mu \delta_{su} + 
    (W_{m}^+)_{su}\Big)\,G^R_{u,t}\,P_+ -\Big(\ii\sum_\mu\gamma_\mu\sin p_\mu \delta_{su} + 
    (W_{m}^-)_{su}\Big)\,G^L_{u,t}\,P_-\,,
    $}
\label{eq:DWFoperatormominv}
\end{equation}
where we use the notation
\begin{align}
 (W_m^+)_{st} &= -W(p)\delta_{st}+\delta_{s+1,t} - m\,\delta_{s,L_\mathrm{s}}\delta_{t,1}\,,\\
 (W_m^-)_{st} &= -W(p)\delta_{st}+\delta_{s-1,t} - m\,\delta_{s,1}\delta_{t,L_\mathrm{s}}\,,
\end{align}
and define
\begin{equation}
 G^R_{s,t} = \left(\sum_\mu\sin^2p_\mu + W^-_{m} 
W^+_{m}\right)^{-1}_{s,t}\,,
\quad
  G^L_{s,t} = \left(\sum_\mu\sin^2p_\mu + W^+_{m} 
W^-_{m}\right)^{-1}_{s,t}\,.
 \label{eq:GRGL}
\end{equation}
Following the steps in ref.~\cite{Aoki:1997xg} for the conventions used in this work, the inverses in~\cref{eq:GRGL} can be calculated and are given by
\begin{align}
 G^R_{s,t} = G(s,t) + A_{++} \e^{\alpha (s+t)} + A_{+-} \e^{\alpha (s-t)} + 
 A_{-+} \e^{\alpha (-s+t)} + A_{--} \e^{\alpha (-s-t)}\,,\\
 G^L_{s,t} = G(s,t) + B_{++} \e^{\alpha (s+t)} + B_{+-} \e^{\alpha (s-t)} + 
 B_{-+} \e^{\alpha (-s+t)} + B_{--} \e^{\alpha (-s-t)}\,,
\end{align}
where
\begin{equation}
    G(s,t) = A \left(\e^{\alpha(L_\mathrm{s}-|s-t|)}+\e^{-\alpha(L_\mathrm{s}-|s-t|)}\right)\,,
\end{equation}
with 
 \begin{equation}
 A = \frac{1}{2|W|\sinh{\alpha}}\,\cdot\,\frac{1}{2\sinh{(\alpha 
 L_\mathrm{s})}}\,,
\label{eq:FreePropA}
 \end{equation}
and $\alpha$ can be defined via
\begin{equation}
 \cosh{\alpha} = \frac{1+W^2+\sum_\mu\sin^2p_\mu}{2|W|}\,\,.\label{eq:cosha}
 \end{equation}
The coefficients $A_{\pm\pm}$ and $B_{\pm\pm}$ are determined such that the 
boundary conditions ($s=1,L_\mathrm{s}$) in $D_{s,t}S_{t,u}=\delta_{s,u}$ are 
fulfilled
\begin{equation}
\begin{aligned}
& A_{++}  = \frac{A}{F} \left(\e^{-2\alpha 
L_\mathrm{s}}-1\right)\e^{-\alpha}\left(\e^{-\alpha} - |W|\right)(1-m^2)\,,\\
& A_{--}  = \frac{A}{F} \left(1-\e^{2\alpha 
L_\mathrm{s}}\right)\e^{\alpha}\left(\e^{\alpha} -|W|\right)(1-m^2)\,,\\
& B_{++}  = \frac{A}{F} \left(\e^{-2\alpha 
L_\mathrm{s}}-1\right)\left(1 - \e^{-\alpha}|W|\right)(1-m^2)\,,\\
 &B_{--}  = \frac{A}{F} \left(1-\e^{2\alpha 
 L_\mathrm{s}}\right)\left(1 -\e^{\alpha}|W|\right)(1-m^2)\,,\\
&A_{+-} = A_{-+} = B_{+-} = B_{-+} = \frac{A}{F}\, 
2|W|\sinh(\alpha)\left(1+2m\cosh(\alpha L_\mathrm{s})+m^2\right)\,,
\end{aligned}
\end{equation}
with
\begin{align}
 F &= \e^{\alpha L_\mathrm{s}}\Big[1-|W|\e^\alpha+m^2(|W|\e^{-\alpha}-1)\Big] + \e^{-\alpha 
L_\mathrm{s}}\Big[|W|\e^{-\alpha}-1+m^2(1-|W|\e^\alpha)\Big]\nn \\
&\qquad - 4|W|m\sinh(\alpha) \,.
\label{eq:FreePropF}
\end{align}
\textbf{NB:} The five dimensional free propagator in~\cref{eq:DWFoperatormominv} can be projected to four dimensions by
\begin{equation}
 S^\mathrm{4D}(p) = P_- S_{1,1}(p) P_+ + P_+ S_{L_\mathrm{s}, L_\mathrm{s}}(p) P_- + P_- S_{1,L_\mathrm{s}}(p)
P_- + P_+ S_{L_\mathrm{s},1}(p) P_+\,.
\end{equation}
It can be shown that in the infinite $L_\mathrm{s}$ limit, the four dimensional propagator is given by
\begin{equation}
 S^\mathrm{4D} \longrightarrow \frac{-\ii\sum_\mu\gamma_\mu\sin 
p_\mu + m(1-W\e^{-\alpha}) 
}{-(1-|W|\e^\alpha)-m^2(|W|\e^{-\alpha}-1)}\hspace{1cm}\text{for}\qquad 
L_\mathrm{s}\rightarrow\infty 
\end{equation}
in agreement with the expression given in ref.~\cite{Capitani:2002mp}.

\subsection{Pole mass of the free propagator}
\label{app:freeDWFpolemass}
\begin{figure}[H]
 \centering
 \includegraphics[width=0.49\textwidth]{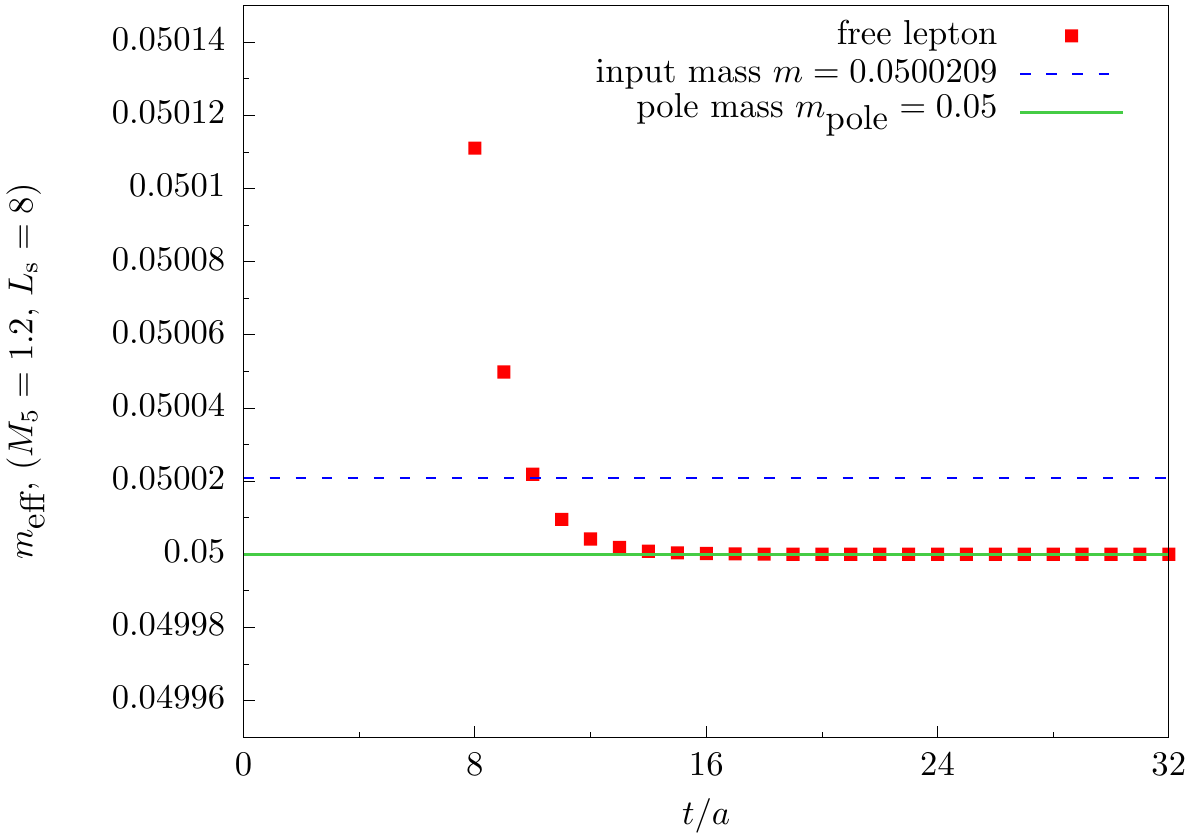}
  \includegraphics[width=0.49\textwidth]{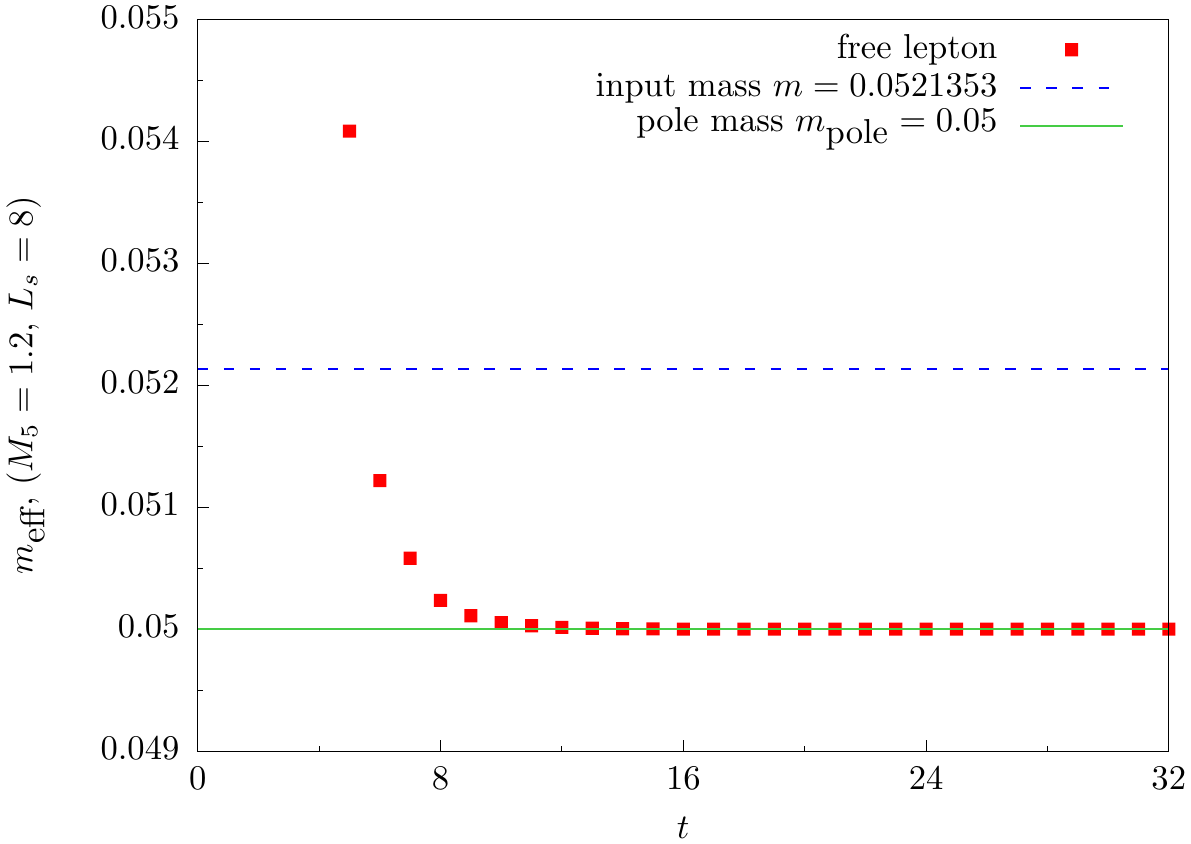}
 \caption{The effective mass of a free propagator calculated from Feynman rules on a $24^3\times64$ lattice for $aM_5=1.0$ (left) and $aM_5=1.2$ (right).}
\label{fig:effmassfreeprop}
\end{figure}

For the calculation of the QED correction from the factorisable diagram, we want to fix the free lepton propagator to the physical muon mass as its pole mass. In the following we describe how to determine the correct input-mass parameter $m$ for the free propagator to reproduce a desired pole mass. The 4D propagator 
can be written with a common denominator $A/F$ (see~\cref{eq:FreePropF,eq:FreePropA})
\begin{equation}
\begin{aligned}
 \left(\frac{A}{F}\right)^{-1} =
 & \,\,2|W|\sinh(\alpha)2\sinh(\alpha L_\mathrm{s})\,\,\Big\{\e^{\alpha 
L_\mathrm{s}}\left[1-|W|\e^\alpha+m^2(|W|\e^{-\alpha}-1)\right] 
\\
&-4|W|m\sinh(\alpha)+ \e^{-\alpha 
L_\mathrm{s}}\left[|W|\e^{-\alpha}-1+m^2(1-|W|\e^\alpha)\right]
\Big\} \,.
\end{aligned}
\end{equation}
We now have to find $m_\text{pole}$ where 
$(A/F)^{-1}|_{p^2=-m^2_\text{pole}}=0$.
$(A/F)^{-1}$ has some trivial zeros, where $\sinh(\alpha)=0$, which we are not 
interested in. We are interested in the case $F=0$, i.e.
\begin{equation}
F\big|_{p^2=-m^2_\text{pole}}\equiv F_\text{pole}
=0\,.
\label{eq:Fequals0}
\end{equation}
In practice, we want to choose a desired pole mass 
$m^2_\text{pole}$ (e.g.\ the muon mass) and determine the input mass $m$ that corresponds to this pole mass, i.e.\ we have to solve with respect to $m$: 
\begin{equation}
\begin{aligned}
F_\text{pole}&=\e^{\alpha_\text{pole} 
L_\mathrm{s}}\Big[1-|W_\text{pole}|\e^{\alpha_\text{pole}}+m^2\big(|W_\text{pole}|\e^{
-\alpha_\text{pole}} -1\big)\Big] 
- 4|W_\text{pole}|m\sinh(\alpha_\text{pole}) 
\\ &\quad + \e^{-\alpha_\text{pole} 
L_\mathrm{s}}\Big[|W_\text{pole}|\e^{-\alpha_\text{pole}}-1+m^2\big(1-|W_\text{pole}|
\e^{\alpha_\text{pole}}\big)\Big]=0\,,
\end{aligned}
\end{equation}
with
\begin{equation}
 W_\text{pole} = 1-M_5+2\sum_\mu\sin^2\left(\frac{p^\mu_\text{pole}}{2} 
\right)\,,\ \ 
\cosh \alpha_\text{pole}= 
\frac{1+W_\text{pole}^2+\sum_\mu\sin^2\left(p^\mu_\text{pole}\right)}{2|
W_\text{pole} | }
 \,.
\end{equation}
This is a simple quadratic equation in $m$ and the solutions are easily obtained from
\begin{equation}
    m = -\frac{p}{2} \pm \sqrt{\left(\frac{p}{2}\right)^2-q}\,,
\label{eq:FreeProppolemass1}
\end{equation}
with
\begin{align}
 p&=\frac{- 
4|W_\text{pole}|\sinh(\alpha_\text{pole})}{\e^{\alpha_\text{pole} 
L_\mathrm{s}}(|W_\text{pole}|\e^{-\alpha_\text{pole}} -1)+\e^{-\alpha_\text{pole} 
L_\mathrm{s}}(1-|W_\text{pole}|\e^{\alpha_\text{pole}} )}\label{eq:FreeProppolemass2}\,,\\
q&=\frac{\e^{\alpha_\text{pole} 
L_\mathrm{s}}(1-|W_\text{pole}|\e^{\alpha_\text{pole}} 
)+\e^{-\alpha_\text{pole} L_\mathrm{s}}(|W_\text{pole}\e^{-\alpha_\text{pole}} 
-1)}{\e^{\alpha_\text{pole} L_\mathrm{s}}(|W_\text{pole}|\e^{-\alpha_\text{pole}} 
-1)+\e^{-\alpha_\text{pole} L_\mathrm{s}}(1-|W_\text{pole}|\e^{\alpha_\text{pole}} )}\,.
\label{eq:FreeProppolemass3}
\end{align}
For large $L_\mathrm{s}$ one finds
\begin{equation}
 m \longrightarrow
\pm 
\sqrt{-\frac{1-|W_\text{pole}|\e^{\alpha_\text{pole}}}{|W_\text{pole}|\e^{ 
-\alpha_\text{pole}}-1}}  \hspace{1cm}\text{for}\qquad L_\mathrm{s}\rightarrow \infty\,.
\end{equation}
In~\cref{fig:effmassfreeprop} we show the effective mass of a free propagator calculated using our implementation in Grid from Feynman rules (see~\cref{app:freeDWFfeynmanrules}) on a $24^3\times64$ lattice. The plot on the left corresponds to a Domain Wall height of $aM_5=1.0$, while the plot on the right to $aM_5=1.2$, both with length $L_\mathrm{s}/a=8$ in the fifth dimension. Red points show the numerical results for the effective mass, the solid green line shows the target pole mass of $am_\text{pole}=0.05$, while the dashed blue line is the required input mass determined according to~\cref{eq:FreeProppolemass1,eq:FreeProppolemass2,eq:FreeProppolemass3}.  For large-enough times $t$ the effective mass of the free propagator plateaus at the desired target value of the pole mass $am_\text{pole}=0.05$. The deviation from the plateau at small $t$ is due to unphysical poles in the free Domain Wall Fermion propagator (see, e.g., the discussion in ref.~\cite{Tomii:2017lyo}).

\subsection{Projectors on definite spinor structure}
\label{app:projectors}

Euclidean free Dirac spinors satisfy the following on-shell Dirac equations
\begin{equation}
    D(\ii \,\omega, \pvec) \, u(\pvec) = 0~, \qquad D(-\ii \,\omega, -\pvec) \, v(\pvec) = 0~,
\end{equation}
where $D(p)$ is the Dirac operator in momentum space and $\omega$ is the energy satisfying the dispersion relation $D(\ii \,\omega,\pvec)D(-\ii \,\omega,-\pvec)=0$.
The spinors also respect the following completeness relations
\begin{equation}
        \sum_r u^r(\pvec)\bar{u}^r(\pvec) = D(-\ii \,\omega,-\pvec)~,\qquad
        \sum_r v^r(\pvec)\bar{v}^r(\pvec) = -D(\ii \,\omega,\pvec)~,
\end{equation}
and orthogonality relations
\begin{equation}
    \begin{split}
        & \bar{v}^r(-\pvec) \, \Gamma_0 \, u^s(\pvec) = 0~,\\
        & \bar{u}^r(-\pvec) \, \Gamma_0 \, v^s(\pvec) = 0~,
    \end{split}\qquad
    \begin{split}
        & \bar{u}^r(\pvec) \, \Gamma_0 \, u^s(\pvec) = 2 \mathcal{E} \, \delta^{rs}~, \\
        & \bar{v}^r(\pvec) \, \Gamma_0 \, v^s(\pvec) = 2 \mathcal{E} \, \delta^{rs}~,
    \end{split}\qquad
    \begin{split}
        & \bar{u}^r(\pvec) \, u^s(\pvec) = 2 \mathcal{M} \, \delta^{rs}~, \\
        & \bar{v}^r(\pvec) \, v^s(\pvec) = -2 \mathcal{M} \, \delta^{rs}~,
    \end{split}\qquad
\end{equation}
where $\mathcal{E}$ and $\mathcal{M}$ are quantities that in the continuum limit reduce to $\lim\limits_{a\to 0}\mathcal{E}=\sqrt{m^2+|\pvec|^2}$ and ${\lim\limits_{a\to 0}\mathcal{M}=m}$, respectively.

As discussed in~\cref{sec:latt_correlators}, the external anti-lepton propagator projected on  momentum $\pvec_\ell$, when evaluated on the lattice with finite time $T$ and anti-periodic boundary conditions takes the following form (neglecting possible contact terms)
\begin{align}
    S_\ell(0|t_\ell,\pvec_\ell) 
    &=\left\{\e^{-\omega_\ell t_\ell }\, \frac{D(\ii \,\omega_\ell,\pvec_\ell)}{2\Omega_\ell}-\e^{-\omega_\ell(T-t_\ell)}\, \frac{D(-\ii \,\omega_\ell,\pvec_\ell)}{2\Omega_\ell}\right\}\times \frac{1}{1+\e^{-\omega_\ell T}}\\
    &=  -\sum_r\left\{\e^{-\omega_\ell t_\ell }\, \frac{ v_r(\pvec)\bar{v}_r(\pvec)}{2 \Omega_\ell}+\,\e^{-\omega_\ell(T-t_\ell)}\, \frac{ u_r(-\pvec)\bar{u}_r(-\pvec)}{2\Omega_\ell}\right\}\times \frac{1}{1+\e^{-\omega_\ell T}}\,,\nn
\end{align}
where we observe that the backward signal has a different Dirac structure compared to the forward one. Here $\Omega_\ell$ is a quantity that in the continuum limit gives $\lim\limits_{a\to 0} \Omega_\ell = \sqrt{m_\ell^2+|\pvec_\ell|^2}$~.  

By using the orthogonality relations above and the fact that
\begin{equation}
    D(-\ii \,\omega,\pvec)-D(\ii \,\omega,\pvec) = 2\mathcal{E} \, \Gamma_0 \,,
\end{equation}
we can define two projectors
\begin{equation}
\begin{split}
    \mathcal{P}_{v(\pvec)} &= \left\{D(-\ii E,\pvec)-D(\ii E,\pvec)\right\}^{-1}[- D(\ii E,\pvec)] \\
    &=\left\{
    u_t(-\pvec)\bar{u}_t(-\pvec) +  v_s(\pvec)\bar{v}_s(\pvec)
\right\}^{-1} \left[ v_r(\pvec)\bar{v}_r(\pvec)\right] \,,\\[8pt]
    \mathcal{P}_{u(-\pvec)} &= \left\{D(-\ii E,\pvec)-D(\ii E,\pvec)\right\}^{-1} D(-\ii E,\pvec) \\
    &=\left\{
    u_t(-\pvec)\bar{u}_t(-\pvec) +  v_s(\pvec)\bar{v}_s(\pvec)
\right\}^{-1} \left[u_r(-\pvec)\bar{u}_r(-\pvec)\right]\,,
\end{split}
\label{eq:proj_P}
\end{equation}
such that for the lepton propagator we have
\begin{equation}
\begin{split}
    S_\ell(0|t_\ell,\pvec_\ell) \cdot \mathcal{P}_{v(\pvec_\ell)} &= -\sum_r \,
    \left\{ \e^{-\omega_\ell t_\ell}\, \frac{ v_r(\pvec_\ell)\bar{v}_r(\pvec_\ell)}{2\Omega_\ell}\right\}\times \frac{1}{1+\e^{-\omega_\ell T}}\,,\\
    S_\ell(0|t_\ell,\pvec_\ell) \cdot \mathcal{P}_{u(-\pvec_\ell)} &=-\sum_r\,
    \left\{\e^{-\omega_\ell (T-t_\ell)}\, \frac{ u_r(-\pvec_\ell)\bar{u}_r(-\pvec_\ell)}{2\Omega_\ell}\right\}\times \frac{1}{1+\e^{-\omega_\ell T}}\,.
\end{split}
\end{equation}

In order to construct the projectors $\mathcal{P}_{v(\pvec_\ell)}$ and $\mathcal{P}_{u(-\pvec_\ell)}$ we then compute on the lattice the free domain-wall lepton propagator $S_\ell(t_\ell,-\pvec_\ell|0)$, projected on definite momentum $-\pvec_\ell$, having the following temporal behaviour
\begin{equation}
    S_\ell(t_\ell,-\pvec_\ell|0) = \sum_r\left\{\e^{-\omega_\ell t_\ell }\, \frac{ u_r(-\pvec)\bar{u}_r(-\pvec)}{2 \Omega_\ell}+\,\e^{-\omega_\ell(T-t_\ell)}\, \frac{ v_r(\pvec)\bar{v}_r(\pvec)}{2\Omega_\ell}\right\}\times \frac{1}{1+\e^{-\omega_\ell T}}.
\end{equation}
Since the definitions of the projectors in~\cref{eq:proj_P} do not depend on the spinor normalization $2\Omega_\ell(1+\e^{-\omega_\ell T})$, because it cancels out in the matrix multiplications, they can be easily obtained from the free lattice lepton propagator just by extracting and combining the coefficients of the forward and backward exponentials. 

\begin{figure}[H]
    \centering
    \begin{subfigure}[b]{0.48\textwidth}
        \includegraphics[width=\textwidth]{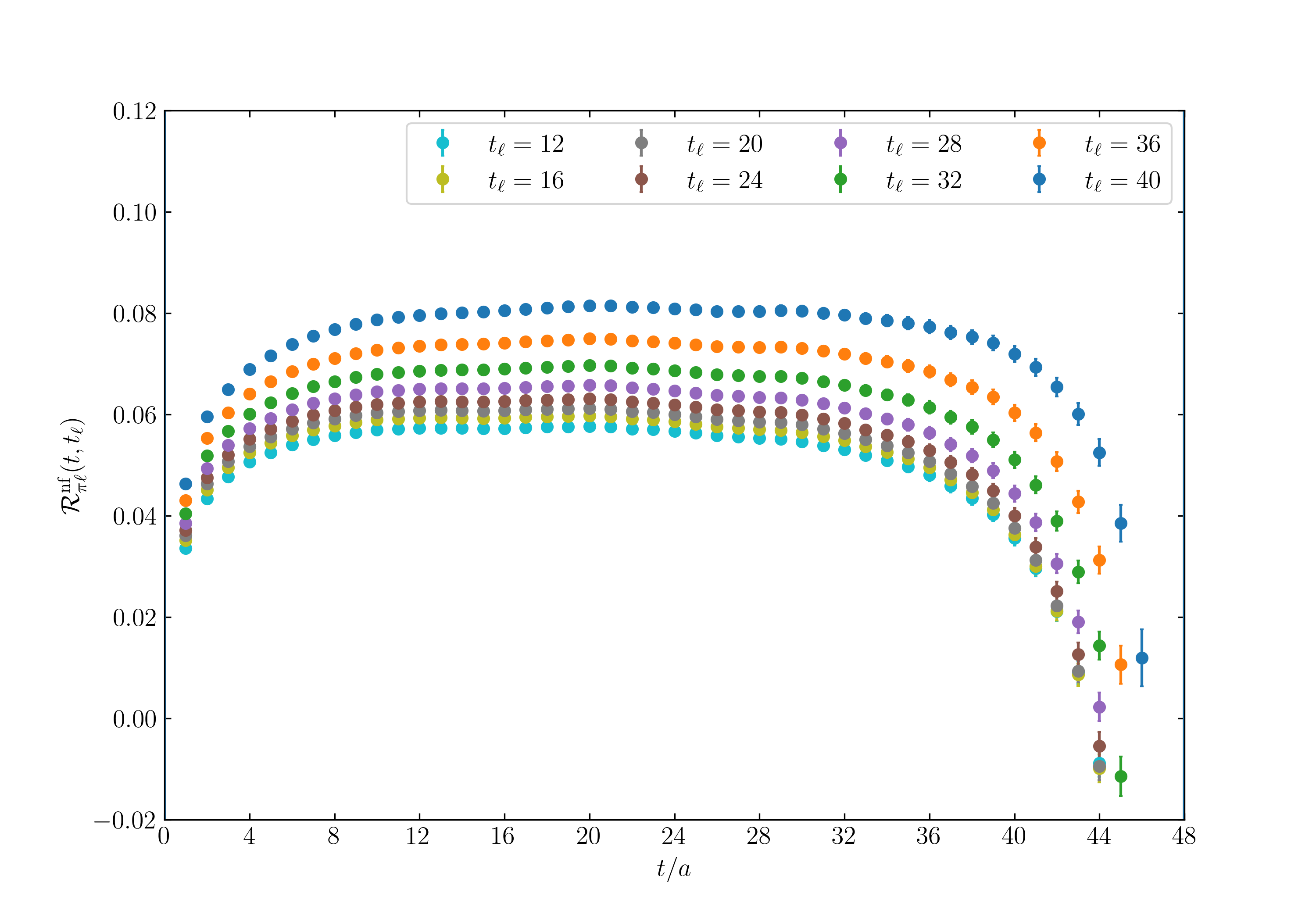} 
        \caption{without projection}
    \end{subfigure}
    \begin{subfigure}[b]{0.48\textwidth}
        \includegraphics[width=\textwidth]{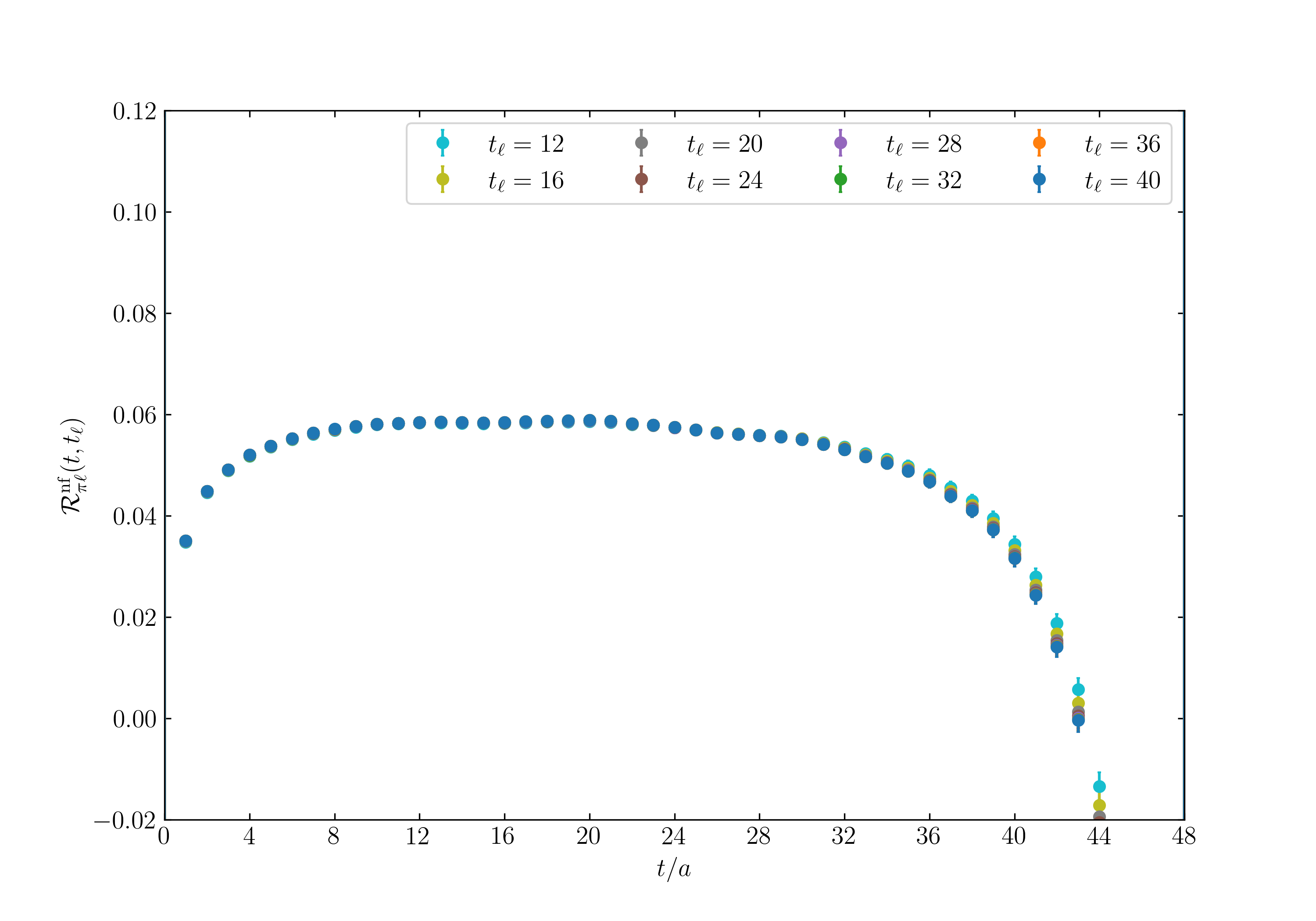} 
        \caption{with projection}
    \end{subfigure}
 \caption{Comparison of the pion non-factorizable correlator $\mathcal{R}_{\pi\ell}^\mathrm{nf}$ obtained without the use of the projector $\mathcal{P}_{v(\pvec_\ell)}$ (a) and with the backward-propagating signal removed (b). }
 \label{fig:proj_nonfact_pion}
\end{figure}

\begin{figure}[t]
    \centering
    \begin{subfigure}[b]{0.48\textwidth}
        \includegraphics[width=\textwidth]{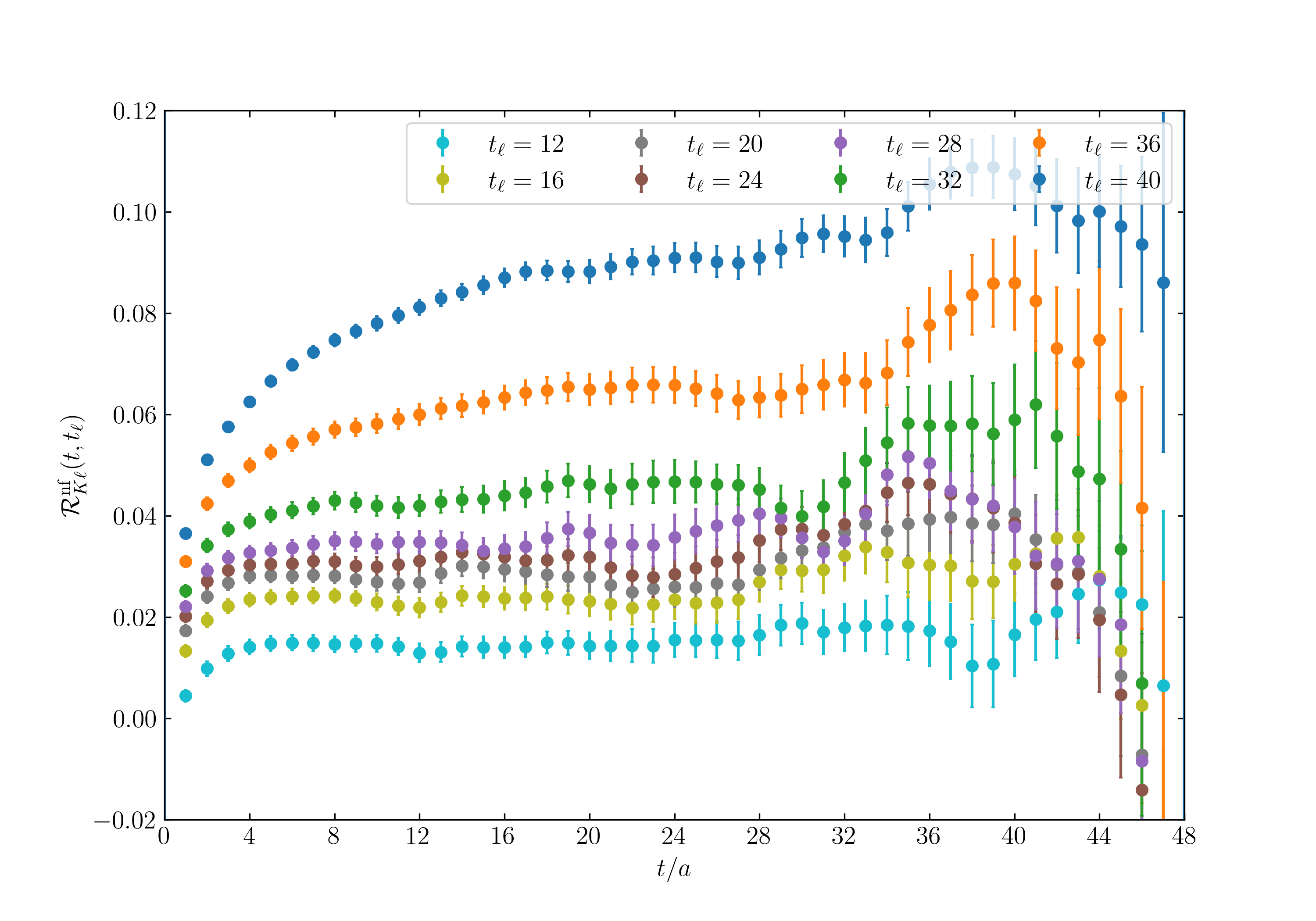} 
        \caption{without projection}
    \end{subfigure}
    \begin{subfigure}[b]{0.48\textwidth}
        \includegraphics[width=\textwidth]{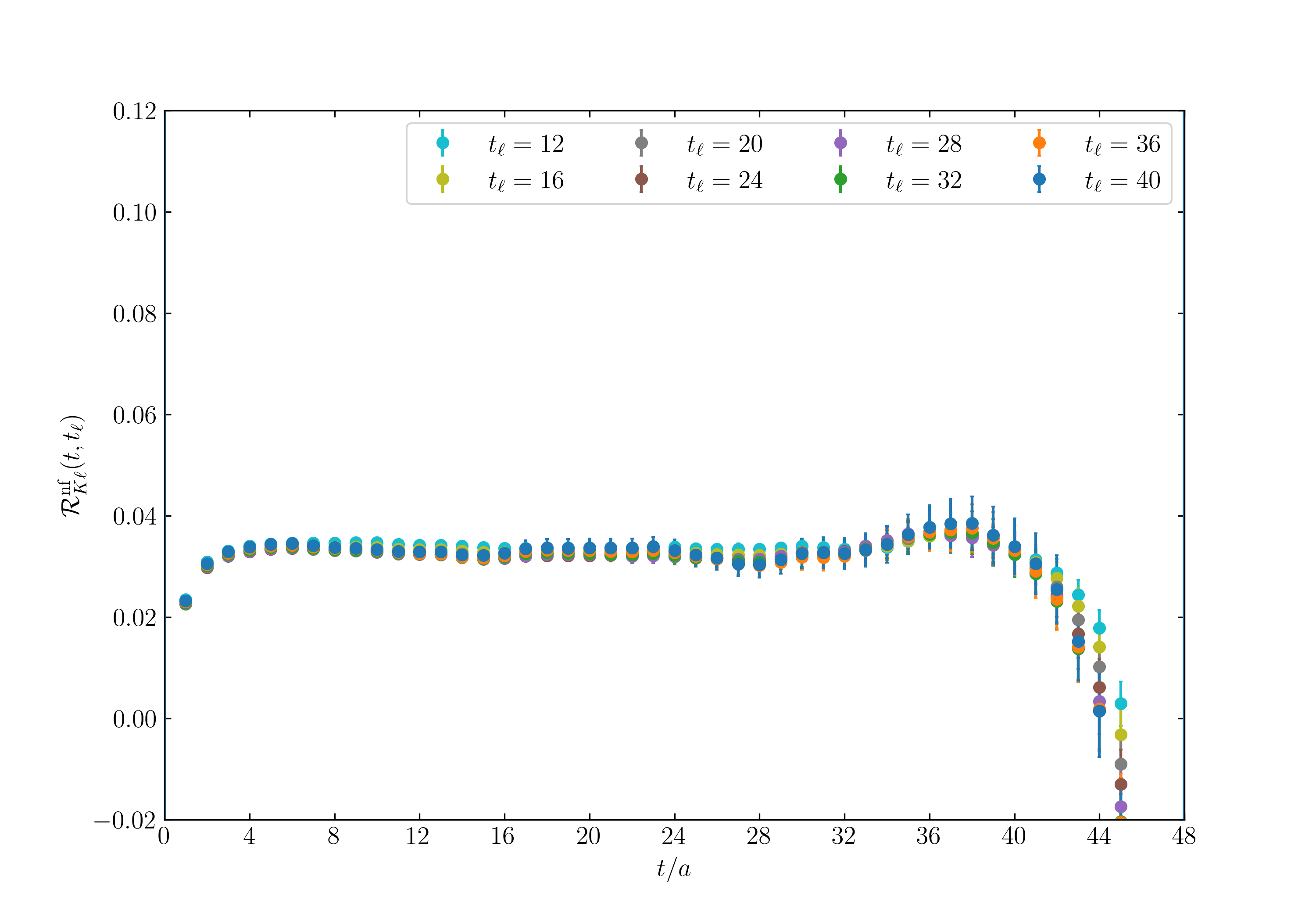} 
        \caption{with projection}
    \end{subfigure}
 \caption{Comparison of the kaon non-factorizable correlator $\mathcal{R}_{K\ell}^\mathrm{nf}$ obtained without the use of the projector $\mathcal{P}_{v(\pvec_\ell)}$ (a) and with the backward propagating signal removed (b).}
 \label{fig:proj_nonfact_kaon}
\end{figure}

The effect of using the projector $\mathcal{P}_{v(\pvec_\ell)}$ on the  non-factorizable correlator $\mathcal{R}_{P\ell}^\mathrm{nf}$ defined in~\cref{eq:non_fact_fitansatz} is shown in~\cref{fig:proj_nonfact_pion,fig:proj_nonfact_kaon} for the pion and kaon decay, respectively (computed with zMöbius fermions). We note that the backward signal is drastically suppressed and the dependence on the lepton source-sink separation $t_\ell$ is barely visible for $t\ll T/2$. The use of these projectors makes then a crucial difference in the extraction of a clear signal from the lattice data.

\section{Correlation functions for the \texorpdfstring{$\Omega$}{Omega} baryon}
\label{app:omega}
\begin{figure}[h]
	\centering
	\includegraphics[width=1\textwidth]{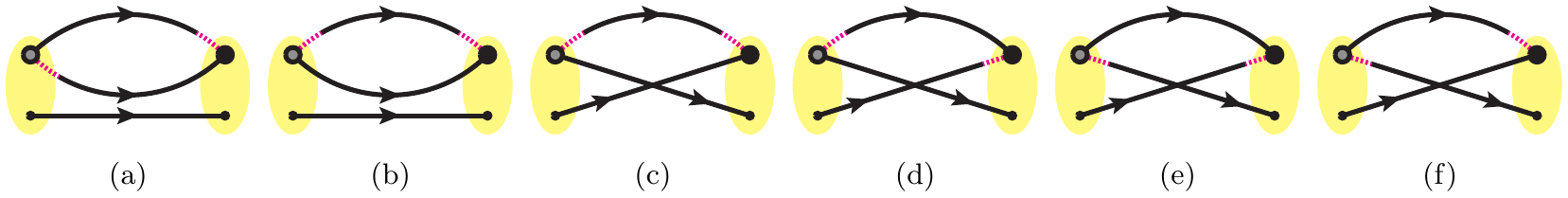}
	\caption{All Feynman diagrams corresponding to the tree-level correlation function $\widetilde{C}_{\Omega\Omega}(t)$. Points connecting two propagators are contractions of a diquark pair, and dashed magenta portions of a propagator indicate contraction with a transposed quark field.}
	\label{fig:omega_tree_contr}
\end{figure}

In this appendix we discuss the construction of the $\Omega^-$ baryon correlation functions used in this work, as well as their spectral representation.
We begin by considering the Wick contractions for the tree-level iso-QCD correlator given in~\cref{eq:Omega_isoQCD_2pt}, of which there are 6 contributions. These are shown diagrammatically in~\cref{fig:omega_tree_contr}, where the points connecting two propagators are contractions of a diquark pair in~\cref{eq:Omega_interpolator}, and dashed magenta portions of a propagator indicate contraction with a transposed quark field. The colour structure of these contractions is not represented in these diagrams.

\begin{figure}[h]
	\centering
	\includegraphics[width=1\textwidth]{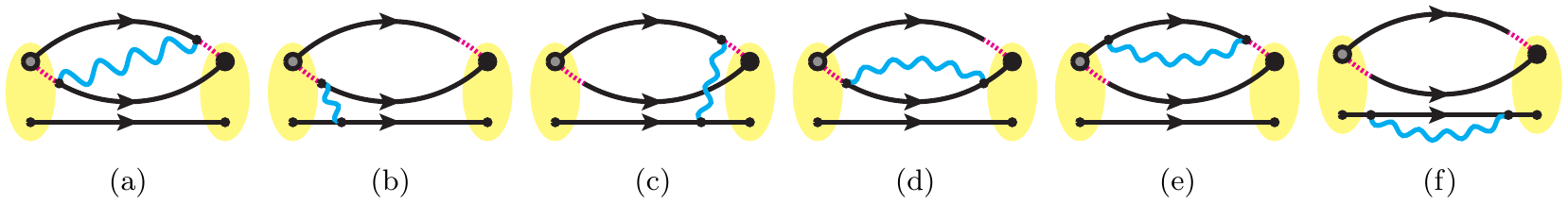}
	\caption{All (connected) Feynman diagrams contributing to $\mathcal{R}_{\Omega\Omega}^\mathrm{e.m.}(t)$ originating from the tree-level contribution shown in~\cref{fig:omega_tree_contr} (a). Similar diagrams exist for the other contributions.}
	\label{fig:omega_qed_contr}
\end{figure}

\begin{figure}[h]
	\centering
	\includegraphics[width=0.5\textwidth]{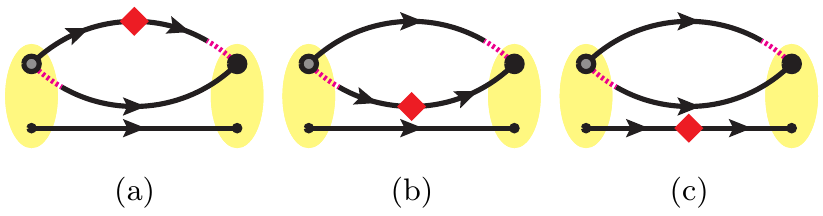}
	\caption{All (connected) Feynman diagrams contributing to $\mathcal{R}_{\Omega\Omega}^\mathrm{\mathcal{S},s}(t)$ originating from the tree-level contribution shown in~\cref{fig:omega_tree_contr} (a). Similar diagrams exist for the other contributions.}
	\label{fig:omega_ms_contr}
\end{figure}

For the QED corrections to this correlator, we require two insertions of the quark-photon interaction $\ii Z_V \sum_x \widebar{s} \slashed{A} s$ (see~\cref{eq:SqS_SqA}) which corresponds to a photon propagator connecting the quark legs, as well as a quark-disconnected contribution that is omitted in this work. Taking for example diagram (a) in~\cref{fig:omega_tree_contr}, the corresponding QED corrections are shown in~\cref{fig:omega_qed_contr} where (a), (b) and (c) are the exchange diagrams contributing to $\delta \widetilde{C}^\mathrm{exch}_{\Omega\Omega}(t)$ and (d), (e) and (f) are the self energy diagrams contributing to $\delta \widetilde{C}^\mathrm{self,s}_{\Omega\Omega}(t)$ in~\cref{eq:Omega_corrections}.
Similarly, quark-mass corrections are given by the insertion of the scalar density $\sum_x \bar{s}s$  (see~\cref{eq:SqS_SqA}). Again taking diagram (a) in~\cref{fig:omega_tree_contr} as an example, the mass corrections are given by the diagrams in~\cref{fig:omega_ms_contr}, as well as a disconnected contribution that is also omitted.


Once the $\Omega^-$ correlators have been constructed, their spectral representation must be evaluated. First note that the interpolator in~\cref{eq:Omega_interpolator} contains a parity projector $P_+$ which causes $\widebar{\psi}_\Omega^\mu$ to create states of positive parity, but also annihilate states of negative parity. Therefore, the ground state spectral representation will have the form
\begin{align}
	C_{\Omega \Omega}(t) = & \frac{a^3}{2} \sum_i \sum_s \left[ \frac{\bra{0} \psi_\Omega^i(0) \ket{\Omega,s} \bra{\Omega,s} \widebar{\psi}_\Omega^i(0) \ket{0}}{2 m_\Omega} \e^{-m_\Omega t} \right. \\
	& \left. \hspace{2cm} - \frac{\bra{\widebar{\Omega},s} \psi_\Omega^i(0) \ket{0} \bra{0} \widebar{\psi}_\Omega^i(0) \ket{\widebar{\Omega},s}}{2 \widebar{\omega}_\Omega} \e^{-\widebar{\omega}_\Omega (T-t)} \right] \,,\nonumber
\end{align}
where the relative sign change between the forward and backward-propagating components comes from assuming anti-periodic boundary conditions in time on the quarks, and therefore also on the baryon fields.
We have additionally distinguished the notation of the rest energy of the negative parity state $\widebar{\omega}_\Omega$ from the positive parity one $m_\Omega$ due to the fact that, at the physical point, the negative parity $\Omega^-$ baryon is not simply a single state in the QCD Fock space, but is instead a resonance in the $\Xi K$ channel and therefore there is a whole spectrum of finite volume multi-particle states contributing in the backward time direction. However, this does not complicate our analysis since we are restricted to early times where the backward propagating contributions are negligible.

The operator-state overlaps have the form
\begin{align}
	\bra{0} \psi_\Omega^\mu(0) \ket{\Omega,s} = Z_\Omega u^\mu_s \hspace{0.8cm} & , \hspace{0.5cm} \bra{\Omega,s} \widebar{\psi}_\Omega^\mu(0) \ket{0} = Z_\Omega^{\ast} \widebar{u}^\mu_s\,, \\
	\bra{\bar{\Omega},s} \psi_\Omega^\mu(0) \ket{0} = \widebar{Z}_\Omega \gamma_5 v^\mu_s \hspace{0.5cm} & , \hspace{0.5cm} \bra{0} \widebar{\psi}_\Omega^\mu(0) \ket{\bar{\Omega},s} = \widebar{Z}_\Omega^{\ast} \widebar{v}^\mu_s \gamma_5\,,
\end{align}
where $u^\mu_s$ and $v^\mu_s$ are the positive and negative energy solutions to the spin-$3/2$ Rarita-Schwinger equation respectively (see e.g.~\cite{Shi-Zhong:2003} for a recent review), and the $\gamma_5$ is present in the negative parity matrix elements to obtain the correct transformation properties. Using Euclidean conventions, the completeness relations for zero momentum spinors with mass $m$ are given by
\begin{align}
	\sum_s u^i_s \widebar{u}^j_s = 2 m P_+ \bigg(\delta^{ij} - \frac{1}{3} \gamma^i \gamma^j\bigg)\ , \qquad \sum_s v^i_s \widebar{v}^j_s = -2 m P_- \bigg(\delta^{ij} - \frac{1}{3} \gamma^i \gamma^j\bigg) \,,
\end{align}
which result in the form of the correlator given in~\cref{eq:Omega_no_trace}.
The spectral representation of the QED and quark mass corrections to this tree-level correlator are simply found by expanding the two parameters $Z_\Omega$ and $m_\Omega$ to first order in the respective isospin breaking parameter.

\begin{figure}[h]
	\centering
	\includegraphics[width=0.5\textwidth]{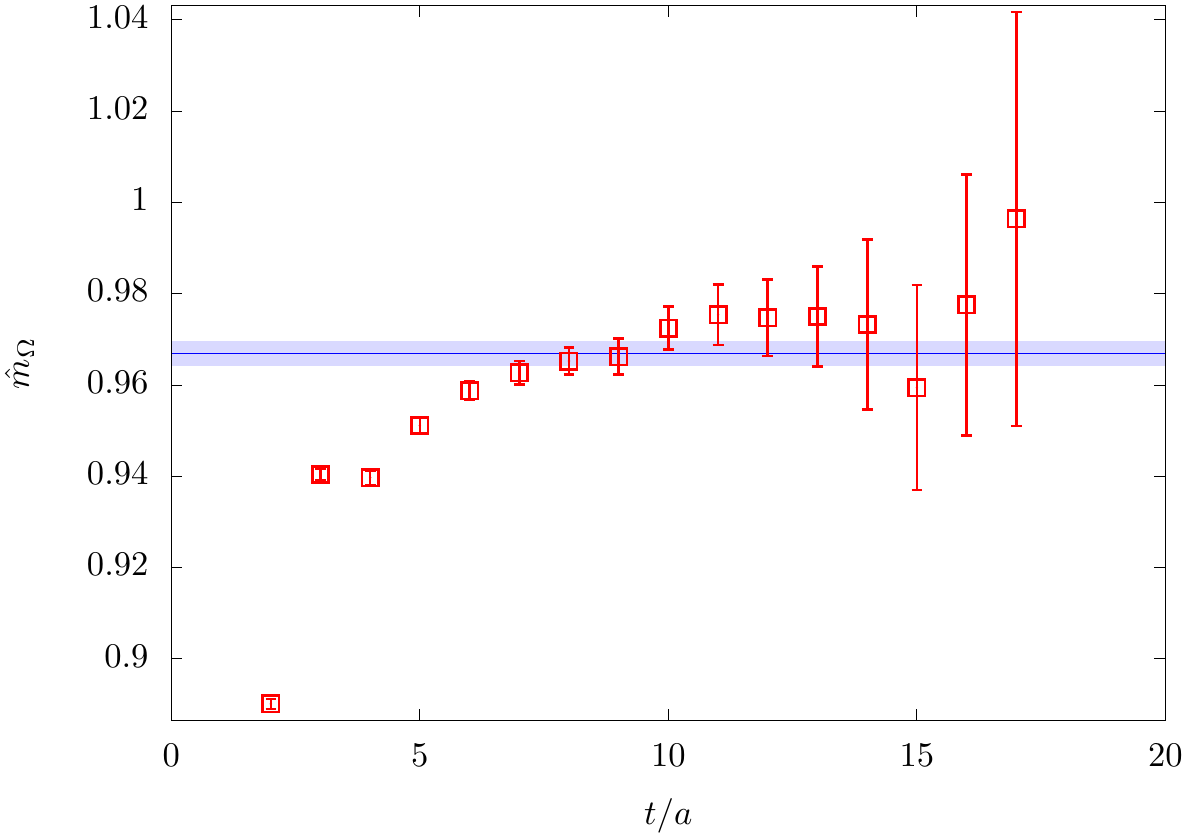}
    \caption{Log effective mass of the tree-level $\Omega^-$ baryon (in lattice units) in red and the fit result of the mass parameter in blue.}
	\label{fig:omega_treelevel_effmass}
\end{figure}

\begin{figure}[h]
	\centering
	\begin{subfigure}[b]{0.48\textwidth}
		\includegraphics[width=0.95\textwidth]{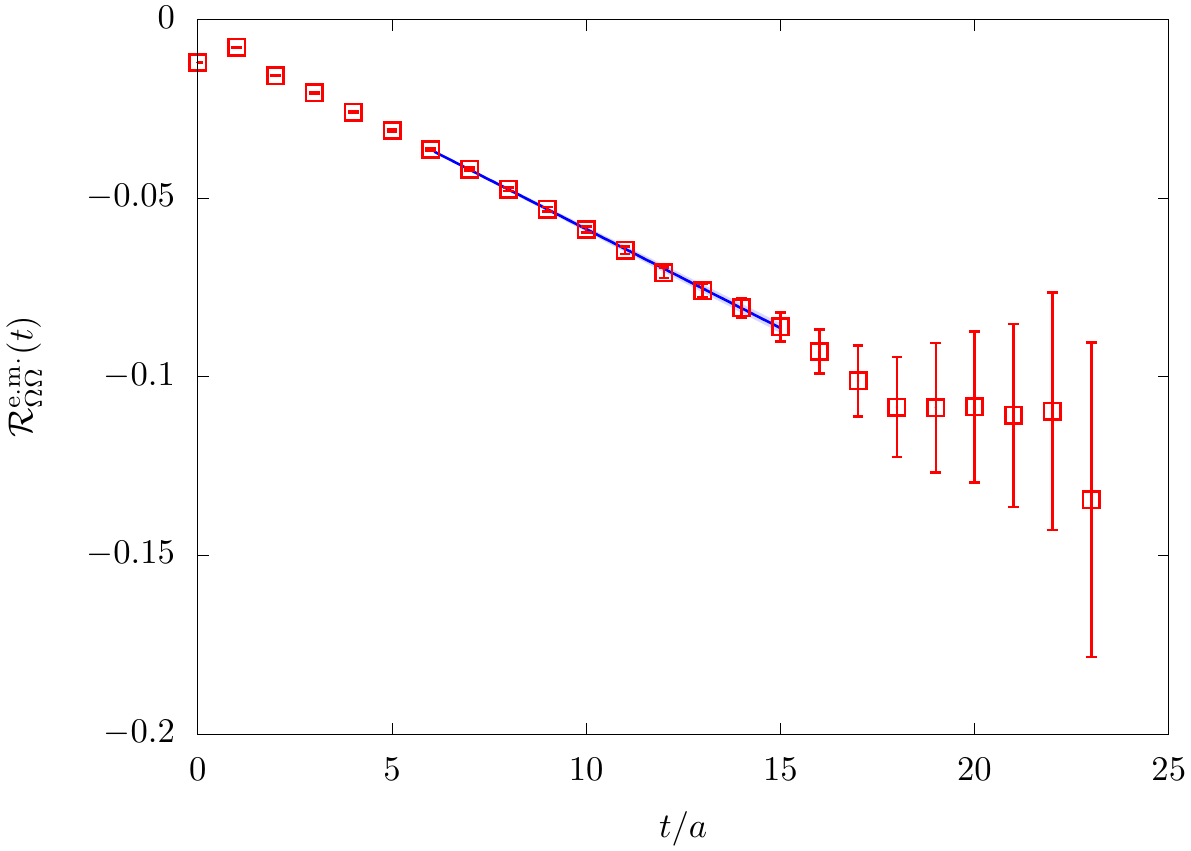} 
		\caption{QED correlator ratio fit}
	\end{subfigure}
	\begin{subfigure}[b]{0.48\textwidth}
		\includegraphics[width=0.95\textwidth]{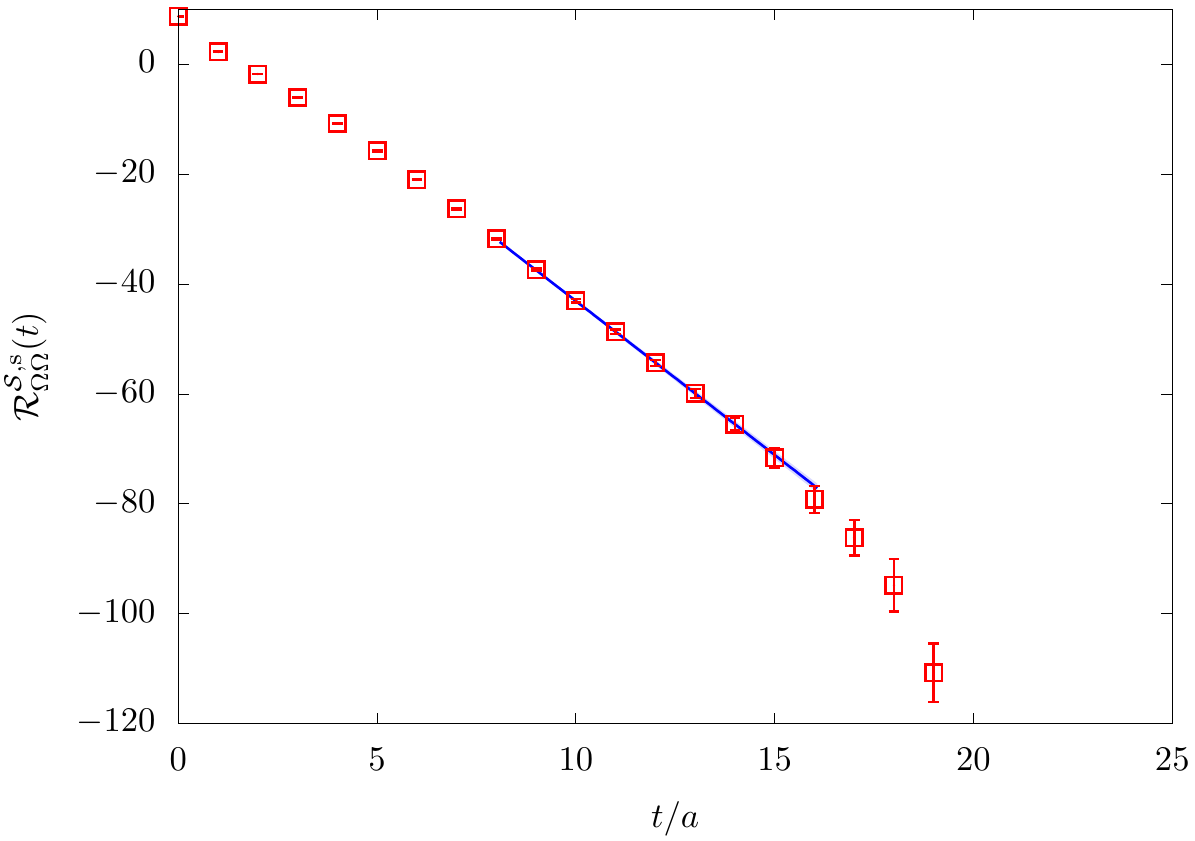} 
		\caption{$m_\rms$ correlator ratio fit}
	\end{subfigure}
	\caption{Ratios of $\Omega^-$ baryon QED corrections (a) and quark mass corrections (b) to the tree-level correlator, $\mathcal{R}_{\Omega\Omega}^\mathrm{e.m.}(t)$ and $\mathcal{R}_{\Omega\Omega}^\mathrm{\mathcal{S},s}(t)$ respectively, in red and the fit to the data in blue.}
	\label{fig:omega_ratios}
\end{figure}

Figure~\ref{fig:omega_treelevel_effmass} shows the log effective mass of the tree-level correlator $\widetilde{C}_{\Omega \Omega}(t)$ along with the fit result of the iso-QCD mass. It should be noted that the fit was performed to the correlator and not directly to the effective mass. The value obtained for the mass of the $\Omega^-$ baryon, $\hat{m}_\Omega=0.967\,(3)$ for the best fit shown in~\cref{fig:omega_treelevel_effmass}, is in agreement with that obtained in ref.~\cite{RBC:2014ntl} using the same gauge ensemble.
Figure~\ref{fig:omega_ratios} shows the ratio of the QED and $m_\rms$ corrections of the $\Omega^-$ correlator to the tree-level result, $\mathcal{R}_{\Omega\Omega}^\mathrm{e.m.}(t)$ and $\mathcal{R}_{\Omega\Omega}^\mathrm{\mathcal{S},s}(t)$ respectively. Included is the fit to this ratio using the linear fit model given in~\cref{eq:Omega_ratio}.

\section{Determining best fits with a genetic algorithm}
\label{app:GA}

In this appendix, we discuss in detail the setup of the genetic algorithm (GA) used in the data
analysis described in~\cref{sec:num_strategy} to select best correlator fits. After summarising the
purpose of a GA, we describe the genetic operators used, and we present the GA hyperparameters used
to produce the factorisable analysis fit results in~\cref{sec:numeval}. 

GAs form a class of global optimizers which stochastically evolve a set of candidate solutions
toward ones which maximize/minimize a given objective function. The evolution process is inspired by
natural selection in biological systems by proposing new solutions attempting to combine best
features from a previous generation of solutions. Because of this analogy, the set of candidate
solutions is generally referred as the \emph{population}, and the state of the population at a given
iteration of the algorithm is called a \emph{generation}. The step between one generation to the
next is done via genetics-inspired operators called \emph{crossover} and \emph{mutation} operators.
The crossover operator aims at producing a better solution to the optimisation problem by combining
two members of the current generation. The mutation operator make arbitrary random changes to
members of the current generation, increasing the space of solutions explored by the algorithm.

In the context of the factorisable analyses, a population member is a vector
$\boldsymbol{\tau}=(\tau^{(1)},\dots,\tau^{(n_\mathrm{corr})})$ where each component is a fit
interval $\tau^{(j)}=[t_\mathrm{min}^{(j)},t_\mathrm{max}^{(j)}]$ for the $j$-th~correlator, and
$n_\mathrm{corr}$ is the total number of correlators to fit. The objective function here is the AIC
weight introduced in~\cref{sec:num_strategy}, which we aim at maximising. The space of all possible
fits is finite, although it contains a very large number of elements. However, this finiteness
guarantees that at least one solution to the optimisation problem exists. A high-level description
of the algorithm is as follows: 
\begin{enumerate}
    \item in each generation, begin with an initial population of $\{\boldsymbol{\tau}_k\}$ with
    $P_0$ elements;
    \item evolve $\{\boldsymbol{\tau}_k\}$ with genetic operators to produce $n_\mathrm{off}$ new
    elements (called \emph{offspring});
    \item compute the AIC for all population members through $\chi^2$ minimization as described
    in~\cref{sec:num_strategy};
    \item choose among $\{\boldsymbol{\tau}_k\}$ the best $P_0$ fit ranges with the largest AIC
    weights and discard all other elements;
    \item repeat steps 2-4 until a termination condition is satisfied.
\end{enumerate}

Let us now introduce the genetic operators. Consider two candidate fit intervals in the initial population, $\boldsymbol{\tau}_k$ and $\boldsymbol{\tau}_{k'}$. The crossover operator, $X$, generates a new $\boldsymbol{\tau}$ with fit intervals from either of the parent members based on random numbers $0\leq p_j\leq 1$ for $1\leq j \leq n_\mathrm{corr}$ That is, 
\begin{equation}
    \boldsymbol{\tau}_{k''} \equiv X\big(\boldsymbol{\tau}_{k},\boldsymbol{\tau}_{k'}\big) = \left\{X\big(\tau^{(1)}_{k},\tau^{(1)}_{k'}\big),\dots,X\big(\tau^{(n_\mathrm{corr})}_{k},\tau^{(n_\mathrm{corr})}_{k'}\big)\right\},
\end{equation}
where
\begin{equation}
    X\big(\tau^{(j)}_k,\tau^{(j)}_{k'}\big) = 
    \begin{cases}
        \tau^{(j)}_k    \quad &\text{if}\quad p_j<0.5, \\
        \tau^{(j)}_{k'} \quad &\text{otherwise}.
    \end{cases}
\end{equation}
This is repeated until one obtains a population size of $P > P_0$. The mutation operator, $M$, then mutates the population at a given rate $m$. That is, for some randomly drawn value of $p$ ($0\leq p \leq 1$), one has
\begin{equation}
    M\big(\boldsymbol{\tau}_k\big)=
    \begin{cases}
    \boldsymbol{\tau}_{P+1},\quad&\text{if}\quad p < m\\        
    \boldsymbol{\tau}_k\quad&\text{otherwise},
    \end{cases}
    \label{eq:mutation-operator}
\end{equation}
with
\begin{equation}
    \boldsymbol{\tau}_{P+1} \equiv \left\{\tau^{(1)}_{P+1},\dots,M\big(\tau^{(j)}_{P+1}\big),\dots,\tau^{(n_\mathrm{corr})}_{P+1}\right\}
\end{equation}
where the index $j$ ($1\leq j \leq n_\mathrm{corr}$), is also randomly drawn. The mutation $M\big(\tau^{(j)}_{P+1}\big)$ is a fit interval where either $t_\mathrm{min}$ or $t_\mathrm{max}$ or both have been modified randomly. 

With the operators defined, we can discuss the GA parameters used for this work. To begin, the free parameters in a GA are: the size of the initial population, $P_0$; the crossover rate, which is parametrised in this work by the population size after crossover $P$; the mutation rate $m$; the weight function to optimize $w$; the maximum number of generations $G_\mathrm{max}$ and the termination condition. We studied three GA setups to check the validity of our fit conclusions. These are summarised in~\cref{tab:GA-setups}.
First, let \begin{equation}
    \bar{w}_N = \frac{1}{N}\sum^N_{i=1} w_i
    \label{eq: aic average}
\end{equation}
be the average of the top $N$ weights in each generation. In all three setups, we aim to maximize
the AIC in each GA run, which terminates when the average of the $N=5$ top fit, $\bar{w}_5$, does
not improve over 1000 successive generations. If this condition cannot be satisfied, we impose a
cut-off when a GA run exceeds $G_\mathrm{max}$ generations. In practice, however, none of the runs
hit this cut-off limit. To accelerate the GA in its exploration of the $\boldsymbol{\tau}$-space, we
cache all fit results during the process.

\begin{figure}[t]
    \centering
    \begin{minipage}{0.5\textwidth}
        \centering
        \includegraphics[width=1\textwidth]{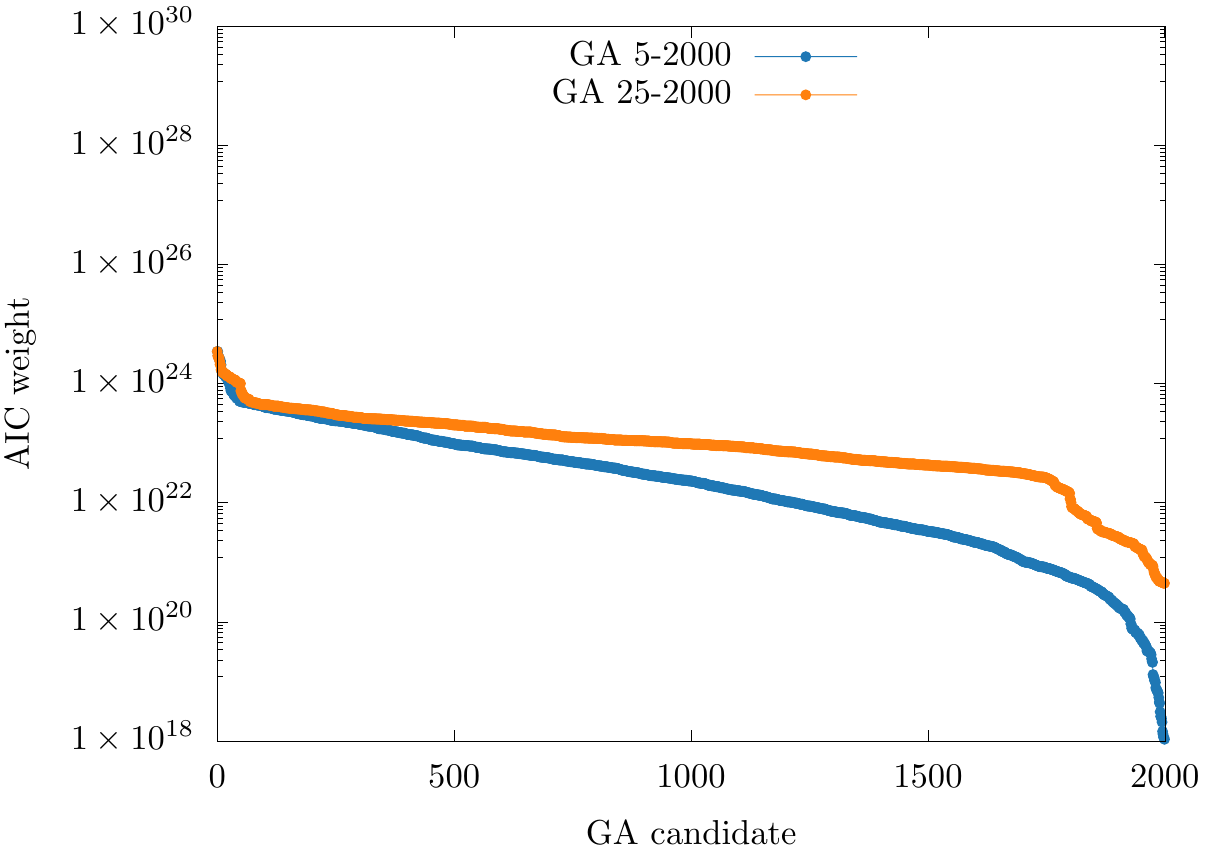} 
    \end{minipage}\hfill
    \begin{minipage}{0.5\textwidth}
        \centering
        \includegraphics[width=1\textwidth]{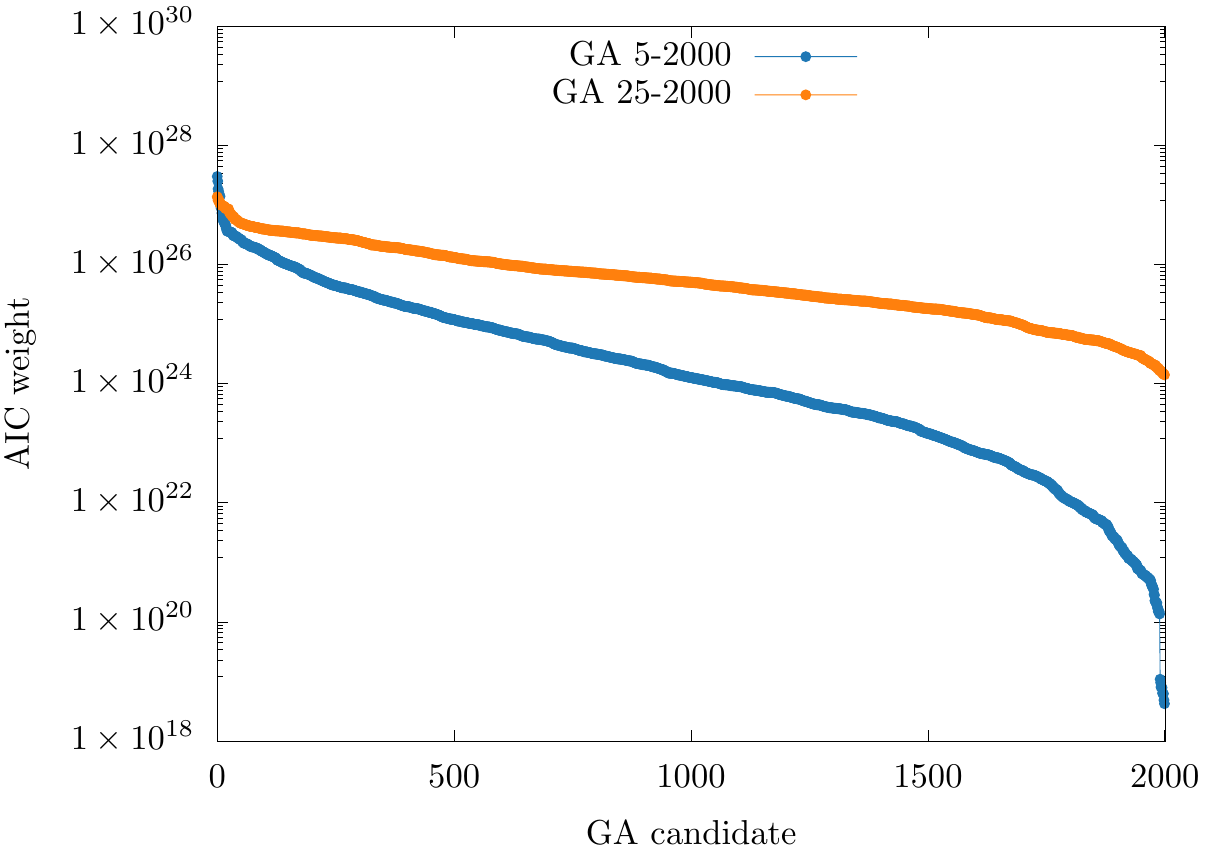} 
    \end{minipage}
    \caption{Comparison of the 2000 GA candidates between two setups GA 5-2000 and GA 25-2000, sorted in descending order of AIC weights, for $\pi$ (left) and $K$ (right) meson correlator analysis.}
    \label{fig:GA5-vs-GA25-2k}
\end{figure}

Let the label `GA $X$-$Y$' refer to a GA setup with a population size $P_0=X$ which has been run
multiple times until obtaining a total of $Y$ candidates. Across the different runs are varied the
initial condition of the algorithm and the random number sequence used in the genetic operators. We
consider $3$ different setups GA 5-2000, GA 25-2000, and GA 25-5000, summarised
in~\cref{tab:GA-setups}. Additionally, \cref{fig:GA5-vs-GA25-2k} compare the AICs of the 2000
outcomes, sorted by their weights in descending order. It is worth commenting on two features. First,
the best AIC fits in both setups are very similar, demonstrating some level of independence between
the optimal solution found and the hyperparameters of the GA. Beyond that, it is clear that the
range of AIC weights is narrower in GA 25-2000 than GA 5-2000 for both pion and kaon. This suggests
that a population size of $P_0=25$ allows one to discover more optimal solutions than $P_0=5$ for a
given target number of candidates. This is expected as the algorithm will try more candidates at
each generation. However, this also suggests that the $P_0=5$ set of runs is not saturating its
exploration of the best AIC fit space. To address this, the GA 25-5000 setup was designed to check
that such saturation was achieved for GA 25-2000. In~\cref{tab:average-ancestor-per-run}, we give
the average number of distinct fits explored by each setup for each analysis. As we can see,
considerably increasing the number of GA runs in the $P_0=25$ case does not lead to a significant
volume of new fits tried, meaning that the additional runs were to a large extent redundant in terms of optimal solutions found. 

Ultimately, as it is generally the case with GAs in this type of context, it is not possible to
demonstrate with absolute certainty that the GA found the best fits without knowing the exact
solution to the problem. However, we remind the reader that the main aim here is to establish a representative
spread in our final result for $\dRKPi$ in order to assign a systematic error related to the
selection of fit ranges. To check the stability of our systematic error under variations of the GA setup,
we take the top 5 fits of both GA 25-2000 and GA 25-5000 setups and generate an AIC-weighted
histogram of $\delta R_{K\pi}^\mathrm{latt}$ for each setup. These are shown
in~\cref{fig:GA25-2k-vs-5k}, along with their median and the fit systematic errors as defined
in~\cref{sec:model_uncertainties}. We see that they lead to very similar conclusions in terms of
median and systematic spread, and we consider that as a compelling evidence that the GA is converging on a set of fit candidates which is representative enough to estimate the fit range selection systematic uncertainty. Finally, the GA 25-5000 setup was used to produce the final result of this work.

\begin{table}[t]
    \centering
    \begin{tabular}{| c | c | c | c |} 
    \hline
    & GA 5-2000  & GA 25-2000 & GA 25-5000 \\ \hline \hline
    number of runs      & 400    & 80    & 200 \\ \hline
    \hline
    $P_0$                & 5     & 25    & 25 \\ \hline
    $P$                  & 20    & 100   & 100 \\ \hline
    $m$                  & 0.3   & 0.3   & 0.3 \\ \hline
    $G_\mathrm{max}$            & 25000 & 10000 & 10000 \\ \hline
    termination cond. n. 1 & \multicolumn{3}{c|}{$\bar{w}_5$ unchanged for 1000 generations } \\ \hline
    termination cond. n. 2 & \multicolumn{3}{c|}{GA exceeds $G_\mathrm{max}$ } \\ \hline
    \hline
    total GA candidates & 2000 & 2000 & 5000 \\ \hline
    \end{tabular}
    \caption{Table of three different GA setups used in this work. All setups maximize the AIC weight. $\bar{w}_5$ is defined in the text below~\cref{eq: aic average}.}
    \label{tab:GA-setups}
\end{table}
\begin{figure}[t]
    \centering
    \begin{minipage}{0.5\textwidth}
        \centering
        \includegraphics[width=1\textwidth]{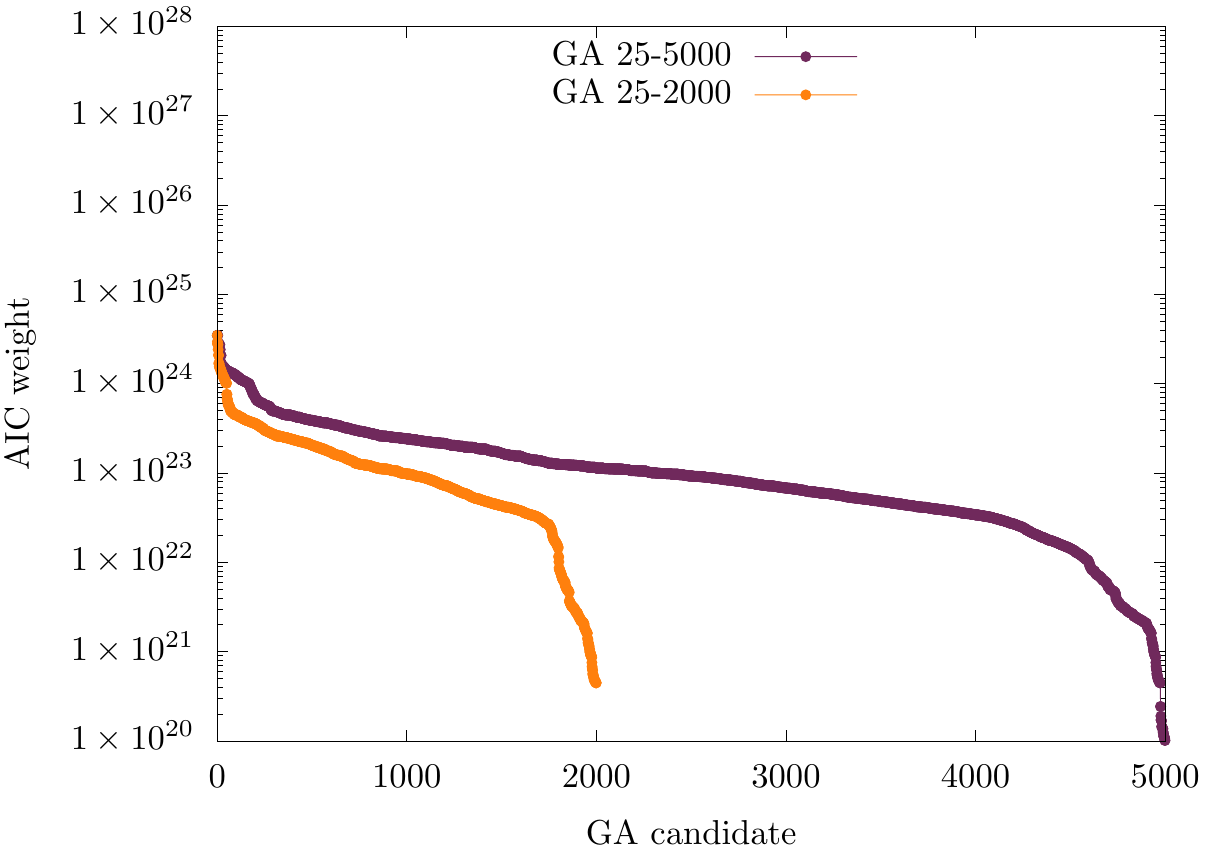}
    \end{minipage}\hfill
    \begin{minipage}{0.5\textwidth}
        \centering
        \includegraphics[width=1\textwidth]{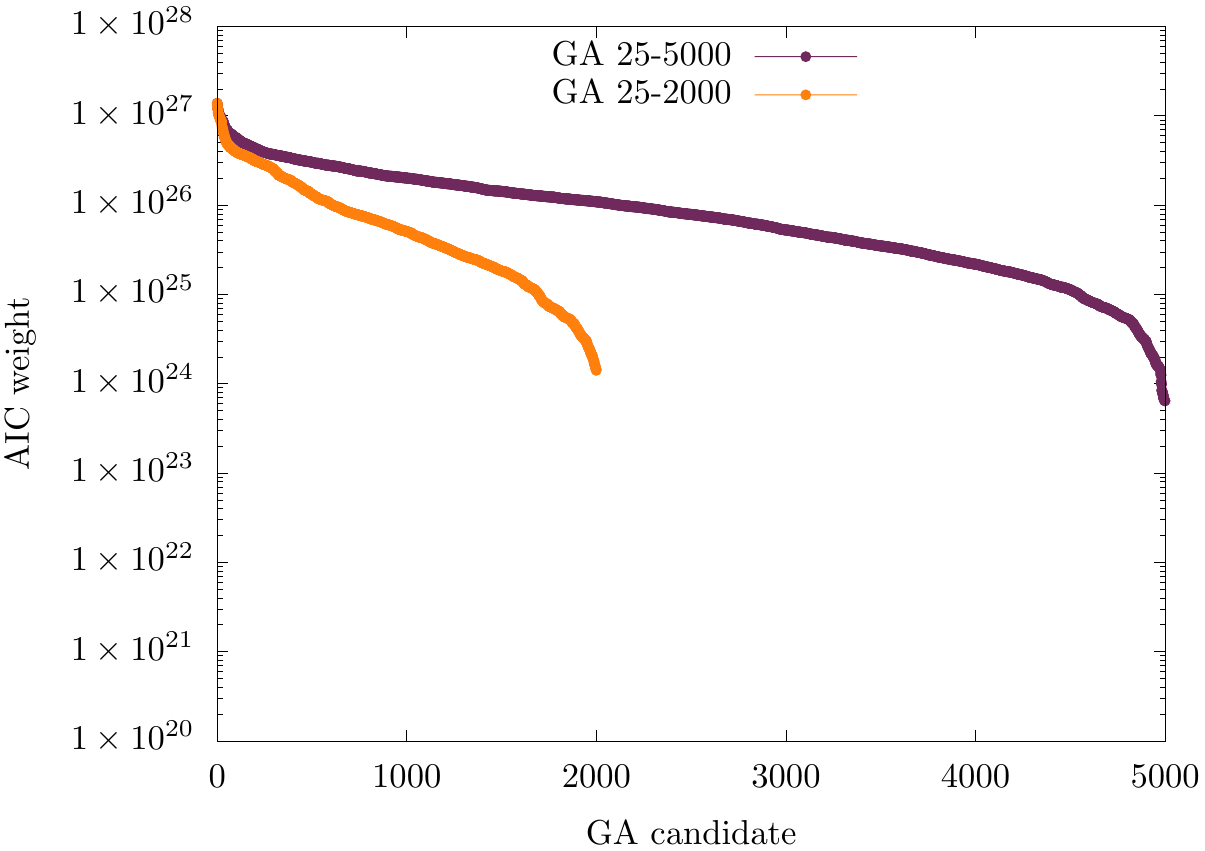}
    \end{minipage}
    \caption{Comparison of the GA candidates between setups GA 25-5000 and GA 25-2000, sorted in descending order of AIC weights, for the $\pi$ (left) and $K$ (right) meson correlator analysis.}
    \label{fig:GA25-5k-2k}
\end{figure}
\begin{table}[t]
    \centering
    \begin{tabular}{| c | c | c | c |} 
    \hline
    Analysis & GA 5-2000 & GA 25-2000 & GA 25-5000\\ \hline \hline
    $\pi$ & 4085.915  & 32854.5375 & 33767.365  \\ \hline
    $K$   & 4726.1425 & 34430.625  & 33975.085 \\ \hline
    \end{tabular}
    \caption{The average total number of fits tried per GA setup for each meson correlator analysis.}
    \label{tab:average-ancestor-per-run}
\end{table}
\begin{figure}[t]
\centering
\includegraphics[width=0.75\textwidth]{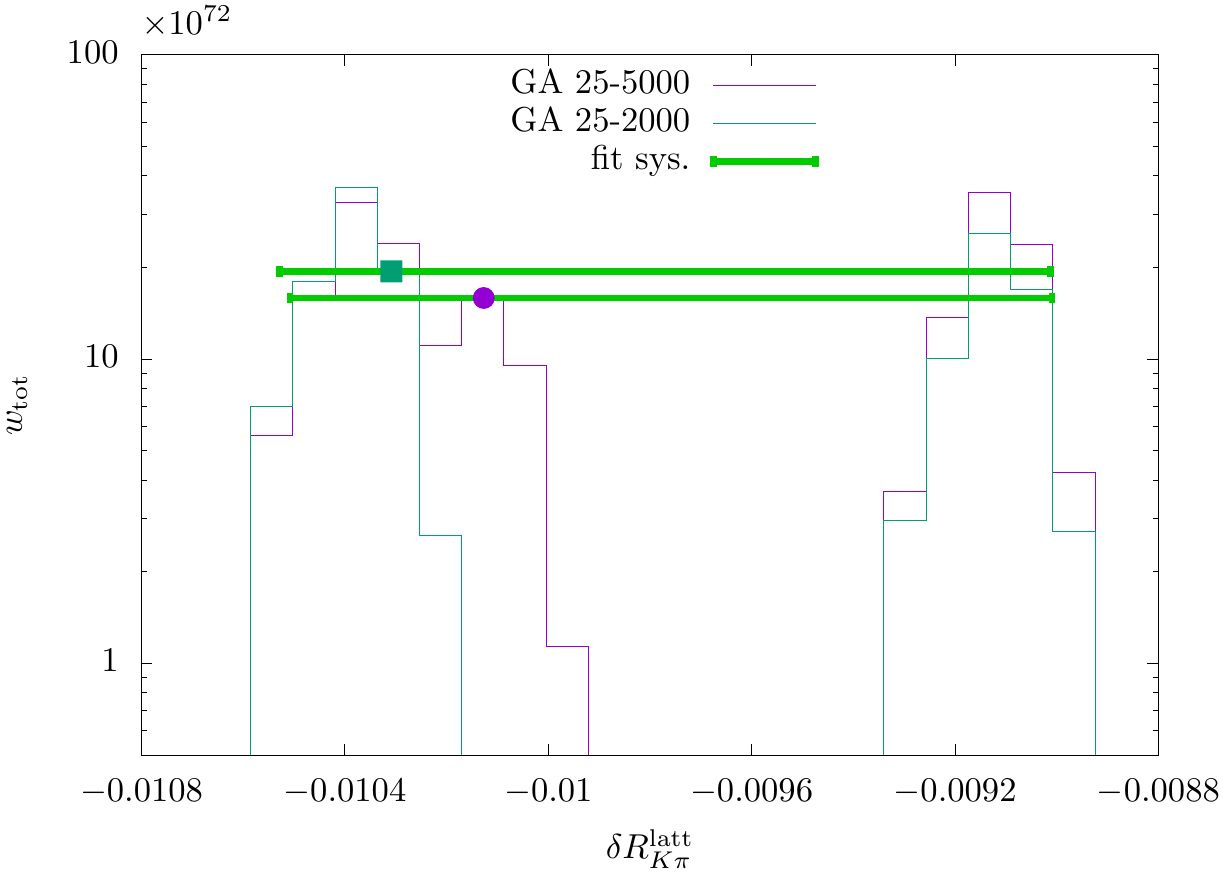}
\caption{AIC-weighted histogram generated from top 5 fits of each analysis as described in~\cref{sec:model_uncertainties}. The median of each histogram and the fit systematics are superimposed on the histogram.}
\label{fig:GA25-2k-vs-5k}
\end{figure}

\newpage
\bibliography{pl2}

\end{document}